\documentclass[12pt]{article}
\usepackage{graphicx} 
\usepackage{feynmf}
\usepackage{amssymb}
\usepackage{amsmath}
\usepackage{xcolor}
\usepackage{dsfont}
\usepackage{cite}
\usepackage{xr}
\newcommand{\cals}{\text{$\cal S$}}

\newcommand{\calf}{\mbox{${\cal F}$}}

\newcommand{\dilog}{\mbox{Li}_2}

\newcommand{\bbar}{\overline{b}}
\newcommand{\detg}{\det{(G)}}
\newcommand{\dets}{\det{(\cals)}}
\newcommand{\sign}{\mbox{sign}}

\newcommand{\baru}{\bar{u}}

\newcommand{\tD}{\widetilde{D}}
\newcommand{\bbj}[2]{\overline{b}_{#1}^{\{#2\}}}
\newcommand{\bbjsq}[2]{\overline{b}_{#1}^{\{#2\} 2}}
\newcommand{\detgj}[1]{\det{(G^{\{#1\}})}}
\newcommand{\detsj}[1]{\det{(\cals^{\{#1\}})}}
\newcommand{\myref}[1]{(\ref{#1})}

\renewcommand\Re{\operatorname{Re}}
\renewcommand\Im{\operatorname{Im}}

\newcommand{\contij}{\mbox{$\widehat{(0,1)}_{i,j}$}}
\newsavebox{\Gammap}
\newsavebox{\Gammam}
\numberwithin{equation}{section}

\begin{document}

\setlength{\unitlength}{1mm}
\begin{fmffile}{samplepics}

\begin{titlepage}

\vspace{1.cm}

\long\def\symbolfootnote[#1]#2{\begingroup%
\def\thefootnote{\fnsymbol{footnote}}\footnote[#1]{#2}\endgroup} 

\begin{center}

{\large \bf A novel approach to \\
the computation of one-loop three- and four-point functions. \\
\vspace{0.1cm}
II - The complex mass case}\\[2cm]

{\large  J.~Ph.~Guillet$^{a}$, E.~Pilon$^{a}$, 
Y.~Shimizu$^{b}$ and M. S. Zidi$^{c}$ } \\[.5cm]

\normalsize
{$^{a}$ Univ. Grenoble Alpes, Univ. Savoie Mont Blanc, CNRS, LAPTH, F-74000 Annecy, France}\\
{$^{b}$ KEK, Oho 1-1, Tsukuba, Ibaraki 305-0801, Japan\symbolfootnote[2]{Y. Shimizu passed away during the completion of this series of articles.}}\\
{$^{c}$ LPTh, Universit\'e de Jijel, B.P. 98 Ouled-Aissa, 18000 Jijel, Alg\'erie}\\
      
\today
\end{center}

\vspace{2cm}

\begin{abstract} 
\noindent
This article is the second of a series of three presenting an alternative 
method to compute the one-loop scalar integrals. It extends the results of the first 
article to general complex masses. Let us remind the main features enjoyed by this method.
It directly proceeds in terms of the quantities driving algebraic 
reduction methods. It applies to the four-point functions in the same way as 
to the three-point functions. Lastly, it extends to kinematics more general than the one of
physical e.g. collider processes relevant at one loop.
\end{abstract}

\vspace{1cm}

\begin{flushright}
LAPTH-44/18\\
\end{flushright}

\vspace{2cm}

\end{titlepage}
\savebox{\Gammap}[7mm][r]{%
  \begin{fmfgraph*}(7,7)
  \fmfset{arrow_len}{3mm}
  \fmfset{arrow_ang}{10}
  \fmfipair{o,xm,xp,ym,yp}
  \fmfipath{c[]}
  \fmfipair{r[]}
  \fmfiequ{o}{(.1w,-.2h)}
  \fmfiequ{xm}{(0,-.2h)}
  \fmfiequ{ym}{(.1w,0)}
  \fmfiequ{r3}{(.35w,-.2h)}
  \fmfiset{c1}{fullcircle scaled 1.5w shifted o}
  \fmfi{fermion}{subpath (length(c1)/4,0) of c1}
  \fmfiequ{r1}{point 0 of c1}
  \fmfiequ{r2}{point length(c1)/4 of c1}
  \fmfi{fermion}{r1--r3}
  \fmfi{fermion}{o--r2}
  \fmfiv{l={\tiny 0},l.a=-120,l.d=0.05w}{o}
  \fmfiv{l={\tiny 1},l.a=-70,l.d=0.06w}{r3}
\end{fmfgraph*}}

\savebox{\Gammam}[7mm][r]{%
  \begin{fmfgraph*}(7,7)
  \fmfset{arrow_len}{3mm}
  \fmfset{arrow_ang}{10}
  \fmfipair{o,xm,xp,ym,yp}
  \fmfipath{c[]}
  \fmfipair{r[]}
  \fmfiequ{o}{(.1w,.5h)}
  \fmfiequ{xm}{(0,.5h)}
  \fmfiequ{ym}{(.1w,0)}
  \fmfiequ{r3}{(.35w,.5h)}
  \fmfiset{c1}{fullcircle scaled 1.5w shifted o}
  \fmfiequ{r1}{point 0 of c1}
  \fmfiequ{r2}{point 3length(c1)/4 of c1}
  \fmfi{fermion}{subpath (3length(c1)/4,length(c1)) of c1}
  \fmfi{fermion}{r1--r3}
  \fmfi{fermion}{o--r2}
  \fmfiv{l={\tiny 0},l.a=90,l.d=0.05w}{o}
  \fmfiv{l={\tiny 1},l.a=70,l.d=0.06w}{r3}
\end{fmfgraph*}}

\newpage

\section{Introduction}\label{intro}

In ref. \cite{letter}, we show that any two-loop scalar function can be written as a two dimensional 
integral of a ``generalised'' one-loop function weighted by a rational function of the two integration 
variables, the present articles addresses the computation of these ``generalised'' one-loop 
functions\footnote{We consider, in this article, only the generalisation concerning the underlying 
kinematics not the one about the integration domain spanned by the Feynman parameters 
c.f. \cite{letter}.} for three- and four-points in the case of general complex masses.
This article is the second of a triptych. 
The first one \cite{paper1} presents 
a method exploiting a Stokes-type identity to compute ``generalised'' one-loop three- and 
four-point scalar integrals in the real mass case in a four dimension spacetime. 
 
\vspace{0.3cm}

\noindent
We refer to
the introduction of \cite{paper1} for more details on the motivation and on the general features of the method.  
Let us stress an important difference with respect to the real mass case.
In the latter,
the imaginary part of the ratio of the kinematical $\cals$ matrix determinant 
over the Gram matrix one (or the various determinants of the pinched matrices 
formed from $\cals$ over their related Gram matrix determinants) was always positive and related to
the Feynman prescription coming from the propagators. 
In the complex mass case, the signs of the imaginary part of these ratios depend on the kinematics
and may be positive or negative.
Despite this difference, the method, developed in \cite{paper1}, to perform analytical integration 
over the remaining parameters after the application of the Stokes-like identity
can be applied in a systematic way for the various cases with slight adaptations.
When expressed in terms of contour integrals, the different cases share a common structure supplemented by logarithmic terms
which are case dependent.
 


\vspace{0.3cm}

\noindent
A couple 
of interesting features compared to the methods of \cite{Guillet:2013mta} and \cite{tHooft:1978jhc,Nhung:2009pm,Denner:1991qq,Denner:2010tr} 
enjoyed by the method presented was stressed in \cite{paper1}.
Namely, it directly proceeds in terms of the algebraic quantities $\dets$, 
$\detg$, $b_i$ etc.
and it also applies to
kinematical configurations beyond those  relevant for collider processes at the
one-loop order
\footnote{We acknowledge that the result for the four-point function with complex masses given in ref. \cite{Denner:2010tr} also holds for kinematics beyond one-loop.}
This novel method suffers from the same drawbacks as those mentioned in \cite{paper1}, namely, an inherent 
increase of the number of dilogarithms compared to the 't Hooft-Veltman 
results or the Denner-Dittmaier ones. This point deserves further discussions but there exists ways to reduce this number.

\vspace{0.3cm}

\noindent
The outline of this article follows closely the one of our preceding article \cite{paper1}.
We start by considering the three-point function $I_{3}^{4}$ 
with complex internal masses considered as a warm-up in sec. \ref{sectthreepoint}.
After having reminded the necessary notations and definitions, we consider the two variants 
of the method presented in the real mass case, namely the ``direct way'' and the ``indirect way''.
The formulas for these two variants obtained in \cite{paper1} still holds for the case of 
complex masses and so their derivations will not be reproduced in this article.
Nevertheless, the equivalence between these two ways is more complicated to show and will be discussed in detail.
We end this section by commenting on the apparent doubling of 
dilogarithms, already there in the real mass case. 
We then apply the 
``indirect way'' to the four-point function with internal masses all complex
in sec. \ref{sectfourpoint}. It results from this application eight formulas
depending on the sign of the imaginary parts of the determinants of the $\cals$ matrix
as well as its pinched ones.
Various appendices gather a number of utilities: tools, proofs of steps, etc. 
we removed them from the main text to facilitate its reading but we 
consider them useful to supply. Accordingly, 
appendix~\ref{appendJ} extends the companion appendix~\ref{P1-appendJ} of \cite{paper1}. 
The so called ``second type'' integral is computed for the case where the complex numbers involved have a finite imaginary part. 
Appendix~\ref{appF} is closely related to the appendix~\ref{P1-appF} of \cite{paper1}. 
It adds to the latter the case where the parameters of the integrand are true complex 
numbers and also the cases where the integral has different bounds required for the treatment of complex masses. 
Appendix~\ref{ImofdetS} widens the discussion started in the appendix~\ref{P1-ImofdetS} of \cite{paper1} 
about the sign of the imaginary part of $\dets$ for general complex masses. 
Lastly, appendix~\ref{cut} gives the conditions on the two complex numbers $A$ and $B$ for 
which one of the cuts of the logarithm $\ln \left( A \, z^2 + B \right)$ crosses the real 
segment $[0,1]$ when $z$ spans the complex plane.

\section{Warm-up: $I_{3}^{4}$}\label{sectthreepoint}

In the previous companion paper \cite{paper1}, we show how to compute the 
three point function using Stokes-type identity (cf.\ section~\ref{P1-sectthreepoint} 
of ref.\ \cite{paper1}) for real mass case. We want to extend these results 
for complex masses. To facilitate the reading, we recap the notations and 
some necessary definitions.

\vspace{0.3cm}

\noindent
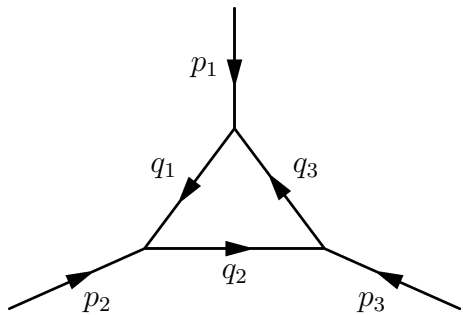
\begin{figure}[h]
\centering
\parbox[c][43mm][t]{80mm}{\begin{fmfgraph*}(60,80)
  \fmfleftn{i}{1} \fmfrightn{o}{1} \fmftopn{t}{1}
  \fmf{fermion,label=$p_1$}{t1,v1}
  \fmf{fermion,label=$p_2$}{i1,v2}
  \fmf{fermion,label=$p_3$}{o1,v3}
  \fmf{fermion,tension=0.5,label=$q_1$}{v1,v2}
  \fmf{fermion,tension=0.5,label=$q_2$}{v2,v3}
  \fmf{fermion,tension=0.5,label=$q_3$,label.side=right}{v3,v1}
\end{fmfgraph*}}
\caption{\footnotesize 
The triangle picturing the one-loop three-point function.}
\label{fig1} 
\end{figure}

\noindent
The usual Feynman integral representation of the three-point function in four dimensions $I_3^4$ is:
\begin{eqnarray}
I^4_3 
& = & 
-\int_0^1 \, \prod_{i=1}^3 \, d z_i \, 
\delta(1- \sum_{i=1}^3 \, z_i) 
\left( 
 - \, \frac{1}{2} \, Z^{\;T} \cdot \cals \cdot Z
\right)^{-1}
\label{eqSTARTINGPOINT3}
\end{eqnarray}
Here $Z$ stands for a column 3-vector whose components are the $z_{i}$, 
$\cals$ is the $3 \times 3$ kinematic matrix associated to the diagram of fig.
\ref{fig1} encoding all the information on the kinematics associated to this diagram by:
\begin{equation}
\cals_{i \, j} = (q_i-q_j)^2 - m_i^2 - m_j^2
\label{eqDEFCALS}
\end{equation}
Each internal line with momentum $q_i$ stands for the propagator of a particle
of mass $m_i$. 
Lastly, the superscript ``$^{T}$" stands for the matrix transpose.
Note that in eq. (\ref{eqSTARTINGPOINT3}) the infinitesimal prescription $- i \, \lambda$,
there in \cite{paper1}, is overcome by the finite 
imaginary parts of the complex masses: it is irrelevant and we drop it in this
article.
Let us single out the subscript value 
$a$ ($ a \in S_3 =\{1,2,3\}$) and write $z_a$
as $1 - \sum_{i \ne a} z_i$. We find:
\begin{eqnarray}
- \, Z^{\;T} \cdot \cals \cdot Z 
& = & 
\sum_{i,j \in S_3 \setminus \{a\}} G_{i\,j}^{(a)} \, z_i \, z_j  - 
2 \sum_{j \in S_3 \setminus \{a\}} V_j^{(a)} \, z_j  -  C^{(a)}
\label{eqVECZT3}
\end{eqnarray}
where the $2 \times 2$ Gram matrix $G^{(a)}$ and the column 2-vector $V^{(a)}$
are defined by
\begin{eqnarray}
G_{i\,j}^{(a)} 
& = & 
-  \, (\cals_{i\,j}-\cals_{a\,j}-\cals_{i\,a}+\cals_{a\,a}), \;\; i,j \neq a
\nonumber
\\
V_j^{(a)} 
& = & 
\;\;\;\;\;  \cals_{a\,j} - \cals_{a \, a}, \;\;  j \neq a
\label{eqVJA3}\\
C^{(a)}& = & 
\;\;\;\;\;  \cals_{a \, a} 
\nonumber
\end{eqnarray}
Labelling $b$ and $c$ the two elements of $S_3 \setminus \{a\}$ with $b < c$, 
the polynomial (\ref{eqVECZT3}) can be written as:
\begin{equation}
D^{(a)}(z_b,z_c) 
=  
X^{(a)\;T} \cdot G^{(a)} \cdot X^{(a)} - 
2 \; V^{(a) \, T} \cdot X^{(a)} - C^{(a)}
\;\; , \;\; 
X^{(a)} =
 \left[
 \begin{array}{c}
   z_b \\
   z_c 
 \end{array}
 \right]
\label{eqDsupa}
\end{equation}
In ref.\ \cite{paper1}, we then applied once the Stokes-type identity 
presented in the appendix~\ref{P1-ap1} of this reference to transform 
an integration over a Feynman parameter into a sum of integrals over 
$[0,\infty[$. The derivation of this transformation 
  is valid also in the complex mass case and will not be reproduced here, we refer the reader to ref.\ \cite{paper1} 
for more details. At the end of this transformation, we could perform 
the integration over the half real line and got the result coined 
``direct way'':
\begin{equation}
I_3^4 
= 
- \, \sum_{i \in S_3} \, 
\frac{\bbar_i}{\detg} \,  \int_0^1
\frac{d x}{D^{\{i\}(l)}(x) + \Delta_2} \, 
\left[ 
 \ln \left( D^{\{i\}(l)}(x) \right)
 - 
 \ln \left( - \, \Delta_2 \right)
\right]
\label{eqI346}
\end{equation}
with $l \equiv 1 + (i \; \text{modulo} \; 3)$ and
where\footnote{The different matrices $G^{(a)}$ have the same determinant represented by $\detg$.}
\begin{align}
\Delta_2 &= \frac{\dets}{\detg}
\label{eqtruc4} \\
\frac{\bbar_i}{\detg} &= \left( \left( G^{(a)} \right)^{-1} \cdot V^{(a)} \right)_i \quad i \ne a \label{eqrecapbbarj} \\
\frac{\bbar_a}{\detg} &= 1 - \sum_{j \in S_3 \setminus \{a\}} \, \left( \left( G^{(a)} \right)^{-1} \cdot V^{(a)} \right)_j
\end{align}
The second degree polynomial $D^{\{i\}(l)}(x)$ is defined with the one-pinched $\cals^{\{i\}}$ matrix as follows:
\begin{equation}
  D^{\{i\}(l)}(x) 
=  
G^{\{i\}(l)} \, x^2 - 2 \, V^{\{i\}(l)} \, x - C^{\{i\}(l)}
\label{eqDsupa1}
\end{equation}
with
\begin{align}
  G^{\{i\}(l)}_{jk} &= - \left( \cals^{\{i\}}_{jk} - \cals^{\{i\}}_{jl} - \cals^{\{i\}}_{lk} + \cals^{\{i\}}_{ll} \right) \quad j,k \ne i,l
  \label{eqrappgmati1} \\
  V^{\{i\}(l)}_j &= \cals^{\{i\}}_{lj} - \cals^{\{i\}}_{ll} \quad j \ne i,l \label{eqrappgmati2} \\
  C^{\{i\}(l)} &= \cals^{\{i\}}_{ll} \label{eqrappgmati3}
\end{align}
Note that in the case of the three-point function, since the set $S_3$ has only three elements, $j$ has to be equal to $k$ in eq.~(\ref{eqrappgmati1}) so the matrix $G^{\{i\}(l)}$ is a $1 \times 1$ matrix and the vector $V^{\{i\}(l)}$ has only one component, hence the notation used in eq.~(\ref{eqDsupa1}).

\vspace{0.3cm}

\noindent
Unfortunately, in the case of
the four-point function we did not succeed in proceeding as simply. 
We have therefore formulated an alternative to the ``direct way'', henceforth 
coined ``indirect''. 
In this formulation, the Stokes-type identity is applied twice and
the three-point function is written as a sum over the 
coefficients $\bbar$ and $\bbj{}{i}$ weighted by a two dimensional integral 
over the first quadrant\footnote{As for the ``direct way'', the derivation presented in \cite{paper1} still holds for complex mass case.}
(see ref.\ \cite{paper1} for more details):
\begin{align}
&I_3^4  = \, \sum_{i \in S_3} \, 
\sum_{j \in S_3 \setminus \{i\}} \, \frac{\bbar_i}{\detg} \, 
\frac{\bbj{j}{i}}{\detgj{i}} \; 
L_3^4 \left( \Delta_2, \Delta_{1}^{\{i\}}, \tD_{ij} \right)
\label{eqI349}
\end{align}
with
\begin{eqnarray}
L_3^4 \left( \Delta_2, \Delta_{1}^{\{i\}}, \tD_{ij} \right)
& = &
\int_0^{+\infty} \frac{d \xi}{(-\Delta_2 +  \, \xi)}
\nonumber\\
&&  \times
 \int_0^{+\infty} \frac{d \rho}{ 
 (- \Delta_1^{\{i\}} +  \, \xi +  \, \rho^2) \,
 (\tD_{ij} + \xi + \rho^2)^{1/2}
}
\label{eqdeflij1}
\end{eqnarray}
and
\begin{align}
  \frac{\bbj{j}{i}}{\detgj{i}} &= - \left( G^{\{i\}(l)} \right)^{-1} \,  V^{\{i\}(l)}, \quad j \ne i,l \label{eqrecapbbarij1} \\
  \frac{\bbj{l}{i}}{\detgj{i}} &= - 1 + \left( G^{\{i\}(l)} \right)^{-1} \,  V^{\{i\}(l)} \label{eqrecapbbarij2}\\
- \, \Delta_1^{\{i\}} 
&=  
\frac{\detsj{i}}{\detgj{i}}
\label{eqtruc3} \\
\widetilde{D}_{ij} &= 2 \, m_l^2, \quad j \in S_3 \setminus \{i,l\} \label{eqrecapdtildeij}
\end{align}

\noindent
Some linear combinations of $\Delta_2$, $\Delta_1^{\{i\}}$ and  $\widetilde{D}_{ij}$ are
expressed in terms of the various determinants and $\bar{b}$ coefficients (cf.\ the identities~(\ref{P1-magicid1}) and (\ref{P1-magicid2}) of ref.\ \cite{paper1}):
\begin{eqnarray}
\widetilde{D}_{ij} + \Delta_1^{\{i\}} 
& = & 
\frac{\bbjsq{j}{i}}{\detgj{i}} 
\label{eqtruc1}\\
\Delta_2 - \Delta_1^{\{i\}} 
& = & 
\frac{\bbar_i^2}{\detg \, \detgj{i}} 
\label{eqtruc2}
\end{eqnarray}

\vspace{0.3cm}

\noindent
To show that the two ways ``direct'' and ``indirect'' are equivalent is more tricky in the complex mass case than in the real mass one.
Let us discuss this point now.
We have to distinguish according to the 
sign of $\Im(\Delta_1^{\{i\}})$ only. Indeed, after having performed the $\rho$ integration, the $\xi$ integration is always of the type 
\begin{equation}
\int^{\infty}_0 \frac{d \xi}{(\xi - \Delta_2) \, (\xi+A)}
\label{eqdefk1II}
\end{equation}
where $A$ is a complex number. It has been shown in appendix \ref{P1-appendJ} of \cite{paper1} that the 
result of this integral does not depend on the sign of $\Im(\Delta_2)$, neither on the sign of $\Im(A)$. Furthermore, $\tD_{ij}$ which is equal to twice an 
internal mass squared has a negative imaginary part. 

\vspace{0.3cm}

\noindent
{\bf 1)} $\Im(\Delta_1^{\{i\}}) > 0$

\vspace{0.3cm}

\noindent
This case is a straightforward continuation of the real mass case. The result 
is readily given by:
\begin{align}
L_3^4 \left( \Delta_2, \Delta_{1}^{\{i\}}, \tD_{ij} \right)
&= 
\int_0^{1} 
\frac{d z}{(\tD_{ij} + \Delta_1^{\{i\}}) \, z^2 + \Delta_2 - \Delta_1^{\{i\}}}
\notag \\
& \quad {}\quad {}
\left[  
 \ln 
  \left( 
   (\tD_{ij} + \Delta_1^{\{i\}}) \, z^2  - \Delta_1^{\{i\}} 
  \right) 
  - 
  \ln \left( -\Delta_2 \right) 
\right]
\label{eqdeflij3}
\end{align}
(cf.\ eq.~(\ref{P1-eqdeflij3}) of ref.~\cite{paper1}).
Note that the apparent pole in the integrand is fake and the argument of the 
first logarithm never becomes real negative when $z$ spans $[0,1]$.

\vspace{0.3cm}

\noindent
{\bf 2)} $\Im(\Delta_1^{\{i\}}) < 0$

\vspace{0.3cm}

\noindent
Let us come back to eq. (\ref{eqdeflij1}). 
Instead of relying on eq. (\ref{eqdeffuncjp2}) of appendix \ref{appendJ} 
to get rid of the square-root, we have to use  eq. (\ref{eqdeffuncj8}) of this appendix. 
We are left with a $\xi$ integration of the type:
\begin{align}
L_3^4 \left( \Delta_2, \Delta_{1}^{\{i\}}, \tD_{ij} \right)
& = 
\int_0^{+\infty} \frac{d \xi}{\xi-\Delta_2} 
\left[ 
\, i \, \int_0^{+\infty} \, 
  \frac{dz}{\xi - (1+z^2) \, \Delta_1^{\{i\}} - z^2 \, \tD_{ij} } 
\right. 
\notag \\
&  
\quad {}\quad {}\quad {}\quad {}\quad {}\quad {}\quad {}
\left.
 + \int_1^{+\infty} \, 
  \frac{dz}{\xi - (1-z^2) \, \Delta_1^{\{i\}} + z^2 \, \tD_{ij} } \;
\right] 
\label{eqdeflij5}
\end{align}
The $\xi$ integration can be performed first, using eq. (\ref{P1-eqdefk4}) of appendix \ref{P1-appendJ} in \cite{paper1}
and we get: 
\begin{align}
&L_3^4 \left( \Delta_2, \Delta_{1}^{\{i\}}, \tD_{ij} \right)
\notag\\
&= - i \int_0^{+\infty} \, 
 \frac{dz}{(\tD_{ij} + \Delta_1^{\{i\}}) \, z^2 - \Delta_2 + \Delta_1^{\{i\}}} 
 \, 
 \left[  
  \ln \left( -(\tD_{ij} + \Delta_1^{\{i\}}) \, z^2  - \Delta_1^{\{i\}} \right) 
  - 
  \ln \left( -\Delta_2 \right) 
 \right] 
\notag \\
&\quad {} 
 - \int_1^{+\infty} \, 
 \frac{dz}{(\tD_{ij} + \Delta_1^{\{i\}}) \, z^2 + \Delta_2 - \Delta_1^{\{i\}}} 
 \, 
 \left[  
  \ln \left( (\tD_{ij} + \Delta_1^{\{i\}}) \, z^2  - \Delta_1^{\{i\}} \right) 
  - 
  \ln \left( -\Delta_2 \right) 
 \right]
\label{eqdeflij6}
\end{align}
where each of the two integrals converges at $\infty$, the apparent pole in the 
integrand of each term is fake again, and the arguments of logarithms never 
become real negative along the integration paths of none of the two 
integrals. 

\vspace{0.3cm}

\begin{figure}[h]
\begin{center}
\includegraphics[scale=0.8]{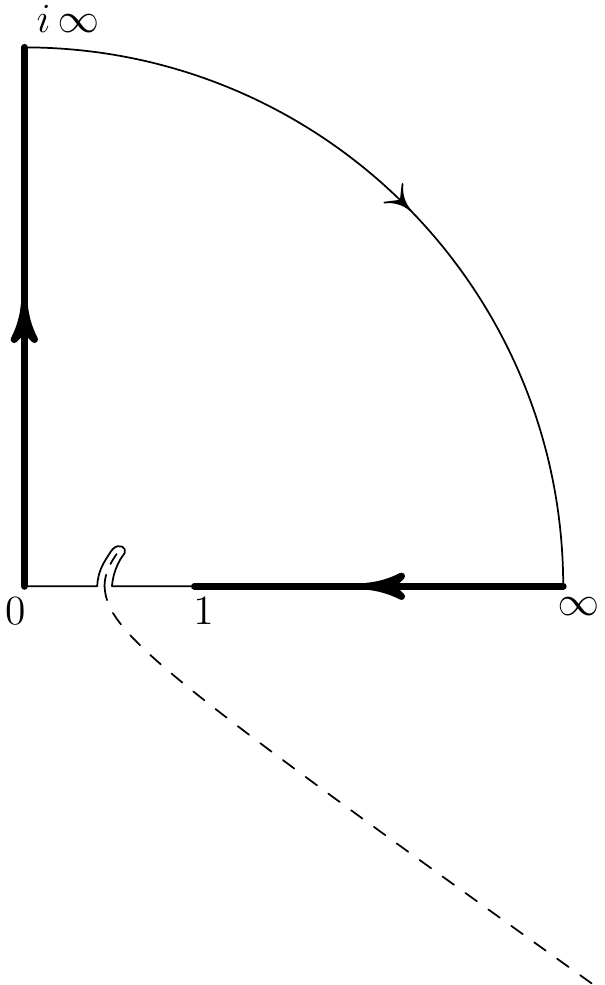}
\end{center}
\caption{\footnotesize Location of the relevant discontinuity cut 
${\cal C}_{ij}$ with respect to the two half straight lines $[0, + i \infty[$
and $[1,+\infty[$ and deformation of the contour $\widehat{(0,1)}$ partly 
wrapping the extremity of the cut.}\label{contour}
\end{figure}

\noindent
The two cases 1) vs. 2) disentangled above can be reunified by seeing 
eq. (\ref{eqdeflij6})
as an analytic continuation in $\Delta_1^{\{i\}}$ of eq. (\ref{eqdeflij3}) 
which possibly requires a deformation of the contour $[0,1]$ originally drawn
along the real axis in eq. (\ref{eqdeflij3}). This feature is discussed in
appendix \ref{appendJ} on eqs. (\ref{eqdeffuncj2}) and (\ref{eqdeffuncj7}) vs. 
(\ref{eqdeffuncj8}). It is interesting to formulate it on eq. (\ref{eqdeflij6}) 
directly as follows. Let us alternatively rewrite eq. (\ref{eqdeflij6}) as:
\begin{align}
L_3^4 \left( \Delta_2, \Delta_{1}^{\{i\}}, \tD_{ij} \right)
&= \left\{ \int_0^{+ i\infty}  + \int_{+\infty}^{1} \right\}
\frac{dz}{(\tD_{ij} + \Delta_1^{\{i\}}) \, z^2 + \Delta_2 - \Delta_1^{\{i\}}}  
\notag\\
& \quad {} \quad {}
\times \left[  
  \ln \left( (\tD_{ij} + \Delta_1^{\{i\}}) \, z^2  - \Delta_1^{\{i\}} \right) 
  - 
  \ln \left( -\Delta_2 \right) 
 \right]
\label{eqdeflij6bis}
\end{align}
The logarithm $\ln((\tD_{ij} + \Delta_1^{\{i\}}) \, z^2  - \Delta_1^{\{i\}})$ 
has two discontinuity cuts supported by one and the other branch of the 
hyperbola 
$\{\Im[(\tD_{ij} + \Delta_1^{\{i\}}) \, z^2  - \Delta_1^{\{i\}}]= 0\}$ 
in the complex $z$-plane respectively.
One of the two cuts\footnote{The other cut is the symmetric of 
${\cal C}_{ij}$ under parity: located in the left half plane 
$\{\Re(z) < 0\}$ it is irrelevant for our concern.}, let us label it 
${\cal C}_{ij}$, lies in the right half $z$-plane 
$\{\Re(z) > 0\}$. It originates at the point 
$z_{ij} = [\Delta_1^{\{i\}}/(\tD_{ij} + \Delta_1^{\{i\}})]^{1/2}$
and slashes the right half plane away to $\infty$ through the lower right 
quadrant 
$\{ \Re(z) > 0, \, \Im(z)<0 \}$. In case $z_{ij}$ belongs to
the upper right quadrant $\{ \Re(z) > 0, \, \Im(z)>0 \}$, this cut 
runs from $z_{ij}$ away to $\infty$
by crossing the real segment $[0,1]$ at the value 
$\Re(z) = [\Im(\Delta_1^{\{i\}})/
\Im(\tD_{ij} + \Delta_1^{\{i\}})]^{1/2}$.
The integration contour of the r.h.s.\ of eq. (\ref{eqdeflij6bis}) can be 
closed by drawing an arc 
$\contij$ between 
0 and 1, the extra arc at $\infty$ also involved by the Cauchy theorem
to close the contour yields a vanishing contribution $\sim {\cal O}(\ln R/R)$
where $R$ is ``$|z|$ on the contour at $\infty$". 
\begin{description}
  \item[(i)] If ${\cal C}_{ij}$ entirely belongs to the quadrant 
$\{ \Re(z) > 0, \, \Im(z)<0 \}$ this extra
arc $\contij$ can be taken along the real segment $[0,1]$.
\item[(ii)] However if $z_{ij}$ belongs 
to the quadrant $\{ \Re(z) > 0, \, \Im(z)>0 \}$, 
the extra arc $\contij$ shall wrap the bit of 
${\cal C}_{ij}$ inside the upper right 
quadrant from above as if ${\cal C}_{ij}$ were locally pushing 
up $(0,1)$ away from the real segment $[0,1]$ inside this quadrant as pictured 
on figure \ref{contour}.
\end{description}
In either case, $L_3^4 \left( \Delta_2, \Delta_{1}^{\{i\}}, \tD_{ij} \right)$ 
can be represented also when $\Im(\Delta_{1}^{\{i\}}) < 0$ by an 
integral along the contour $\contij$ whether along $[0,1]$ in case (i) or 
deformed as described above in case (ii) according to the Cauchy theorem:
\begin{align}
L_3^4 \left( \Delta_2, \Delta_{1}^{\{i\}}, \tD_{ij} \right)
&= 
\int_{\widehat{(0,1)}_{i,j}}
 \frac{dz}{(\tD_{ij} + \Delta_1^{\{i\}}) \, z^2 + \Delta_2 - \Delta_1^{\{i\}}} 
\notag\\
& \quad {} \quad {} \quad {}
\times 
\left[  
  \ln \left( (\tD_{ij} + \Delta_1^{\{i\}}) \, z^2  - \Delta_1^{\{i\}} \right) 
  - 
  \ln \left( -\Delta_2 \right) 
 \right]
\label{eqdeflij6ter}
\end{align}
which is the argued analytic continuation in $\Delta_1^{\{i\}}$ of eq. 
(\ref{eqdeflij3}). When the contour deformation is required, the split form 
(\ref{eqdeflij6}) is more convenient from a computational point of view.
However the alternative
form (\ref{eqdeflij6ter}) proves more convenient to extend to the complex mass 
case the recasting of the expression of $I_{3}^{4}$ obtained via
the indirect way into the one obtained via the direct way. 

\vspace{0.3cm}

\begin{figure}[h]
\begin{center}
\includegraphics[scale=0.4]{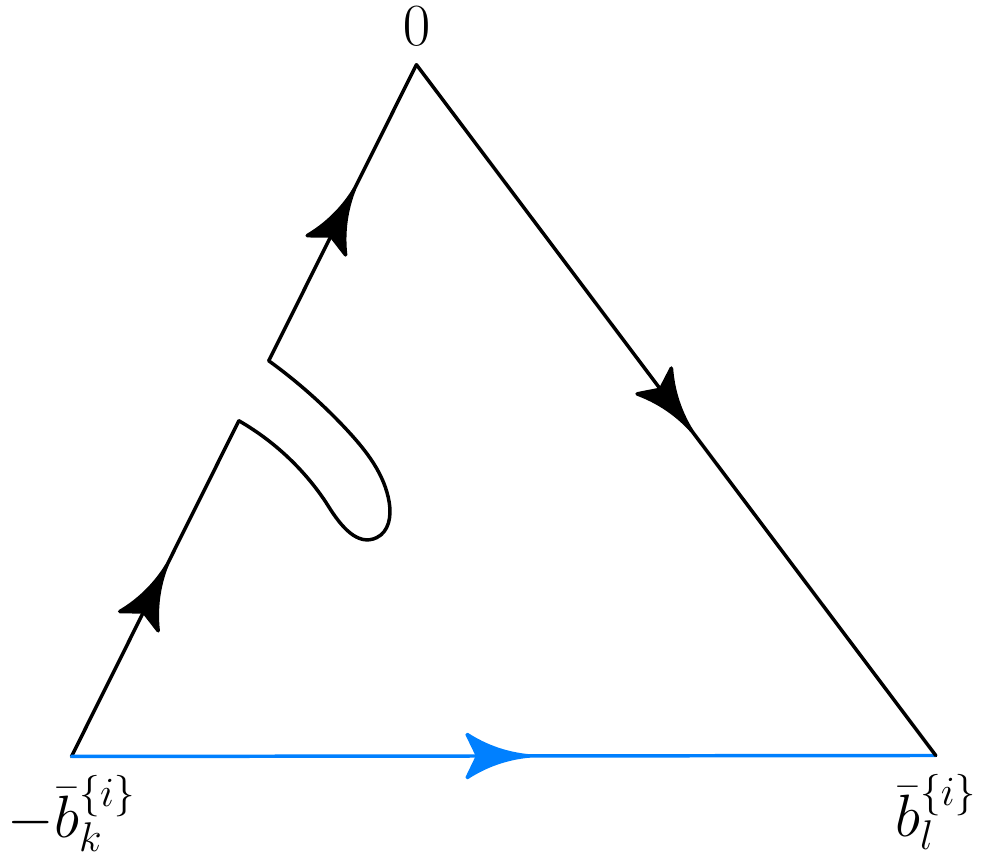}
\end{center}
\caption{\footnotesize Example of a contour deformation involving a 
triangle with one distorted side, for which no cut crosses the straight base 
$[ -\bbj{k}{i},\bbj{l}{i}]$.}\label{triangle1}
\end{figure}

\begin{figure}[h]
\begin{center}
\includegraphics[scale=0.4]{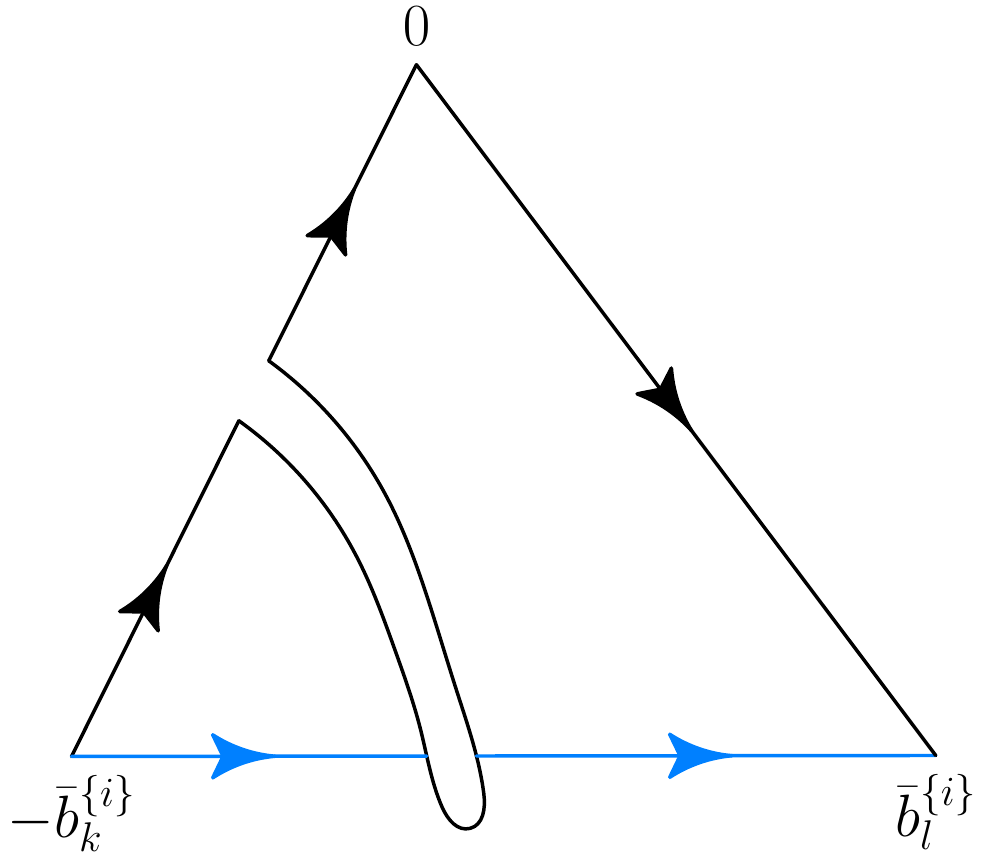}
\end{center}
\caption{\footnotesize Hypothetical case of a contour deformation involving a 
triangle with one distorted side such that
the contribution along the distorted side $-\bbj{k}{i}\widehat{(0,1)}_{i,k}$ 
wraps a cut which would cross the straight line 
$[ -\bbj{k}{i},\bbj{l}{i}]$: this would force one to distort the base
$( -\bbj{k}{i},\bbj{l}{i})$ of the triangle accordingly. }\label{triangle2}
\end{figure}

\noindent
Putting eq. (\ref{eqdeflij6ter}) into eq. (\ref{eqdeflij1}) results in a modification of
eq. (\ref{P1-eqI3414}) of ref. \cite{paper1} in the following form:
\begin{eqnarray}
I_3^4 
& = & 
- \sum_{i \in S_3} \, \sum_{j \in S_3 \setminus \{i\}} \, 
\bbar_i \, \bbj{j}{i} \, 
\int_{\widehat{(0,1)}_{i,j}} \frac{dz}{\detg \, \bbjsq{j}{i} \, z^2 + \bbar_i^2} \,
\nonumber\\
&&
\;\;\;\;\;\;\;\;\;\;\;\;\;\;\;\;\;\;\;\;\;\;\;\;\;\;\;\;\;\;\;
\times 
\left[
 \ln 
 \left( \frac{\bbjsq{j}{i} \, z^2  + \detsj{i}}{\detgj{i}} \right)
 -
 \ln 
 \left(- \, \frac{\dets}{\detg} \right)
\right]
\label{eqI3414cx}
\end{eqnarray}
i.e. in eq.(\ref{eqI3414cx}) each integral ``from 0 to 1" is now understood
in the sense of eq. (\ref{eqdeflij6ter}) as an integral along a contour
$\contij$ specific to each $i,j$. 
As we did in the real mass case, for each $i$, we perform two operations: 1) the change of 
variable $s = \bbj{j}{i} \, z$ in the integrals 
corresponding to the two values of $j \in S_{3} \setminus \{i\}$, so that the
integrands become identical in the two integrals; 2) the two integrals are joined
end-to-end into a single one integrated along 
the contour ${\cal I}^{(i)}_{k,l} \equiv 
-\bbj{k}{i}\widehat{(0,1)}_{i,k} \cup \bbj{l}{i}\widehat{(0,1)}_{i,l}$ in 
the complex $s$-plane. We again specify the two 
elements of $S_{3} \setminus \{i\}$ to be $k \equiv 1 + ((i+1)$ modulo $3)$ 
and $l \equiv 1 + (i$ modulo $3)$. 
In the real mass case, these two operations yield
the following result:
\begin{eqnarray}
I_3^4 
& = & 
\sum_{i \in S_3} \, \bbar_i  \, \int_{-\bbj{k}{i}}^{\bbj{l}{i}} 
\frac{d s}{s^2  \detg + \bbar_i^2} \,
\nonumber\\
&&
\;\;\;\;\;\;\;\;\;\;\;\;
\times
\left[
 \ln 
 \left( 
   \frac{s^2  + \detsj{i}}{\detgj{i}}  - i \, \lambda
 \right)
 - 
 \ln \left( - \, \frac{\dets}{\detg} - i \, \lambda \right)
\right]
\label{eqI3416}
\end{eqnarray}
and the following change of variable
\begin{equation}\label{cvar1}
s= - \, \bbj{k}{i} + \left( \bbj{k}{i} + \bbj{l}{i}\right) \, u
= - \, \bbj{k}{i} - \detgj{i} \, u
\end{equation}
leads to the same formula obtained in the case of the ``direct way''.
For general complex masses the three
points $-\bbj{k}{i},0,\bbj{l}{i}$ are no longer aligned in general.
Furthermore, either of the paths $-\bbj{k}{i}\widehat{(0,1)}_{i,k}$ and 
$\bbj{l}{i}\widehat{(0,1)}_{i,l}$ (or both) may not be straight any more.  
Let us instead consider the extension of eq. (\ref{eqI3416}) to the
complex mass case: the integration contour in eq. (\ref{eqI3416}) shall still 
be understood as the {\em straight} line stretched from $-\bbj{k}{i}$ to 
$\bbj{l}{i}$ and running parallel to the real axis cf. eq. (\ref{cvar1}). 
In this latter case, these two operations give:
\begin{eqnarray}
I_3^4 
& = & 
- \sum_{i \in S_3} \, \bbar_i  \, 
\int_{{\cal I}^{(i)}_{k,l}}
\frac{d s}{s^2  \detg + \bbar_i^2} \,
\nonumber \\ 
& & 
\;\;\;\;\;\;\;\;\;\;\;\;\;\;\;\;\;\;\;\;\;\;\;\;\;\;\;\;\;\;\;\;
\times 
\left[ 
 \ln \left( \frac{s^2 + \detsj{i}}{\detgj{i}} \right)
 - \, \ln \left( \frac{\dets}{\detg} \right)
\right]
\label{eqI3415extend}
\end{eqnarray}
Eq. (\ref{eqI3415extend}) is the extension of eq. (\ref{P1-eqI3415}) of ref. \cite{paper1}.  
Going from eq. (\ref{eqI3415extend})
to the extension of eq. (\ref{eqI3416}) which reads:
\begin{align}
&\mbox{extension of r.h.s. (\ref{eqI3416})}
\notag\\
&= 
- \sum_{i \in S_3} \, \bbar_i  \, \int_{-\bbj{k}{i}}^{\bbj{l}{i}} 
\frac{d s}{s^2  \detg + \bbar_i^2} 
\left[ 
 \ln \left( \frac{s^2 + \detsj{i}}{\detgj{i}} \right)
 - \, \ln \left( \frac{\dets}{\detg} \right)
\right]
\label{eqI3416extend}
\end{align}
thus involves a contour deformation by means of the (possibly distorted) 
``triangle" in the complex $s$-plane whose sides are 
$-\bbj{k}{i}\widehat{(0,1)}_{i,k}$ and 
$\bbj{l}{i}\widehat{(0,1)}_{i,l}$ (either or both being possibly non straight) 
and $[ -\bbj{k}{i},\bbj{l}{i}]$ (straight), as illustrated on figures 
\ref{triangle1} and \ref{triangle2}.

\noindent
This raises two issues.  
The first issue concerns the possible presence of poles in the integrand of 
eqs.\ \myref{eqI3415extend}, \myref{eqI3416extend} inside the distorted ``triangle". 
Yet the poles are fake as their residues vanish by construction: this issue is 
thus irrelevant. The second issue concerns the respective location of the 
discontinuity cuts of the logarithm $\ln[ s^2 + \detsj{i}/\detgj{i}]$ w.r.t. 
the side $[- \bbj{k}{i}, \bbj{l}{i}]$ of the triangle, namely whether the side 
$[- \bbj{k}{i}, \bbj{l}{i}]$ crosses a discontinuity cut of 
$\ln[ s^2 + \detsj{i}/\detgj{i}]$ as illustrated by fig. \ref{triangle2}. 
By means of the change of variable 
(\ref{cvar1}) this is equivalent to the issue whereby the real interval $[0,1]$ 
would cross a discontinuity cut of $\ln [D^{\{i\}(l)}(u)]$ in the 
complex $u$-plane (cf.\ eq.~(\ref{P1-Db}) of ref.\ \cite{paper1} for the definition of $D^{\{i\}(l)}(u)$). In this respect we shall note that the imaginary part of 
$D^{\{i\}(l)}(u)$ is a convex combination of the form 
$\Im(m_{k}^{2}) \, u + \Im(m_{l}^{2}) \, (1-u)$ where the 
imaginary parts of the masses $m_{k,l}^{2}$ have the same (negative) sign and 
$u$ spans $[0,1]$, so that $\Im(D^{\{i\}(l)}(u))$ keeps a 
constant (negative) sign on $[0,1]$, which implies no cut crossing: the case at
hand is the one pictured by fig. \ref{triangle1} whereas the case of fig. 
\ref{triangle2} does not occur. We can therefore safely deform 
${\cal I}^{(i)}_{k,l}$
into the straight line 
$[-\bbj{k}{i},\bbj{l}{i}]$, map the latter onto $[0,1]$ using eq. (\ref{cvar1})
and we finally recover the same expression as obtained according to the direct 
way with complex masses.

\vspace{0.3cm}

\noindent
To finish this section let us comment about the proliferation of dilogarithms. 
To cover the case of general complex masses for the scalar three-point function, 
the integration contour has to be modified depending on the imaginary part of $\Delta_1^{\{i\}}$
(cf.\ eqs.~(\ref{eqdeflij3}) and~(\ref{eqdeflij6})). But even if the contour is not 
on the real axis, it can be decomposed on a part along one of the half imaginary 
axes and another part on the real axis between $1$ and $+\infty$. As shown in 
appendix~\ref{appF}, the contribution along the half imaginary axis gives only 
logarithms and the one on the real axis between $1$ and $+\infty$ yields the same 
combination of dilogarithms as an integration between $0$ and $1$ on the real axis (irrespectively of the 
fact that the cut of the integrand may cross the real axis between these bounds!). 
This is due to the fact that the integrand is even with respect to the integration variable 
and so, only the bound $1$ produces dilogarithms.
To sum up, whatever the sign of the imaginary part of $\Delta_1^{\{i\}}$ is, the 
dilogarithms obtained after the last integration are the same
\footnote{Up to the fact that, in the case of general complex masses, the $\cals$ matrix elements 
entering into the arguments of the dilogarithms are 
complex numbers with a non vanishing imaginary part.} to those of the real mass case. 
So the discussion of subsec.~\ref{P1-sss223} of \cite{paper1} about the number of 
dilogarithms versus \cite{tHooft:1978jhc} is still valid and 
leads to the same conclusion.

\section{Leg up: $I_4^4$}\label{sectfourpoint}

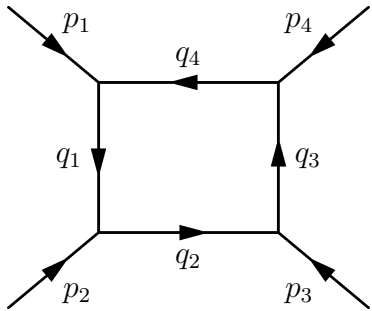
\begin{figure}[h]
\centering
\parbox[c][43mm][t]{40mm}{\begin{fmfgraph*}(60,40)
  \fmfleftn{i}{2} \fmfrightn{o}{2}
  \fmf{fermion,label=$p_1$}{i2,v1}
  \fmf{fermion,label=$p_2$}{i1,v2}
  \fmf{fermion,label=$p_3$}{o1,v3}
  \fmf{fermion,label=$p_4$}{o2,v4}
  \fmf{fermion,tension=0.5,label=$q_1$}{v1,v2}
  \fmf{fermion,tension=0.5,label=$q_2$}{v2,v3}
  \fmf{fermion,tension=0.5,label=$q_3$}{v3,v4}
  \fmf{fermion,tension=0.5,label=$q_4$}{v4,v1}
\end{fmfgraph*}}
\caption{The box picturing the one-loop four point function.}
\label{fig2} 
\end{figure}

\noindent
Let us start this section by recapping the definitions and notations required
for the extension to general complex masses. This section complements the section \ref{P1-sectfourpoint} of the companion article \cite{paper1}.
The usual integral representation of $I_4^4$ in terms of Feynman parameters is 
given by:
\begin{eqnarray}
I_4^4 & = & \int_0^1 \, \prod_{i=1}^4 \, dz_{i} \, \delta(1- \sum_{i=1}^4 z_i) 
\left( 
 - \, \frac{1}{2} \, Z^{\;T} \cdot \cals \cdot Z  
\right)^{-2}
\label{eqSTARTINGPOINT}
\end{eqnarray}
where $Z$ is now a column 4-vector whose components are the $z_{i}$.
Singling out arbitrarily the
subscript value $a$ ($a \in S_{4}=\{1,2,3,4\}$), and writing $z_a$ as 
$1 - \sum_{j \neq a} z_j$, we find:
\begin{eqnarray}
- \, Z^{\;T} \cdot \cals \cdot Z 
& = & 
\sum_{i,j \in S_4 \setminus \{a\}} G_{i\,j}^{(a)} \, z_i \, z_j  - 
2 \sum_{j \in S_4 \setminus \{a\}} V_j^{(a)} \, z_j  -  C^{(a)}
\label{eqVECZT4}
\end{eqnarray}
where the $3 \times 3$ Gram matrix $G^{(a)}$ and the column 3-vector $V^{(a)}$
are defined by
\begin{eqnarray}
G_{i\,j}^{(a)} 
& = & 
-  \, (\cals_{i\,j}-\cals_{a\,j}-\cals_{i\,a}+\cals_{a\,a}), \quad i,j \ne a
\nonumber
\\
V_j^{(a)} 
& = & 
\;\;\;\;\;  \cals_{a\,j} - \cals_{a \, a}, \quad j \ne a 
\label{eqVJA4}\\
C^{(a)}& = & 
\;\;\;\;\;  \cals_{a \, a} 
\nonumber
\end{eqnarray}
Labelling $b$, $c$ and $d$ the three elements of $S_4 \setminus \{a\}$ with 
$b < c < d$, the polynomial (\ref{eqVECZT4}) reads:
\begin{eqnarray}
  D^{(a)}(X) 
= 
X^{\;T} \cdot G^{(a)} \cdot X - 2 V^{(a) \, T} \cdot X - C^{(a)}
& , &  X =
 \left[
 \begin{array}{c}
   z_b \\
   z_c \\
   z_d
 \end{array}
 \right]
  \label{eqDEFD}
\end{eqnarray}
Again the dependencies on $G^{(a)}$, $V^{(a)}$ and $C^{(a)}$ will arise through 
quantities independent of the actual choice of $a$. 
In ref. \cite{paper1}, we applied three time the Stokes-type identity and traded the
three dimensional Feynman parameter integral over the simplex against a sum of
three dimensional integrals over the first octant of $\mathds{R}^3$. 
The four-point function was written as a sum over the 
coefficients $\bbar$, $\bbj{}{i}$ and $\bbj{}{i,j}$ weighted by a three dimensional integral 
over the first octant
(see ref.\ \cite{paper1} for more details):
\begin{align}
I_4^4 
& =  
\sum_{i \in S_{4}} \, \sum_{j \in S_{4} \setminus \{i\}} \, 
\sum_{k \in S_{4} \setminus \{i,j\}} 
\frac{\bbar_i}{\detg} \, \frac{\bbj{j}{i}}{\detgj{i}} \, 
\frac{\bbj{k}{i,j}}{\detgj{i,j}} \, L_4^4(\Delta_3,\Delta_2^{\{i\}},\Delta_1^{\{ij\}},\tD_{ijk})
\label{eqI449new}
\end{align}
with
\begin{align}
L_4^4(\Delta_3,\Delta_2^{\{i\}},\Delta_1^{\{ij\}},\tD_{ijk})
& = \kappa \, 
\int^{+\infty}_0 
\frac{d \xi }{(\xi^2 - \Delta_3)} 
\int^{+\infty}_0 
\frac{d \rho}{(\rho^2 + \xi^2 - \Delta_{2}^{\{i\}})} 
\label{eqDEFK2} \\
&
\;\;\;\;
\mbox{} \times 
\int^{+\infty}_0
\frac{d \sigma}
{(\sigma^2 + \rho^2 + \xi^2 - \Delta_1^{\{i,j\}}) \, 
(\sigma^2 + \rho^2 + \xi^2 + \tD_{ijk})^{1/2}}
\notag \\
\kappa & = \frac{16}{3 \, B(2,1/2) \, B(3/2,1/2) \, B(1,1/2)}
\notag
\end{align}
The quantities $\Delta_3$, $\Delta_2^{\{i\}}$ and $\Delta_1^{\{i,j\}}$,
involved in 
eq. (\ref{eqDEFK2}), are expressed in terms of the determinants of the $\cals$ matrix just as the one-pinched and two-pinched $\cals$ matrices and the associated Gram matrices: 
\begin{align}
\Delta_3 
&= 
- \, \frac{\dets}{\detg} \label{eqdefdelta3} \\
\Delta_{2}^{\{i\}} 
&= 
\frac{\det \, ({\cal S}^{\{i\}})}{\det \,(G^{\{i\}})} 
 \label{eqdefdelta2}\\
\Delta_{1}^{\{i,j\}} 
&= 
- \; \frac{\det \, ({\cal S}^{\{i,j\}})}{\det \,(G^{\{i,j\}})} 
 \label{eqdefdelta1}
\end{align}
As for $\widetilde{D}_{ijk}$, it is proportional to an internal mass squared:
\begin{align}
  \widetilde{D}_{ijk} &=  2 \, m^{2}_{l} \quad l \in S_{4} \setminus \{i,j,k\}
  \label{dijk}
\end{align}
The coefficients $\bbar$, $\bbj{}{i}$ and $\bbj{}{i,j}$ can be built with respectively the Gram matrix $G^{(a)}$ and the 3-vector $V^{(a)}$, the one-pinched Gram matrix $G^{\{i\}(a)}$ and the 2-vector $V^{\{i\}(a)}$ and the two-pinched Gram matrix\footnote{``$\det \, (G^{\{i,j\}})$" is merely a fancy notation to
keep some unity in formulas, as $G^{\{i,j\}}$ reduces to one single scalar.} $G^{\{i,j\}(a)}$ and the 1-vector $V^{\{i,j\}(a)}$\footnote{The coefficients $\bbar$ do not depend on the subscript which has been used to build the Gram matrix, it is the same for the coefficients $\bbj{}{i}$ and $\bbj{}{i,j}$, cf.\ ref.\ \cite{paper1}.}: 
\begin{align}
\frac{\bbar_j}{\detg} 
&= 
- \, ((G^{(a)})^{-1} \cdot V^{(a)})_{j}, \;\; j \in S_{4} \setminus \{a\} \label{eqrecapbbj40} \\
\frac{\bbar_a}{\detg}
&= 
- \, 1 + \sum_{j \in S_{4} \setminus \{a\}}((G^{(a)})^{-1} \cdot V^{(a)})_{j} \label{eqrecapbbj41}
\end{align}
\begin{align}
\frac{\bbar_j^{\{i\}}}{\det \,(G^{\{i\}})} 
&= 
((G^{\{i\}(a)})^{-1} \cdot V^{\{i\}(a)})_{j}, \;\; 
j \in S_{4} \setminus \{i,a\} 
\label{eqrecapbbj42} \\
\frac{\bbar_{a}^{\{i\}}}{\det \,(G^{\{i\}})}
&= 
1 - 
\sum_{j \in S_{4} \setminus \{i,a\}}
((G^{\{i\}(a)})^{-1} \cdot V^{\{i\}(a)})_{j} \label{eqrecapbbj43}
\end{align}
\begin{align}
\frac{\bbj{l}{i,j}}{\detgj{i,j}} 
&= 
- \; (G^{\{i,j\}(a)})^{-1} \; V^{\{i,j\}(a)}, \;\;
l \in S_{4} \setminus \{i,j,a\} 
\label{eqrecapbbj44} \\
\frac{\bbar_{a}^{\{i,j\}}}{\det \,(G^{\{i,j\}})} 
&= 
- 1 \, + \, (G^{\{i,j\}(a)})^{-1} \; V^{\{i,j\}(a)}
 \label{eqrecapbbj45}
\end{align}
To finish the recap, we have introduced for convenience in ref.\ \cite{paper1} the following quantities which will be used in the rest of this section:
\begin{eqnarray}
\left.
\begin{array}{lclcl}
 P_{ijk} & = & \;\;\; \tD_{ijk}            & + & \Delta_1^{\{i,j\}} \\
 R_{ij}  & = & \;\;\;\Delta_{2}^{\{i\}} & - & \Delta_1^{\{i,j\}}   \\
 Q_i     & = & \;\;\; \Delta_3       & - & \Delta_{2}^{\{i\}}       \\
 T       & = & - \, \Delta_3 &&
\end{array}
\right\}
& \Leftrightarrow &
\left\{
\begin{array}{lcl}
 P_{ijk} + R_{ij} + Q_i + T & = & \;\;\; \tD_{ijk} \\
 R_{ij} \; + \, Q_i\; + \, T  & = & - \, \Delta_1^{\{i,j\}}   \\
 Q_i \;\; + T & = & - \, \Delta_{2}^{\{i\}}      \\
 T       & = & - \, \Delta_3
\end{array}
\right.
\label{newparamdefPQRT}
\end{eqnarray}
 
\subsection{Extension to the general complex mass case}\label{sectfourpointcomp}

We now extend the above results to the general complex mass case.
Coming back to eq. (\ref{eqDEFK2}), $\Delta_3$, $\Delta_2^{\{i\}}$,
$\Delta_1^{\{i,j\}}$ and $\tD_{ijk}$ now assume finite i.e. non vanishing 
imaginary parts and the infinitesimal parameter $\lambda$ specifying the 
Feynman contour prescription becomes irrelevant and can be put 
equal to zero. Whereas $\Im(\tD_{ijk})$ is always $< 0$ we have to distinguish 
between $2^{3}=8$ cases according to the signs of $\Im(\Delta_3)$, 
$\Im(\Delta_2^{\{i\}})$ and $\Im(\Delta_1^{\{i,j\}})$.
 
\vspace{0.3cm}

\noindent
{\bf 1.(a)} $\Im(\Delta_3) > 0$, $\Im(\Delta_2^{\{i\}}) > 0$, 
$\Im(\Delta_1^{\{i,j\}}) > 0$

\vspace{0.3cm}

\noindent
This case is a trivial extension of the real mass case. The expression
of $L_4^4(\Delta_3,\Delta_2^{\{i\}},\Delta_1^{\{i,j\}},\tD_{ijk})$ is provided by:
\begin{align}
L_4^4(\Delta_3,\Delta_2^{\{i\}},\Delta_1^{\{i,j\}},\tD_{ijk})
&= - \, \int^1_0 \frac{d u}{u^2 \, P_{ijk} \, Q_i - R_{ij} \, T} \,
\label{eqcas1} \\
& \;\;\;\;\;\;\;\;\;\;\;\;\; {} \times
\left[  
\ln 
\left(
 \frac{\; u^2 \, P_{ijk} + (R_{ij} + Q_i \; + T)}
{u^2 \, (P_{ijk} + R_{ij} + Q_i) + T} 
\right) 
-
\ln 
\left(\frac{\;\;\;\, Q_i + T}{u^2 \, Q_i + T} \right)  
\right]
\notag
\end{align}
which is the eq. (\ref{P1-eqlijk14}) of ref. \cite{paper1} with $\lambda$ sets to 0.
The result (\ref{eqcas1}) is cast in a form such that the contributions of 
the two logarithms to the residue of the pole 
$1/(u^2 \, P_{ijk} \, Q_i - R_{ij} \, T)$ cancel each other. 
This pole is fake, it is an artefact of partial fraction
decomposition, cf.\ eq.~(\ref{P1-pfdx}) of ref. \cite{paper1}. In each logarithm, the imaginary parts of the 
numerator and of the denominator of the argument have the same sign and this 
common sign is kept constant. Logarithms of ratios can all be safely split into 
differences of logarithms, and the integration contour considered does not cross 
any discontinuity cut of any of the logarithms, so that 
eq. (\ref{eqcas1}) takes the alternative form:
\begin{align}
L_4^4(\Delta_3,\Delta_2^{\{i\}},\Delta_1^{\{i,j\}},\tD_{ijk})
&= - \, \int^1_0 \frac{d u}{u^2 \, P_{ijk} \, Q_i - R_{ij} \, T} \,
\label{eqcas1eclate} \\
& \;\;\;\;\;\;\;\;\;\;\;\;\; {} \times
\Big[  
 \ln \left( \; u^2 \, P_{ijk} + (R_{ij} + Q_i \; + T) \right) 
 -
\ln \left( Q_i + T \right)
\notag\\
& \quad \quad \quad \quad \quad \; {}
 -
 \ln \left( u^2 \, (P_{ijk} + R_{ij} + Q_i) + T \right)
 + 
 \ln \left( u^2 \, Q_i + T \right)
 \Big]
\notag
\end{align}
On the alternative form (\ref{eqcas1eclate}) it is no longer manifest that 
the residue of the fake pole vanishes. Subtracting and 
adding the value taken at the pole by the split combination of logarithms 
leads to:
\begin{align}
\hspace{2em}&\hspace{-2em}L_4^4(\Delta_3,\Delta_2^{\{i\}},\Delta_1^{\{i,j\}},\tD_{ijk}) \notag \\
&= - \, \int^1_0 \frac{d u}{u^2 \, P_{ijk} \, Q_i - R_{ij} \, T} \,
\notag \\
& \;\;\;\;\;\;\;\;\;\;\;\;\;\;
\left[  \;\;
 \ln \left( \; u^2 \, P_{ijk} + (R_{ij} + Q_i \; + T) \right) \;
 -
 \;
 \ln \left( \frac{(R_{ij} + Q_i)}{Q_i} \, (Q_i \; + T) \right)
\right.
\notag\\
& \;\;\;\;\;\;\;\;\;\;\;\;\;\;\;
 -
 \ln \left( u^2 \, (P_{ijk} + R_{ij} + Q_i) + T \right)
 \;\;+\;
 \ln 
 \left( 
  T \, \frac{(P_{ijk} + R_{ij})}{P_{ijk}}\frac{(R_{ij} + Q_i)}{Q_i} 
 \right)
\notag\\
&  \;\;\;\;\;\;\;\;\;\;\;\;\;\;\;
 + \ln \left( u^2 \, Q_i + T \right) 
   \quad {} \quad {} \quad {}\quad {} \quad {} \quad {} \;\;\;
 -  \;
  \ln \left( \frac{T}{P_{ijk}} \, (P_{ijk} + R_{ij}) \right)
\notag\\
&  \;\;\;\;\;\;\;\;\;\;\;\;\;\;
\left. 
 + \eta \left( \frac{(R_{ij} + Q_i)}{Q_i}, (Q_i + T) \right)
 - 
 \eta 
   \left( 
    T \, \frac{(P_{ijk} + R_{ij})}{P_{ijk}}, \,  \frac{(R_{ij} + Q_i)}{Q_i}
   \right)
\right]
\label{eqcas1soustraite}
\end{align}
Whereas the first three lines now manifestly vanish at the pole, the presence 
of the two extra $\eta$ functions\footnote{The $\eta$ function is defined by 
eq.~(\ref{P1-eqdefeta01}) in appendix~\ref{appF} of \cite{paper1}} in the last line of 
eq.~(\ref{eqcas1soustraite}) might suggest that the pole residue no longer vanish. 
This paradox is solved as one realises that the splitting of the
logarithms of ratios into differences of logarithms holds on the interval of 
integration but does not hold in general in the vicinity of the pole when the
latter is remote from the integration contour. The splitting shall in general 
be supplemented by $u$-dependent $\eta$ functions. These $\eta$ functions 
vanish on the integration contour thus are not explicitly written in eq. 
(\ref{eqcas1soustraite}). Yet these $\eta$ functions take in general non 
vanishing values at the pole and these values combine into minus the last line 
of eq. (\ref{eqcas1soustraite}). Let us note however that 1) if the pole happens
to be close enough to - or even on - the segment $[0,1]$, the last line of 
eq. (\ref{eqcas1soustraite}) does vanish and the pole residue is
manifestly zero indeed 2) if otherwise the pole is remote from the segment 
$[0,1]$ the issue of subtraction of pole residue is irrelevant insofar as the 
fake pole generates no numerical instability whatsoever.

\vspace{0.3cm}

\noindent
For the seven other cases we follow the same strategy for step 4 as in the 
real mass case (cf.\ subsec.~\ref{P1-fourpointstep4} of \cite{paper1}). 
Two slight complications arise, though. One is induced when the 
variant (\ref{eqdeffuncj8}) instead of (\ref{eqdeffuncjp2}) is at work 
for the integral (\ref{eqdefj1}) for $\nu=2$, which now involves two integrals 
both ranging to $\infty$, instead of one on $[0,1]$ only. 
At substep 4a. of \cite{paper1}, when recasting the integral representation of 
$M_{1}(\xi^2+\rho^2)$ the extension itemises 
into $2$ cases, depending on the sign of $\Im(\Delta_1^{\{i,j\}}) > 0$ w.r.t. 
the negative sign of $\Im(\tD_{ijk})$. 
Then at substep 4d. of \cite{paper1} the extension itemises into $2 \times 2 = 4$ subcases, 
depending of the relative signs of 
$\Im(\Delta_3) > 0$ vs. $\Im(\Delta_2^{\{i\}})$, and of 
$\Im(\Delta_3) > 0$ vs. $\Im(\Delta_1^{\{i,j\}})$. The process of extension thus
goes as follows:
\begin{enumerate}
\item
if $\Im(\Delta_1^{\{i,j\}}) > 0$, at substep 4a. of \cite{paper1} identity (\ref{eqdeffuncjp2}) 
is applied to the term involving $\{\Delta_1^{\{i,j\}},\tD_{ijk}\}$ 
which yields one term. Then at substep 4d. of \cite{paper1},
\begin{enumerate}
 \item 
 if $\Im(\Delta_3) > 0$ and $\Im(\Delta_2^{\{i\}}) > 0$ then \\
 identity (\ref{eqdeffuncjp2}) is applied
 to the term involving $\{\Delta_3,\Delta_2^{\{i\}}\}$\\ 
 identity (\ref{eqdeffuncjp2}) is applied
 to the term involving $\{\Delta_3,\Delta_1^{\{i,j\}}\}$ 
 \item
 if $\Im(\Delta_3) > 0$ and $\Im(\Delta_2^{\{i\}}) < 0$ then \\
 identity (\ref{eqdeffuncj8}) is applied
 to the term involving $\{\Delta_3,\Delta_2^{\{i\}}\}$ \\
 identity (\ref{eqdeffuncjp2}) is applied
 to the term involving $\{\Delta_3,\Delta_1^{\{i,j\}}\}$ 
 \item
 if $\Im(\Delta_3) < 0$ and $\Im(\Delta_2^{\{i\}}) > 0$ then \\
 identity (\ref{eqdeffuncj8}) is applied 
 to the term involving $\{\Delta_3,\Delta_2^{\{i\}}\}$ \\
 identity (\ref{eqdeffuncj8}) is applied to the term involving
 $\{\Delta_3,\Delta_1^{\{i,j\}}\}$ 
 \item
 if $\Im(\Delta_3) < 0$ and $\Im(\Delta_2^{\{i\}}) < 0$ then \\
 identity (\ref{eqdeffuncjp2}) is applied 
 to the term involving $\{\Delta_3,\Delta_2^{\{i\}}\}$ \\
 identity (\ref{eqdeffuncj8}) is applied 
 to the term involving $\{\Delta_3,\Delta_1^{\{i,j\}}\}$; 
 \end{enumerate}
\item
else $\Im(\Delta_1^{\{i,j\}}) < 0$, thus at substep 4a. of \cite{paper1} identity 
(\ref{eqdeffuncj8}) is applied to the term  involving 
$\{\Delta_1^{\{i,j\}},\tD_{ijk}\}$, which yields two terms
in which the $\sigma$ integration has been traded for a $z$ integration ranging for one term from $0$ to $\infty$ and for the other between $1$ to $\infty$. 
Then at substep 4d. of \cite{paper1},
for the term having a $z$ integration range between $1$ to $\infty$ the different splittings are the same as for the case $\Im(\Delta_1^{\{i,j\}}) < 0$, while for the other term
\begin{enumerate}
 \item 
 if $\Im(\Delta_3) > 0$ and $\Im(\Delta_2^{\{i\}}) > 0$ then \\
 identity (\ref{eqdeffuncjp2}) is applied  
 to the terms involving $\{\Delta_3,\Delta_2^{\{i\}}\}$\\ 
 identity (\ref{eqdeffuncj8}) is applied to the terms involving 
$\{\Delta_3,\Delta_1^{\{i,j\}}\}$ 
 \item
 if $\Im(\Delta_3) > 0$ and $\Im(\Delta_2^{\{i\}}) < 0$ then \\
 identity (\ref{eqdeffuncj8}) is applied  
 to the terms involving $\{\Delta_3,\Delta_2^{\{i\}}\}$ \\
 identity (\ref{eqdeffuncj8}) is applied to the terms involving 
$\{\Delta_3,\Delta_1^{\{i,j\}}\}$ 
 \item
 if $\Im(\Delta_3) < 0$ and $\Im(\Delta_2^{\{i\}}) > 0$ then \\
 identity (\ref{eqdeffuncj8}) is applied  
 to the terms involving $\{\Delta_3,\Delta_2^{\{i\}}\}$ \\
 identity (\ref{eqdeffuncjp2}) is applied to the terms involving
 $\{\Delta_3,\Delta_1^{\{i,j\}}\}$ 
 \item
 if $\Im(\Delta_3) < 0$ and $\Im(\Delta_2^{\{i\}}) < 0$ then \\
 identity (\ref{eqdeffuncjp2}) is applied  
 to the terms involving $\{\Delta_3,\Delta_2^{\{i\}}\}$ \\
 identity (\ref{eqdeffuncjp2}) is applied to the terms involving
 $\{\Delta_3,\Delta_1^{\{i,j\}}\}$. 
 \end{enumerate}
\end{enumerate}
\noindent
Another complication comes from the 
exchange of the orders of integrations over $y$ and $u$ while going through 
the counterparts of eqs. (\ref{P1-eqlijk11}) to (\ref{P1-eqlijk13}) of ref. \cite{paper1}, whenever either 
of two integrations (or both) is (are) not performed between 0 and 1 any more.
A splitting into two or more integrals may then be required. 
These two sources of 
complications thereby generate both a proliferation and a diversification of 
integral contributions, resulting into as many final forms as there are cases 
faced. Notwithstanding, further simplifications and rearrangements lead to a 
somewhat common pattern, as will be described below.
These complications let aside, the extension of the 
derivation can be worked through without trouble and we quote the results 
for each case, presented in the order in which they are met during the 
extension process. 
As observed once the calculations have been done, the $2^3$
cases all involve the same three logarithms of second degree polynomials 
$\ln( u^2 \, P_{ijk} + (R_{ij}+Q_i + T))$, $\ln( u^2 Q_i + T)$ and
$\ln(u^2 \, (P_{ijk}+R_{ij}+Q_i) + T)$ 
integrated along contours stretched from 0 to 1, though not necessarily 
along the real axis. Some of these contours may have to be deformed so as 
to partly wrap cuts of the logarithms considered whenever some
cut emerging from some branch point at finite distance from the origin 
slashes across the real interval $[0,1]$.
 
\vspace{0.3cm}

\noindent
{\bf 1.(b)} $\Im(\Delta_3) > 0$, $\Im(\Delta_2^{\{i\}}) < 0$, 
$\Im(\Delta_1^{\{i,j\}}) > 0$
\begin{align}
\hspace{2em}&\hspace{-2em}L_4^4(\Delta_3,\Delta_2^{\{i\}},\Delta_1^{\{i,j\}},\tD_{ijk}) \notag \\
&=  
\left\{ 
i \, \int^{+\infty}_0 \frac{du}{u^2 \, P_{ijk} \, Q_i + R_{ij} \, T} 
\left[ 
 \ln \left( \frac{-Q_i}{-u^2 \, Q_i + T} \right) 
 - 
 \ln \left( \frac{R_{ij}}{u^2 \, (P_{ijk}+R_{ij})} \right) 
\right]  
\right.
\notag \\
&\quad {} \;\;\;\;\; + 
\int^{+\infty}_1 \frac{du}{u^2 \, P_{ijk} \, Q_i - R_{ij} \, T} \, 
\left[ 
 \ln \left( \frac{R_{ij}}{u^2 \, (P_{ijk}+R_{ij})} \right)
 - 
 \ln \left( \frac{Q_i}{u^2 \, Q_i + T} \right) 
\right] 
\notag \\
&\quad {} \;\;\;\;\; + 
\int^{1}_0 \;\;\;\;
\frac{du}{u^2 \, P_{ijk} \, Q_i - R_{ij} \, T} 
\notag\\
& \;\;\;\;\;\;\;\;\;\;\;\;\;\;\;\;\;\;\;\;\;
\left[ 
 \ln \left( \frac{u^2 \, P_{ijk} + R_{ij}}{u^2 \, (P_{ijk} + R_{ij})} \right) 
 -  
 \ln 
 \left( 
  \frac{u^2 \, P_{ijk} + (R_{ij} + Q_i + T)}
  {u^2 \, (P_{ijk} + R_{ij} + Q_i) + T} 
 \right) 
\right] 
\notag \\
&\quad {} \;\;\;\;\; +  \left.
\int^{1}_0 \;\;\;\;
\frac{du}{u^2 \, P_{ijk} \, Q_i - R_{ij} \, T} \, 
\left[ 
 \ln \left( \frac{R_{ij}}{u^2 \, P_{ijk}+R_{ij}} \right)
 - 
 \ln \left( \frac{Q_i}{Q_i + T} \right) 
\right] 
\right\}
\label{eqcas3}
\end{align}
In this case, we have $\Im(R_{ij}) < 0$, $\Im(P_{ijk} + R_{ij}) < 0$, $\Im(Q_{i}+T) > 0$ and
$\Im(Q_i) > 0$.
Furthermore, 
\begin{itemize}
\item
$u^2 \, Q_i - T = (1+u^2) \, \Delta_3 - u^2 \, \Delta_2^{\{i\}}$ \\
thus $\Im(u^2 \, Q_i - T) > 0$ when $u \in [0,\infty[$, 
\item
$u^2 \, Q_i + T =  (u^2-1) \, \Delta_3 - u^2 \, \Delta_2^{\{i\}}$ \\
thus $\Im(u^2 \, Q_i + T) > 0$ when $u \in [1,\infty[$, 
\item
$u^2 \, P_{ijk} + R_{ij} 
= u^2 \, \tD_{ijk} - (1 - u^2) \, \Delta_1^{\{i,j\}} + \Delta_{2}^{\{i\}}$ \\
thus $\Im(u^2 \, P_{ijk} + R_{ij}) < 0$ when $u \in [0,1]$, 
\item
$u^2 \, P_{ijk} + (R_{ij} + Q_i + T)
= u^2 \, \tD_{ijk} - \Delta_1^{\{i,j\}} \, (1-u^2)$ \\
thus $\Im(u^2 \, P_{ijk} + (R_{ij} + Q_i + T)) < 0$ when $u \in [0,1]$,
\item
$u^2 \, (P_{ijk}+R_{ij}+Q_i) + T = u^2 \, \tD_{ijk} - (1-u^2) \, \Delta_3$ \\
thus $\Im(u^2 \, (P_{ijk}+R_{ij}+Q_i) + T) < 0$ when $u \in [0,1]$. 
\end{itemize}
The result (\ref{eqcas3}) is cast in a form such that, in each of the four 
integrals separately, the contributions of the logarithms to the residues of 
the fake pole $1/(u^2 \, P_{ijk} \, Q_i \pm R_{ij} \, T)$ cancel 
each other. This manifest and separate cancellation of residues is favoured at 
the expense of the economy of terms. Partial recombinations of integrals allow 
cancellations which reduce the number of terms. Let us showcase how the 
simplifications and rearrangements proceed on the case at hand. Similar 
handlings hold for the other cases listed further on, we will then only give the
alternative form which they lead to in every other case.

\vspace{0.3cm}

\noindent
In every logarithm in eq. (\ref{eqcas3}), the imaginary parts of the numerator 
and of the denominator of the argument have the same sign which is kept 
constant over the integration interval considered. Logarithms of ratios 
can thus all be safely split into differences of logarithms in each integral, 
and in each integral the integration contour considered never crosses any 
discontinuity cut of any of the logarithms.
\vspace{0.3cm}

\noindent
{\bf i)} A first simplification occurs as the $\ln(u^2 \, P_{ijk} + R_{ij})$ terms 
cancel out among the last two integrals on $[0,1]$ in eq. (\ref{eqcas3}). \\
{\bf ii)} The terms $\ln(-u^2 Q_i+T)$ in the first
integral and $\ln(u^2 Q_i+T)$ in the second integral in eq. (\ref{eqcas3}) 
can be combined into a single contour integral in the ``south-east" 
quadrant $\{\Re(u) > 0, \, \Im(u) < 0\}$ as follows.
As detailed in appendix \ref{cut}, the cut of $\ln(u^2 Q_i+T)$ in 
the right half plane $\{\Re(u) > 0\}$ emerges from $\sqrt{-T/Q_i}$ and runs
towards $\infty$ across the ``north-east" quadrant 
$\{\Re(u) > 0, \,\Im(u) >0\}$.
On the other hand this cut does not extend towards $\infty$ in the 
``south-east" quadrant. We can therefore make the 
change of variable $v = - i \, u$ and rewrite
\[
i \int_{0}^{+ \infty} \frac{du}{u^2 P_{ijk}Q_i + R_{ij}T} \, \ln(- u^2 Q_i+T)
=
\int_{0}^{-i \, \infty} \frac{dv}{v^2 P_{ijk}Q_i - R_{ij}T} \, \ln(v^2 Q_i+T)
\]
and concatenate\footnote{A contribution ``at $\infty$", vanishing 
as ${\cal O}(R^{-1} \, \ln (R))$ when the radius $R$ of the added contour 
$\to \infty$, is added.} the latter with minus the integral of the same integrand on 
$[1,+\infty[$ as the single contour integral
\begin{align}
&\left\{ 
 \int_{0}^{-i \, \infty} + \int_{+ \infty}^{1} 
\right\}
\frac{du}{u^2 P_{ijk}Q_i - R_{ij}T} \, \ln(u^2 Q_i+T)
\notag\\
&\quad {}\quad {}\quad {}\quad {}\quad {}\quad {}\quad {}
\quad {}\quad {}\quad {}\quad {}\quad {}
=
\int_{\usebox{\Gammam}} \frac{du}{u^2 P_{ijk}Q_i - R_{ij}T} \, \ln(u^2 Q_i+T)
\label{recombuQT}
\end{align} 
{\bf iii)} In eq. (\ref{recombuQT}) we then subtract and add to $\ln(u^2 \, Q_i + T)$ its 
value at the pole $\ln((P_{ijk} + R_{ij})\, T/P_{ijk})$, so as to deform the 
integral
\begin{equation}\label{termuQTpolesubtrated}
\int_{\usebox{\Gammam}}
 \frac{du}{u^2 \, P_{ijk} \, Q_i - R_{ij} \, T} \, 
 \left[ 
  \ln(u^2 \, Q_i + T) -
  \ln \left( T \, \frac{(P_{ijk} + R_{ij})}{P_{ijk}} \right) 
\right]
\end{equation}
into an integral along a finite contour $\widehat{(0,1)}_{-}$ stretched 
from 0 to 1. The logarithm $\ln(u^2 \, Q_i + T)$ has a cut in the half complex 
plane $\{ \Re(u) > 0\}$ which extends towards infinity only through the 
``north-east" quadrant. Yet the branch point 
$\sqrt{-T/Q_i}$, which the cut emerges from, may be located inside the 
``south-east" quadrant, so that the cut runs outside this quadrant
crossing the segment $[0,1]$ to further slash the ``north-east" quadrant. 
Whenever this occurs, 
$\widehat{(0,1)}_{-}$ shall differ from the straight line $[0,1]$. It shall instead 
wrap the arc of cut stretched between the branch point $\sqrt{-T/Q_i}$ 
and the real axis, from below inside the ``south-east" quadrant. \\
The left-over contribution of the forced counterterm 
$\ln((P_{ijk} + R_{ij})\, T/P_{ijk})$ can be re\-written:
\begin{align}
  &\int_{\usebox{\Gammam}}
\frac{du}{u^2 \, P_{ijk} \, Q_i - R_{ij} \, T} \, 
\ln \left( T \, \frac{(P_{ijk} + R_{ij})}{P_{ijk}} \right) 
\notag\\
&\quad {}\quad {}\quad {}\quad {}\quad {}\quad {}\quad {}
\quad {}\quad {}
=
\left\{ \int_{\Gamma_{-}} +\int_{0}^{1} \right\}
\frac{du}{u^2 \, P_{ijk} \, Q_i - R_{ij} \, T} \, 
\ln \left( T \, \frac{(P_{ijk} + R_{ij})}{P_{ijk}}  \right) 
\label{poleuQTcontreterme}
\end{align}
where the closed contour ${\Gamma}_{-}$ encircles the ``south-east" quadrant 
counterclockwise. It is also at work in step {\bf iv)} next.\\
{\bf iv)} We group contributions involving $\ln(u^2(P_{ijk}+R_{ij}))$ together with
constant terms $\ln(Q_i)- \ln(R_{ij})$ in eq. (\ref{eqcas3}) in a similar way. 
Contributions from 
integrals on $[0,1]$ and $[1,+\infty[$ can be combined and cast in the form 
\begin{equation}\label{morceau1}
\int_{+\infty}^{0}
\frac{du}{u^2 P_{ijk}Q_i - R_{ij}T} \, 
\left[ \ln(Q_i) - \ln(R_{ij}) + \ln(P_{ijk}+R_{ij}) + \ln(u^2) \right]
\end{equation}
whereas, with the change of variable $v = -i \, u$, the left over contribution 
of the first integral of eq. (\ref{eqcas3}) reads:
\begin{equation}\label{morceau2}
\int_{0}^{-i \infty}
\frac{dv}{v^2 P_{ijk}Q_i - R_{ij}T} \, 
\left[
 \ln(- Q_i) - \ln(R_{ij}) + \ln \left( - v^2 (P_{ijk}+R_{ij}) \right) 
\right]
\end{equation}
We make use of the identity 
$\ln(z) = \ln (-z) + i \, \pi \, \mbox{sign}(\Im(z))$ to write
$\ln(-Q_i) = \ln (Q_i) - i \pi$, and, intending to combine eqs. 
(\ref{morceau1}) and (\ref{morceau2}) into a single integral on a closed 
contour encircling the ``south-east" quadrant, we consider that in eq. 
(\ref{morceau2}) $v$ has an infinitesimal positive real part, so that $-v^2$ 
has an infinitesimal {\em positive} imaginary part. We can thus split 
$\ln(- \, v^2 ( P_{ijk} + R_{ij} ))$
into $\ln (- \, v^2) + \ln(P_{ijk} + R_{ij})$ with
$\ln(-v^2) = \ln(v^2) + i \, \pi$. As anticipated the contributions 
(\ref{morceau1}) and (\ref{morceau2}) are then combined into a single integral 
on the closed contour ${\Gamma}_{-}$ encircling the ``south-east" 
quadrant {\em counterclockwise}\footnote{
The {\em counterclockwise} orientation of the contour ${\Gamma}_{-}$ encircling 
the ``south-east" quadrant is somewhat 
unusual. It is inherited from the construction of ${\Gamma}_{-}$ as the
concatenation of the oriented contours $(0,-i \, \infty)$, $(+\infty, 1)$ and 
$(1,0)$. Similarly, the contour ${\Gamma}^{+}$ encircling the ``north-east" 
quadrant {\em clockwise}, constructed as the
concatenation of the oriented contours $(0,+i \, \infty)$, $(+\infty, 1)$ and 
$(1,0)$ is also used in subsequent cases. Yet this is all matter of
presentation and readers preferring to handle contours with their favourite
orientations can obviously modify the corresponding formulas by appropriate 
sign flips.}:
\begin{align}
&(\ref{morceau1}) + (\ref{morceau2})
\notag\\
& \quad {} \quad {} 
=
\int_{\Gamma_{-}} 
\frac{du}{u^2 P_{ijk}Q_i - R_{ij}T} \, 
\left[ \ln(Q_i) - \ln(R_{ij}) + \ln(P_{ijk}+R_{ij}) + \ln(u^2) \right]
\label{closed2b}
\end{align}
The term $\ln(u^2)$ in eq. (\ref{closed2b}) is then replaced by its residue 
value at the pole $\ln(R_{ij}T/(P_{ijk}Q_i))$.\\
{\bf v)} In the contribution
\begin{align}
&\int^{1}_0 \frac{du}{u^2 \, P_{ijk} \, Q_i - R_{ij} \, T} 
\notag\\
& \quad {} \quad {} 
\left[ 
 \ln \left( u^2 \, P_{ijk} + (R_{ij} + Q_i + T) \right)
 -
 \ln \left( u^2 \, (P_{ijk} + R_{ij} + Q_i) + T \right)
\right] 
\label{2logsur01}
\end{align} 
we subtract and add the pole residue contribution so as to recast eq. 
(\ref{2logsur01}) in the form: 
\begin{align}
(\ref{2logsur01})
&= 
\int_{0}^{1} \frac{du}{u^2 \, P_{ijk}Q_i - R_{ij}T} \, 
\notag\\
& \;\;\;\;\;\;\;\;\;\;
 \left\{ \;\,
  \vphantom{\frac{R \, T}{P_{ijk} \, Q_i}}
  \ln \left( u^2 \, P_{ijk} + (R_{ij} + Q_i + T) \right)
  \, - \,
  \ln \left( \frac{(Q_i+R_{ij})}{Q_i} (Q_i+T)\right)
 \right.
\notag\\
& \;\;\;\;\;\;\;\;\;\;
  \; -
  \ln \left( u^2 \, (P_{ijk} + R_{ij} + Q_i) + T \right)
  \, +
  \ln 
  \left( 
   T \, \frac{(P_{ijk}+R_{ij})}{P_{ijk}} \frac{(Q_i+R_{ij})}{Q_i} 
  \right) 
\notag\\
& \;\;\;\;\;\;\;\;\;\;\;\; 
\left.
+
 \left[
  \ln \left( \frac{(Q_i+R_{ij})}{Q_i} (Q_i+T) \right) 
  -
  \ln 
  \left( 
   T \, \frac{(P_{ijk}+R_{ij})}{P_{ijk}} \frac{(Q_i+R_{ij})}{Q_i} 
  \right) 
 \right]
\right\}
\notag
\end{align}
Putting steps {\bf i)} to {\bf v)} together, $L_4^4(\Delta_3,\Delta_2^{\{i\}},\Delta_1^{\{i,j\}},\tD_{ijk})$ reads:
\begin{align}
\hspace{2em}&\hspace{-2em}L_4^4(\Delta_3,\Delta_2^{\{i\}},\Delta_1^{\{i,j\}},\tD_{ijk}) \notag \\
&= {} -  
\left\{ 
\int_{0}^{1} \frac{du}{u^2 \, P_{ijk}Q_i - R_{ij}T} \, 
\right.
\notag\\
& \;\;\;\;\;\;\;\;\;\;
 \left[ \;\,
  \vphantom{\frac{R \, T}{P_{ijk} \, Q_i}}
  \ln \left( u^2 \, P_{ijk} + (R_{ij} + Q_i + T) \right)
  \, - \,
  \ln \left( \frac{(Q_i+R_{ij})}{Q_i} (Q_i+T) \right)
 \right.
\notag\\
& \;\;\;\;\;\;\;\;\;\;
  \; -
  \ln \left( u^2 \, (P_{ijk} + R_{ij} + Q_i) + T \right)
  \, +
  \ln 
  \left( 
   T \, \frac{(P_{ijk}+R_{ij})}{P_{ijk}} \frac{(Q_i+R_{ij})}{Q_i} 
  \right) 
\notag\\
& \;\;\;\;\;\;\;\;\;\;\;\; 
\left.
+
  \ln \left( \frac{(Q_i+R_{ij})}{Q_i} (Q_i+T) \right) 
  \;\;\;\;\,  - \;\;
  \ln 
   \left( 
   T \, \frac{(P_{ijk}+R_{ij})}{P_{ijk}} \frac{(Q_i+R_{ij})}{Q_i} 
   \right) 
\right.
\notag\\
& \;\;\;\;\;\;\;\;\;\;\;\;\;\;\;\;\;\;\;\;\;\;\;\;
 \left.
  \quad {}   - \ln \left( Q_i + T \right)  
  \quad {} \quad {} \;\;\,  + \;\; \quad {}\quad {} 
  \ln \left( T \, \frac{(P_{ijk} + R_{ij})}{P_{ijk}} \right) 
 \right]
\notag \\
&
\; 
+ 
\int_{\widehat{(0,1)}_{-}} \frac{du}{u^2 \, P_{ijk} \, Q_i - R_{ij} \, T}
\notag\\
& \quad {}\quad {}\quad {}
 \left[ 
  \ln \left( u^2 \, Q_i + T \right) -
  \ln \left( T \, \frac{(P_{ijk} + R_{ij})}{P_{ijk}} \right)
 \right]
\notag\\
&
\; 
-
\int_{{\Gamma}_{-}} \frac{du}{u^2 \, P_{ijk} \, Q_i - R_{ij} \, T} 
\notag\\ 
& \quad {}\quad {}\quad {}
 \left[ 
  \vphantom{\frac{R_{ij} \, T}{P_{ijk} \, Q_i}}
  \ln \left( Q_i \right) - \ln \left( R_{ij} \right) 
  + \ln \left( P_{ijk}+R_{ij} \right) 
\right.
\notag\\
& \;\;\;\;\;\;\;\;\;\;\;\;
\left.
 \left.  
  + \ln \left( \frac{R_{ij} \, T}{P_{ijk} \, Q_i} \right)
  - \ln \left( T \, \frac{(P_{ijk} + R_{ij})}{P_{ijk}} \right) 
 \right]  
\right\}
\label{eqcas3bba}
\end{align}
{\bf vi)} The combination of constant logarithms in the integral on $[0,1]$ in eq. 
(\ref{eqcas3bba}) is the same as the one involved in case {\bf 1.(a)}, it therefore 
takes the same expression in terms of $\eta$ functions as in the last line 
of eq. (\ref{eqcas1soustraite}): 
\begin{align}
\mbox{constant logs}
&=
\eta \left( \frac{(Q_i+R_{ij})}{Q_i}, (Q_i+T)\right) -
\eta 
\left( 
 T \, \frac{(P_{ijk}+R_{ij})}{P_{ijk}}, \, \frac{(Q_i+R_{ij})}{Q_i}
\right) 
\notag
\end{align}
{\bf vii)} To put the combination of constant logarithms involved in the last 
integral in eq. (\ref{eqcas3bba}) in a more compact form we split the 
logarithms as:
\begin{align}
\ln \left( \frac{R_{ij} \, T}{P_{ijk} \, Q_i} \right)
&=
\ln \left( \frac{T}{P_{ijk}} \right) + \ln \left( \frac{R_{ij}}{Q_i} \right) 
+ 
\eta \left( \frac{T}{P_{ijk}}, \frac{R_{ij}}{Q_i} \right)
\notag\\
\ln \left( \frac{R_{ij}}{Q_i} \right)
&=
\ln \left( R_{ij} \right) - \ln \left( Q_i \right) + 
\eta \left( R_{ij}, \frac{1}{Q_i} \right)
\notag\\
\ln \left( \frac{T}{P_{ijk}} (P_{ijk} + R_{ij}) \right)
&=
\ln \left( \frac{T}{P_{ijk}} \right) +
\ln \left( P_{ijk} + R_{ij} \right) + 
\eta \left( \frac{T}{P_{ijk}}, (P_{ijk} + R_{ij})\right)
\notag
\end{align}
so that the combination of five logarithms in the last integral combines into
three $\eta$ functions:
\begin{align}
& 
 \ln \left( Q_i \right) - \ln \left( R_{ij} \right) 
 + \ln \left( P_{ijk}+R_{ij} \right) 
  +  \ln \left( \frac{R_{ij} \, T}{P_{ijk} \, Q_i} \right)
  - \ln \left( T \, \frac{(P_{ijk} + R_{ij})}{P_{ijk}} \right)  
\notag\\
&\quad {} \quad {} \quad {} \quad {} 
= 
\eta \left( \frac{T}{P_{ijk}}, \frac{R_{ij}}{Q_i} \right) +
\eta \left( R_{ij}, \frac{1}{Q_i} \right) -
\eta \left( \frac{T}{P_{ijk}}, (P_{ijk} + R_{ij})\right)
\end{align}
$L_4^4(\Delta_3,\Delta_2^{\{i\}},\Delta_1^{\{i,j\}},\tD_{ijk})$ finally reads:
\begin{align}
\hspace{2em}&\hspace{-2em}L_4^4(\Delta_3,\Delta_2^{\{i\}},\Delta_1^{\{i,j\}},\tD_{ijk}) \notag \\
&= - \,  
\left\{ 
 \int_{0}^{1} \;
 \frac{du}{u^2 \, P_{ijk} \, Q_i - R_{ij} \, T} 
\right.
\notag\\
& \;\;\;\;\;\;\;\;\;\;\;\;\;\;\;\;\;\;\;\;\;
 \left[ \;\;
  \vphantom{\frac{R \, T}{P_{ijk} \, Q_i}}
  \ln \left( u^2 \, P_{ijk} + (R_{ij} + Q_i + T) \right)
  - 
  \ln \left( \frac{(Q_i+R_{ij})}{Q_i} (Q_i+T) \right)
 \right.
\notag\\
& \;\;\;\;\;\;\;\;\;\;\;\;\;\;\;\;\;\;\;\;\;
  \, -
  \ln \left( u^2 \, (P_{ijk} + R_{ij} + Q_i) + T \right)
  \, +
  \ln 
  \left( 
   T \, \frac{(P_{ijk}+R_{ij})}{P_{ijk}}\frac{(Q_i+R_{ij})}{Q_i} 
  \right) 
\notag \\
& \;\;\;\;\;\;\;\;\;\;\;\;\;\;\;\;\;\;\;\;\;
\left.
  +
  \eta \left( \frac{(Q_i+R_{ij})}{Q_i}, (Q_i+T)\right)
  -
 \eta 
 \left( 
  T \, \frac{(P_{ijk} + R_{ij})}{P_{ijk}}, \, \frac{(Q_i+R_{ij})}{Q_i} 
 \right) 
\right]
\notag\\
&\quad {} \;\;\;\;\;\;\; + 
\int_{\widehat{(0,1)}_{-}} \frac{du}{u^2 \, P_{ijk} \, Q_i - R_{ij} \, T}
\notag\\
& \quad {}\quad {}\quad {}\quad {}\quad {}\quad {}
 \left[ 
  \ln \left( u^2 \, Q_i + T \right) -
  \ln \left( T \, \frac{(P_{ijk} + R_{ij})}{P_{ijk}} \right)
 \right]
\notag\\
&\quad {} \;\;\;\;\;\;\; -
\int_{{\Gamma}_{-}} \;\; \frac{du}{u^2 \, P_{ijk} \, Q_i - R_{ij} \, T} 
\notag\\ 
& \quad {}\quad {}\quad {}\quad {}\quad {}\quad {}
\left.
 \left[
  \eta \left( \frac{T}{P_{ijk}}, \frac{R_{ij}}{Q_i} \right) 
  +
  \eta \left( R_{ij}, \frac{1}{Q_i} \right) 
  -
  \eta \left( \frac{T}{P_{ijk}}, (P_{ijk} + R_{ij}) \right)
 \right]  
\right\}
\label{eqcas3eclatesoustrait}
\end{align}
We thereby get an expression reminiscent of eq. (\ref{eqcas1soustraite}) 
of case {\bf 1.(a)}, albeit modified in two ways. Firstly, the integral involving
$\ln( u^2 \, Q_i + T)$ is performed along a contour $\widehat{(0,1)}_{-}$ 
stretched from 0 to 1 which however may differ from $[0,1]$. The cut of 
$\ln( u^2 \, Q_i + T)$ indeed runs towards $\infty$ inside the ``north-east"
quadrant. Yet the branch point $\sqrt{-T/Q_i}$ which this cut emerges from may
lie inside the ``south-east" quadrant, in which case $\widehat{(0,1)}_{-}$ 
shall wrap the branch point and arc of cut, from below inside this quadrant.   
Whereas the contents in terms of dilogarithms is unchanged,
extra logarithmic contributions are generated along the wrapped cut;
this feature is readily observed on $K^C_{1,\infty}$ functions in appendix \ref{appF}.
Besides, the integral on ${\Gamma}_{-}$ provides an extra residue contribution 
involving a combination of $\eta$ functions. This contribution is non 
vanishing only if the pole $u = \sqrt{R_{ij} \, T/(P_{ijk} \, Q_i)}$ lies in 
the ``south-east" quadrant i.e. if $\Im(R_{ij} \, T/(P_{ijk} \, Q_i)) < 0$.

\vspace{0.5cm}

\noindent
{\bf 1.(c)} $\Im(\Delta_3) < 0$, $\Im(\Delta_2^{\{i\}}) > 0$, 
$\Im(\Delta_1^{\{i,j\}}) > 0$
\begin{align}
\hspace{2em}&\hspace{-2em}L_4^4(\Delta_3,\Delta_2^{\{i\}},\Delta_1^{\{i,j\}},\tD_{ijk}) \notag \\
&= 
\left\{ 
i \, \int^{+\infty}_0 \frac{du}{u^2 \, P_{ijk} \, Q_i + R_{ij} \, T} \, 
\right.
\notag\\
& \;\;\;\;\;\;\;\;\;\;\;\;\;\;\;\;\;\;\;\;\;
\left[ 
 \ln \left( \frac{R_{ij}+Q_i}{u^2 \, (P_{ijk}+R_{ij}+Q_i) - T} \right) 
 - 
 \ln \left( \frac{Q_i}{u^2 \, Q_i - T} \right) 
\right]  
\notag \\
&\quad {} \;\;\;\;\; + 
\int^{+\infty}_1 \frac{du}{u^2 \, P_{ijk} \, Q_i - R_{ij} \, T} \, 
\notag\\
& \;\;\;\;\;\;\;\;\;\;\;\;\;\;\;\;\;\;\;\;\;
\left[ 
 \ln \left( \frac{R_{ij}+Q_i}{u^2 \, (P_{ijk}+R_{ij}+Q_i)+T} \right)  
  - 
 \ln \left( \frac{Q_i}{u^2 \, Q_i + T} \right) 
\right] 
\notag \\
&\quad {} \;\;\;\;\; + 
\int^{1}_0 \frac{du}{u^2 \, P_{ijk} \, Q_i - R_{ij} \, T} \, 
\notag\\
& \;\;\;\;\;\;\;\;\;\;\;\;\;\;\;\;\;\;\;\;\;
\left. 
\left[ 
 \ln \left( \frac{R_{ij}+Q_i}{u^2 \, P_{ijk}+(R_{ij}+Q_i+T)} \right)  
 - 
 \ln \left( \frac{Q_i}{Q_i + T} \right) 
\right] 
\right\}
\label{eqcas4}
\end{align}
In this case, we have $\Im(R_{ij}+Q_i) < 0$, $\Im(Q_i+T) < 0$ and $\Im(Q_i) < 0$.
Furthermore,
\begin{itemize}
\item
$u^2 \, (P_{ijk}+R_{ij}+Q_i) - T = \tD_{ijk} \, u^2 + \Delta_3 \, (1+u^2)$ \\
thus $\Im(u^2 \, (P_{ijk}+R_{ij}+Q_i) - T) < 0$ when $u \in [0,+\infty[$, 
\item
$u^2 \, Q_i - T = (u^2+1) \, \Delta_3 - u^2 \, \Delta_2^i$ \\
thus $\Im(u^2 \, Q_i + T) < 0$ when $u \in [0,+\infty[$, 
\item
$u^2 \, (P_{ijk}+R_{ij}+Q_i) + T = u^2 \, \tD_{ijk} + (u^2-1) \, \Delta_3$ \\
thus $\Im(u^2 \, (P_{ijk}+R_{ij}+Q_i) + T) < 0$ when $u \in [1,+\infty[$,
\item
$u^2 \, Q_i + T = (u^2-1) \, \Delta_3 - u^2 \, \Delta_2^i$ \\
thus $\Im(u^2 \, Q_i + T) < 0$ when $u \in [1,+\infty[$, 
\item
$u^2 \, P_{ijk} + R_{ij} + Q_i + T 
= u^2 \, \tD_{ijk} - (1-u^2) \, \Delta_1^{\{i,j\}}$ \\
thus $\Im(u^2 \, P_{ijk} + R_{ij} + Q_i + T) < 0$ when $u \in [0,1]$. 
\end{itemize}
Similar comments as in case {\bf 1.(b)} hold regarding explicitly vanishing residues
in each of the integrals, and further similar simplifications can be carried 
through exploiting the splittings of the logarithms and recombinations of 
integrals. We do not elaborate on their derivation again, we only quote the 
result and comment it:
\begin{align}
\hspace{2em}&\hspace{-2em}L_4^4(\Delta_3,\Delta_2^{\{i\}},\Delta_1^{\{i,j\}},\tD_{ijk}) \notag \\
&= - \, 
\left\{ 
 \int_{\widehat{(0,1)}^{+}_{1}} 
\frac{du}{u^2 \, P_{ijk} \, Q_i - R_{ij} \, T} \,
\right.
\notag\\
& \;\;\;\;\;\;\;\;\;\;\;\;\;\;\;\;\;\;\;\;\;
 \left[
  - \ln \left( u^2 \, (P_{ijk}+R_{ij}+Q_i) + T \right) 
  + \ln 
    \left( 
     T \, \frac{(P_{ijk}+R_{ij})}{P_{ijk}} \frac{(R_{ij}+Q_i)}{Q_i} 
    \right)
\right]
\notag\\
& \;\;\;\;\;\;\;\;\;\;
+ \int_{\widehat{(0,1)}^{+}_{2}} \frac{du}{u^2 \, P_{ijk} \, Q_i - R_{ij} \, T} \,
\notag\\
& \;\;\;\;\;\;\;\;\;\;\;\;\;\;\;\;\;\;\;\;\;
 \left[  
  \ln \left( u^2 \, Q_i + T \right)
   - 
  \ln \left( T \, \frac{(P_{ijk}+R_{ij})}{P_{ijk}} \right)
 \right]
\notag\\
& \;\;\;\;\;\;\;\;\;\;
+ \int_{0}^{1} \frac{du}{u^2 \, P_{ijk} \, Q_i - R_{ij} \, T} \, 
\notag\\
& \;\;\;\;\;\;\;\;\;\;\;\;\;\;\;\;\;\;\;\;\;
 \left[ 
  \ln \left( u^2 \, P_{ijk} + (R_{ij}+Q_i+T) \right)
  -
  \ln \left( \frac{(R_{ij}+Q_i)}{Q_i} \, (Q_i+T) \right)
  \right.
\notag\\
& \;\;\;\;\;\;\;\;\;\;\;\;\;\;\;\;\;\;\;\;\;
 \left.
 \;\;
  + \eta \left( \frac{(R_{ij}+Q_i)}{Q_i}, (Q_i+T) \right)
  - \eta 
    \left( 
     T \, \frac{(P_{ijk}+R_{ij})}{P_{ijk}}, \, \frac{(R_{ij}+Q_i)}{Q_i}
    \right)
 \right]
\notag\\
& \;\;\;\;\;\;\;\;\;\;\;
\left.
- \int_{{\Gamma}^{+}} \frac{du}{u^2 \, P_{ijk} \, Q_i - R_{ij} \, T} \, 
   \eta 
   \left( 
    T \, \frac{(P_{ijk}+R_{ij})}{P_{ijk}}, \, \frac{(R_{ij}+Q_i)}{Q_i}
   \right)
\right\}
\label{eqcas4eclatesoustrait}
\end{align}
Eq. (\ref{eqcas4eclatesoustrait}) has a structure very similar to eq. 
(\ref{eqcas3eclatesoustrait}). The respective cuts of $\ln(u^2 \, Q_i + T)$
and $\ln(u^2 \, (P_{ijk}+R_{ij}+Q_i) + T)$ both run 
towards $\infty$ inside the ``south-east" 
quadrant. Yet either or both branch points which each of these cuts emerge 
from may lie inside the ``north-east" quadrant. Accordingly the contours 
$\widehat{(0,1)}^{+}_{1,2}$ on which the first two integrals 
are performed shall be deformations of $[0,1]$ so as to wrap the 
corresponding branch point and arc of cut from above inside the 
``north-east" quadrant.
The two contours $\widehat{(0,1)}^{+}_{1,2}$ stretched from 0 to 1 may be 
chosen distinct from each other so as to best fit the respective cuts. 
The combination of two constant $\eta$ terms in the integral on $[0,1]$ is 
the same as the one in eqs. (\ref{eqcas1soustraite}) and 
(\ref{eqcas3eclatesoustrait}). 
Lastly, and similarly to eq. (\ref{eqcas3eclatesoustrait}) there is an extra 
``residue" contribution given by the integral of the pole factor on the 
closed contour ${\Gamma}^{+}$ encircling the ``north-east" quadrant clockwise, weighted 
by a constant $\eta$ term specific to the sign case {\bf 1.(c)} at hand. The
integral is non vanishing only if the pole $\sqrt{R_{ij}T/(P_{ijk}Q_i)}$ lies
inside the ``north-east" quadrant.

\vspace{0.3cm}

\noindent
{\bf 1.(d)} $\Im(\Delta_3) < 0$, $\Im(\Delta_2^{\{i\}}) < 0$, 
$\Im(\Delta_1^{\{i,j\}}) > 0$
\begin{align}
\hspace{2em}&\hspace{-2em}L_4^4(\Delta_3,\Delta_2^{\{i\}},\Delta_1^{\{i,j\}},\tD_{ijk}) \notag \\
&= - \,
\left\{ 
i \int^{+\infty}_0 \frac{du}{u^2 \, P_{ijk} \, Q_i + R_{ij} \, T}
\right.
\notag\\
&\;\;\;\;\;\;\;\;\;\;\;\;\;\;\;\;\;\;\;\;\;\;\;\;
\left[ 
 \ln \left( \frac{R_{ij}}{u^2 \, (P_{ijk}+R_{ij})} \right) 
 - 
 \ln \left( \frac{R_{ij} + Q_i}{u^2 \, (P_{ijk}+R_{ij}+Q_i) - T} \right) 
\right] 
\notag \\
&\quad {} \;\;\;\;\;\;\;  + 
\int^{+\infty}_1 \frac{du}{u^2 \, P_{ijk} \, Q_i - R_{ij} \, T} 
\notag\\
&\;\;\;\;\;\;\;\;\;\;\;\;\;\;\;\;\;\;\;\;\;\;\;\;
\left[ 
 \ln \left( \frac{R_{ij}}{u^2 \, (P_{ijk}+R_{ij})} \right)  
 - 
 \ln \left( \frac{R_{ij} + Q_i}{u^2 \, (P_{ijk}+R_{ij}+Q_i) + T} \right) 
\right] 
\notag \\
&\quad {}  \;\;\;\;\;\;\; + 
\int^{1}_0 \;\;\;\; \frac{du}{u^2 \, P_{ijk} \, Q_i - R_{ij} \, T} \, 
\notag\\
&\;\;\;\;\;\;\;\;\;\;\;\;\;\;\;\;\;\;\;\;\;\;\;\;
\left[ 
 \ln \left( \frac{R_{ij}}{u^2 \, P_{ijk}+R_{ij}} \right) 
 - 
 \ln \left( \frac{R_{ij}+Q_i}{u^2 \, P_{ijk}+R_{ij}+Q_i + T} \right) 
\right] 
\notag \\
&\quad {} \;\;\;\;\;\;\; + 
\int^{1}_0 \;\;\;\; \frac{du}{u^2 \, P_{ijk} \, Q_i - R_{ij} \, T} \, 
\notag\\
&\;\;\;\;\;\;\;\;\;\;\;\;\;\;\;\;\;\;\;\;\;\;\;\;
\left. 
\left[ 
 \ln \left( \frac{u^2 \, P_{ijk} + R_{ij}}{u^2 \, (P_{ijk} + R_{ij})} \right) 
 - 
 \ln \left( \frac{Q_i + T}{u^2 \, Q_i + T} \right) 
\right] 
\right\}
\label{eqcas5}
\end{align}
In this case we have $
\Im(R_{ij}) < 0$, $\Im(P_{ijk} + R_{ij}) < 0$, 
$\Im(R_{ij}+Q_i) < 0$, $\Im(Q_i+T) > 0$.
Furthermore, 
\begin{itemize}
\item
$u^2 \, (P_{ijk}+R_{ij}+Q_i) - T = u^2 \, \tD_{ijk} + (1+u^2) \, \Delta_3$ \\
thus $\Im(u^2 \, (P_{ijk}+R_{ij}+Q_i) - T) < 0$ when $u \in [0,+\infty[$, 
\item
$u^2 \, (P_{ijk}+R_{ij}+Q_i) + T = u^2 \, \tD_{ijk} + (u^2-1) \, \Delta_3$ \\
thus $\Im(u^2 \, (P_{ijk}+R_{ij}+Q_i) + T) < 0$ when $u \in [1,+\infty[$,
\item
$u^2 \, P_{ijk} + R_{ij} 
= u^2 \, \tD_{ijk} - (1 - u^2) \, \Delta_1^{\{i,j\}} + \Delta_{2}^{\{i\}}$ \\
thus $\Im(u^2 \, P_{ijk} + R_{ij}) < 0$ when $u \in [0,1]$, 
\item
$u^2 \, Q_i + T = - (1-u^2) \, \Delta_3 - u^2 \, \Delta_2^{\{i\}}$ \\
thus $\Im(u^2 \, Q_i + T) > 0$ when $u \in [0,1]$,
\item 
$u^2 \, P_{ijk} + R_{ij} + Q_i + T 
= u^2 \, \tD_{ijk} - (1-u^2) \, \Delta_1^{\{i,j\}}$ \\
thus $\Im(u^2 \, P_{ijk} + R_{ij} + Q_i +T) < 0$ when $u \in [0,1]$. 
\end{itemize}
The use of the same technics as in {\bf 1.(b)} leads to the following alternative expression:
\begin{align}
\hspace{2em}&\hspace{-2em}L_4^4(\Delta_3,\Delta_2^{\{i\}},\Delta_1^{\{i,j\}},\tD_{ijk}) \notag \\
&= - \, 
\left\{ 
 \int_{\widehat{(0,1)}^{+}} \frac{du}{u^2 \, P_{ijk} \, Q_i - R_{ij} \, T} \,
\right.
\notag\\
& \;\;\;\;\;\;\;\;\;\;\;\;\;\;\;\;\;\;\;\;\;
 \left[
  - \ln \left( u^2 \, (P_{ijk}+R_{ij}+Q_i) + T \right) 
  + \ln 
    \left( 
     T \, \frac{(P_{ijk}+R_{ij})}{P_{ijk}} \frac{(R_{ij}+Q_i)}{Q_i} 
    \right)
\right]
\notag\\
& \;\;\;\;\;\;\;\;\;\;
+ \int_{0}^{1} \frac{du}{u^2 \, P_{ijk} \, Q_i - R_{ij} \, T} \, 
\notag\\
& \;\;\;\;\;\;\;\;\;\;\;\;\;\;\;\;\;\;\;\;\;
 \left[ 
  \ln \left( u^2 \, P_{ijk} + (R_{ij}+Q_i+T) \right)
  -
  \ln \left( \frac{(R_{ij}+Q_i)}{Q_i} \, (Q_i+T) \right)
 \right.
\notag\\
& \;\;\;\;\;\;\;\;\;\;\;\;\;\;\;\;\;\;\;\;\;
  + 
  \;\;\;\;\;\;\;\;
  \ln \left( u^2 \, Q_i + T \right) 
  \;\;\;\;\;\;\;\;\;\;\;\;\;
   - 
  \ln \left( T \, \frac{(P_{ijk}+R_{ij})}{P_{ijk}}\right)
 \notag\\
& \;\;\;\;\;\;\;\;\;\;\;\;\;\;\;\;\;\;\;\;\;
 \left.
  + \eta \left( \frac{(R_{ij}+Q_i)}{Q_i}, (Q_i+T) \right)
  - \eta 
    \left( 
     T \, \frac{(P_{ijk}+R_{ij})}{P_{ijk}}, \, \frac{(R_{ij}+Q_i)}{Q_i}
    \right)
 \right]
\notag\\
& \;\;\;\;\;\;\;\;\;\;\;
\left.
- \int_{{\Gamma}^{+}} \frac{du}{u^2 \, P_{ijk} \, Q_i - R_{ij} \, T} \, 
   \eta 
   \left( 
    T \, \frac{(P_{ijk}+R_{ij})}{P_{ijk}} \frac{R_{ij}}{Q_i},  \,
    \frac{R_{ij}+Q_i}{R_{ij}}
   \right)
\right\}
\label{eqcas5eclatesoustrait}
\end{align}
Again eq. (\ref{eqcas5eclatesoustrait}) has a structure very similar to eqs. 
(\ref{eqcas3eclatesoustrait}) and (\ref{eqcas4eclatesoustrait}). The cut of
$\ln(u^2\, (P_{ijk}+R_{ij}+Q_i) + T)$ runs towards $\infty$ inside the
``south-east'' quadrant yet the branch point 
$\sqrt{-T/(P_{ijk}+R_{ij}+Q_i)}$ which it emerges from may lie inside the
``north-east" quadrant.
Accordingly the contour $\widehat{(0,1)}^{+}$ stretched from 0 to 1 may
wrap the branch point and arc of cut from above inside the ``north-east" 
quadrant.

\vspace{0.3cm}

\noindent
{\bf 2.(a)} $\Im(\Delta_3) > 0$, $\Im(\Delta_2^{\{i\}}) > 0$, 
$\Im(\Delta_1^{\{i,j\}}) < 0$
\begin{align}
\hspace{2em}&\hspace{-2em}L_4^4(\Delta_3,\Delta_2^{\{i\}},\Delta_1^{\{i,j\}},\tD_{ijk}) \notag \\
&=  
\left\{ 
 \;\;\;\, i \int^{+\infty}_0 \frac{du}{u^2 \, P_{ijk} \, Q_i + R_{ij} \, T} \,
\left[ 
 \ln \left( \frac{- u^2 \, P_{ijk} + R_{ij}}{- u^2 \, P_{ijk}} \right) 
 - 
 \ln \left( \frac{Q_i+T}{T} \right) 
\right] 
\right. 
\notag \\
&\quad {} \;\;\;\;\;\;\; + \;\; 
\int^{+\infty}_0 \frac{du}{u^2 \, P_{ijk} \, Q_i - R_{ij} \, T}  
\left[ 
 \ln \left( \frac{ - R_{ij}}{u^2 \, P_{ijk}} \right) 
 - 
 \ln \left( \frac{Q_i+R_{ij}}{-u^2 \, P_{ijk}-T} \right) 
\right] 
\notag \\
&\quad {} \;\;\;\;\;\;\; + 
i \int^{+\infty}_0 \frac{du}{u^2 \, P_{ijk} \, Q_i + R_{ij} \, T} \, 
\notag\\
& \;\;\;\;\;\;\;\;\;\;\;\;\;\;\;\;\;\;\;\;\;\;\;\;\;
\left[ 
 \ln \left( \frac{R_{ij}}{-u^2 \, P_{ijk} + R_{ij}} \right) 
 - 
 \ln \left( \frac{Q_i+R_{ij}}{- u^2 \, P_{ijk} + R_{ij} + Q_i + T} \right) 
\right] 
\notag \\
&\quad {} \;\;\;\;\;\;\; + \;\;
\int^{1}_0 \;\; 
\frac{du}{u^2 \, P_{ijk} \, Q_i - R_{ij} \, T} \, 
\notag\\
& \;\;\;\;\;\;\;\;\;\;\;\;\;\;\;\;\;\;\;\;\;\;\;\;\;
\left[ 
 \ln 
  \left( \frac{u^2 \, (P_{ijk} + R_{ij} + Q_i) + T}{u^2 \, P_{ijk} + T} \right) 
 - \ln \left( \frac{u^2 \, Q_i+T}{T} \right) 
\right] 
\notag \\
&\quad {} \;\;\;\;\;\;\; + 
\int^{+\infty}_1 \frac{du}{u^2 \, P_{ijk} \, Q_i - R_{ij} \, T} \, 
\notag\\
& \;\;\;\;\;\;\;\;\;\;\;\;\;\;\;\;\;\;\;\;\;\;\;\;\;
\left. 
\left[ 
 \ln 
  \left( \frac{u^2 \, P_{ijk} + R_{ij} + Q_i + T}{u^2 \, P_{ijk} + T} \right) 
 - 
\ln \left( \frac{Q_i+T}{T} \right) 
\right] 
\right\}
\label{eqcas2a}
\end{align}
In this case, we have $\Im(P_{ijk}) < 0$, $\Im(R_{ij}) > 0$, 
$\Im(R_{ij}+Q_i) > 0$, $\Im(R_{ij}+Q_i+T) > 0$, $\Im(Q_i+T) < 0$ and 
$\Im(T) < 0$. Furthermore,
\begin{itemize}
\item
  $u^2 \, P_{ijk} - R_{ij} = u^2 \, \tD_{ijk}+ (1+u^2) \, \Delta_1^{\{i,j\}} - \Delta_2^{\{i\}}$ \\
  thus $\Im(u^2 \, P_{ijk} - R_{ij}) < 0$ when $u \in [0,+\infty[$, 
\item 
$u^2 \, (P_{ijk}+R_{ij}+Q_i) + T = u^2 \, \tD_{ijk} - (1-u^2) \, \Delta_3$\\ 
thus $\Im(u^2 \, (P_{ijk}+R_{ij}+Q_i) + T) < 0$ when $u \in [0,1]$, 
\item
$u^2 \, P_{ijk} + T = u^2 \, (\tD_{ijk}+\Delta_1^{\{i,j\}}) - \Delta_3$ \\
thus $\Im(u^2 \, P_{ijk} + T) < 0$ when $u \in [0,+\infty[$, 
\item 
$- u^2 \, P_{ijk} + R_{ij} + Q_i + T 
= - [ u^2 \, \tD_{ijk} + (u^2+1) \, \Delta_1^{\{i,j\}}]$ \\
thus $\Im(- u^2 \, P_{ijk} + R_{ij} + Q_i + T) > 0$ when $u \in [0,+\infty[$.
\item
$u^2 \, Q_i + T = - (1-u^2) \, \Delta_3 - u^2 \, \Delta_2^{\{i\}}$ \\
thus $\Im(u^2 \, Q_i + T) < 0$ when $u \in [0,1]$, 
\item 
$u^2 \, P_{ijk} + R_{ij} + Q_i + T 
= u^2 \, \tD_{ijk} + (u^2-1) \, \Delta_1^{\{i,j\}}$ \\
thus $\Im(u^2 \, P_{ijk} + R_{ij} + Q_i + T) < 0$ when $u \in [1,+\infty[$.
\end{itemize} 
The same tricks as in case {\bf 1.(b)} lead to:
\begin{align}
\hspace{2em}&\hspace{-2em}L_4^4(\Delta_3,\Delta_2^{\{i\}},\Delta_1^{\{i,j\}},\tD_{ijk}) \notag \\
&= - \, 
\left\{
 \int_0^{1} 
 \frac{du}{u^2 \, P_{ijk} \, Q_i - R_{ij} \, T} \, 
\right.
\notag\\
&\quad {} \quad{}\quad {} \quad{} \;\;
 \left[
  \vphantom{\frac{R_{ij} \, T}{Q_{i}}} 
  \;\;\;
  \ln \left( u^2 \, Q_i + T \right) 
  \quad {} \quad {} \quad {} \quad {} \quad {} \quad {} \;
  - \;
  \ln \left( T \, \frac{(P_{ijk} + R_{ij})}{P_{ijk}} \right)
\right.
\notag\\
&\quad {} \quad{}\quad {} \quad{} \quad {}
  - 
  \ln \left( u^2 \, (P_{ijk} + R_{ij} + Q_i) + T \right) 
  + 
  \ln 
  \left( 
   T \, \frac{(P_{ijk} + R_{ij})}{P_{ijk}} \frac{(Q_i + R_{ij})}{Q_{i}} 
  \right)
\notag\\
&\quad {} \quad{}\quad {} \quad{} \quad {}
\left.
  + \;
  \eta \left( \frac{(R_{ij}+Q_i)}{Q_i}, \; (Q_i+T) \right)
  \;\;\;\; - \;\;
 \eta 
  \left( 
   T \, \frac{(P_{ijk} + R_{ij})}{P_{ijk}}, \, \frac{(R_{ij} + Q_i)}{Q_{i}} 
  \right)
\right]
\notag \\
& \quad {} \quad {}\quad {}  +
 \int_{\widehat{(0,1)}^{+}}  
 \frac{du}{u^2 \, P_{ijk} \, Q_i - R_{ij} \, T}  
\notag\\
& \quad {} \quad {}\quad {} \quad {}\quad {}
 \left[
  \ln \left( u^2 \, P_{ijk} + (R_{ij} + Q_i + T) \right)
  - 
  \ln \left( \frac{(Q_i + R_{ij})}{Q_{i}} (Q_i + T) \right)
 \right]
\notag\\
&\quad {} \quad {}\quad {}  + 
 \int_{{\Gamma}^{+}}  \,
 \frac{du}{u^2 \, P_{ijk} \, Q_i - R_{ij} \, T}  
\notag\\
& \quad {} \quad {}\quad {} \quad {}\quad {}
\left.
  \left[ 
   \eta \left( (R_{ij}+Q_i)(Q_i + T), \frac{1}{Q_{i}}  \right)
   - 
   \eta \left( R_{ij} \, T, \frac{1}{Q_{i}} \right) 
  \right]
\right\}
\label{eqcas2aa}
\end{align}
Again eq. (\ref{eqcas2aa}) has a structure very similar to eqs. 
(\ref{eqcas3eclatesoustrait}), (\ref{eqcas4eclatesoustrait}) and 
(\ref{eqcas5eclatesoustrait}). The cut of
$\ln(u^2 \, P_{ijk} + (R_{ij} + Q_i + T))$ run towards $\infty$ inside the
``south-east" quadrant. yet the branch point $\sqrt{-(R_{ij} + Q_i + T)/P_{ijk}}$
may lie inside the ``north-east" quadrant. In this case, the
contour $\widehat{(0,1)}^{+}$ stretched from 0 to 1 shall wrap the branch 
point and the arc of cut located inside the ``north-east" quadrant, from above
inside this quadrant. 

\vspace{0.3cm}

\newpage
\noindent
{\bf 2.(b)} $\Im(\Delta_3) > 0$, $\Im(\Delta_2^{\{i\}}) < 0$, 
$\Im(\Delta_1^{\{i,j\}}) < 0$
\begin{align}
\hspace{2em}&\hspace{-2em}L_4^4(\Delta_3,\Delta_2^{\{i\}},\Delta_1^{\{i,j\}},\tD_{ijk}) \notag \\
&= - \, 
\left\{
 i \, \int^{+\infty}_0 \frac{du}{u^2 \, P_{ijk} \, Q_i + R_{ij} \, T} \, 
\right.
\notag\\
&\;\;\;\;\;\;\;\;\;\;\;\;\;\;\;\;\;\;\;\;\;
 \left[ 
  \;\;
  \ln \left( \frac{Q_i \, u^2-T}{-T} \right) 
  \;\; - \;\;
  \ln \left( \frac{P_{ijk} + R_{ij}}{P_{ijk}} \right) 
 \right.
\notag \\
&\;\;\;\;\;\;\;\;\;\;\;\;\;\;\;\;\;\;\;\;\;\;
 \left.
  - 
  \ln \left( \frac{Q_i}{Q_i + T} \right) 
  \;\;\;\;\; + \;\;\;
  \ln \left( \frac{Q_i + R_{ij}}{ - u^2 \, P_{ijk} + (R_{ij}+Q_i+T)} \right) 
\right] 
\notag \\
&\quad {} \;\;\;\;\;\;\; - 
\int^{+\infty}_0 \frac{du}{u^2 \, P_{ijk} \, Q_i - R_{ij} \, T} \, 
\left[ 
 \ln \left( \frac{Q_i}{-T} \right) 
 - 
\ln \left( \frac{R_{ij}+Q_i}{-u^2 \, P_{ijk} - T} \right) 
\right] 
\notag \\
&\quad {} \;\;\;\;\;\;\; + 
\int^{+\infty}_1 \frac{du}{u^2 \, P_{ijk} \, Q_i - R_{ij} \, T} \, 
\left[ 
 \ln \left( \frac{u^2 \, (P_{ijk} + R_{ij})}{u^2 \, P_{ijk} + R_{ij}} \right)
 - 
 \ln \left( \frac{u^2 \, Q_i+T}{Q_i+T} \right) 
\right] 
\notag \\
&\quad {} \;\;\;\;\;\;\; + 
\int^{+\infty}_1 \frac{du}{u^2 \, P_{ijk} \, Q_i - R_{ij} \, T} \, 
\notag\\
&\;\;\;\;\;\;\;\;\;\;\;\;\;\;\;\;\;\;\;\;\;\;\;\;
\left[ 
 \ln \left( \frac{u^2 \, P_{ijk} + R_{ij}}{u^2 \, P_{ijk}} \right)
 - 
 \ln
 \left( 
  \frac{u^2 \, P_{ijk} + R_{ij} + Q_i + T}{u^2 \, P_{ijk} + T} 
 \right) 
\right] 
\notag \\
&\quad {} \;\;\;\;\;\;\; +  
\int^{1}_0 \frac{du}{u^2 \, P_{ijk} \, Q_i - R_{ij} \, T} \, 
\notag\\
&\;\;\;\;\;\;\;\;\;\;\;\;\;\;\;\;\;\;\;\;\;\;\;\;
\left.
\left[
 \ln \left( \frac{P_{ijk} + R_{ij}}{P_{ijk}} \right) 
 - 
 \ln 
 \left( 
  \frac{u^2 \, (P_{ijk} + R_{ij} + Q_i) + T}{u^2 \, P_{ijk} + T} 
 \right) 
\right] 
\right\}
\label{eqcas2b}
\end{align}
We have here:
$\Im(P_{ijk}) < 0$, $\Im(P_{ijk}+R_{ij}) < 0$, $\Im(R_{ij}+Q_i) > 0$,
$\Im(Q_i) > 0$, $\Im(Q_i+T) > 0$ and $\Im(T) < 0$. Furthermore,
\begin{itemize}
\item
$u^2 \, (P_{ijk}+R_{ij}+Q_i) + T 
=  u^2 \, \tD_{ijk} - \, (1-u^2) \, \Delta_3$ \\
thus $\Im(u^2 \, (P_{ijk}+R_{ij}+Q_i) + T) < 0$ when $u \in [0,1]$, 
\item
$u^2 \, P_{ijk} + T = u^2 \, (\tD_{ijk} + \Delta_1^{\{i,j\}}) - \Delta_3$ \\
thus $\Im(u^2 \, P_{ijk} + T) < 0$ when $u \in [0,+\infty[$, 
\item
$u^2 \, Q_i + T = (u^2 - 1) \, \Delta_3  - u^2 \, \Delta_2^{\{i\}}$ \\
thus $\Im(u^2 \, Q_i + T) > 0$ when $u \in [1,+\infty[$, 
\item
$u^2 \, Q_i - T = 
(1 +u^2) \, \Delta_3 - u^2 \, \Delta_2^{\{i\}}$ \\
thus $\Im(u^2 \, Q_i - T) > 0$ when $u \in [0,+\infty[$, 
\item
$u^2 \, P_{ijk} + R_{ij} = 
u^2 \, \tD_{ijk} + (u^2 - 1) \, \Delta_1^{\{i,j\}} + \Delta_{2}^{\{i\}}$ \\ 
thus $\Im(u^2 \, P_{ijk} + R_{ij}) < 0$ when $u \in [1,+\infty[$, 
\item
$- \, u^2 \, P_{ijk} + (R_{ij} + Q_i + T) = 
- \, u^2 \, \tD_{ijk} -  \, (1+u^2) \, \Delta_1^{\{i,j\}}$  \\
thus $\Im(-u^2 \, P_{ijk} +( R_{ij} + Q_i + T)) > 0$ when $u \in [0,+\infty[$,  
\item
$u^2 \, P_{ijk} + (R_{ij} + Q_i + T) = 
u^2 \, \tD_{ijk} + (u^2 - 1) \, \Delta_1^{\{i,j\}}$ \\  
thus $\Im(u^2 \, P_{ijk} + (R_{ij} + Q_i + T)) < 0$ when $u \in [1,+\infty[$. 
\end{itemize}
Using the same technics as in previous cases yields:
\begin{align}
\hspace{2em}&\hspace{-2em}L_4^4(\Delta_3,\Delta_2^{\{i\}},\Delta_1^{\{i,j\}},\tD_{ijk}) \notag \\
&= - \, 
\left\{
 \int_0^{1} 
 \frac{du}{u^2 \, P_{ijk} \, Q_i - R_{ij} \, T} \, 
\right.
\notag\\
&\quad {} \quad{}\quad {} \quad{} \;\;
 \left[
  \vphantom{\frac{R_{ij} \, T}{Q_{i}}} 
  - 
  \ln \left( u^2 \, (P_{ijk} + R_{ij} + Q_i) + T \right) 
  + 
  \ln 
  \left( 
   T \, \frac{(P_{ijk} + R_{ij})}{P_{ijk}} \frac{(Q_i + R_{ij})}{Q_{i}} 
  \right)
\right.
\notag\\
&\quad {} \quad{}\quad {} \quad{} \quad {}
\left.
  + \;
  \eta \left( \frac{(R_{ij}+Q_i)}{Q_i}, \; (Q_i+T) \right)
  \;\;\;\; - \;\;
 \eta 
  \left( 
   T \, \frac{(P_{ijk} + R_{ij})}{P_{ijk}}, \, \frac{(R_{ij} + Q_i)}{Q_{i}} 
  \right)
\right]
\notag \\
& \quad {} \quad {}\quad {}  +
 \int_{\widehat{(0,1)}^{+}} 
 \frac{du}{u^2 \, P_{ijk} \, Q_i - R_{ij} \, T}  
\notag\\
& \quad {} \quad {}\quad {} \quad {}\quad {}
 \left[
  \ln \left( u^2 \, P_{ijk} + (R_{ij} + Q_i + T) \right)
  - 
  \ln \left( \frac{(Q_i + R_{ij})}{Q_{i}} (Q_i + T) \right)
 \right]
\notag\\
& \quad {} \quad {}\quad {}  +
 \int_{\widehat{(0,1)}_{-}} 
 \frac{du}{u^2 \, P_{ijk} \, Q_i - R_{ij} \, T}  
\notag\\
& \quad {} \quad {}\quad {} \quad {}\quad {}
 \left[
  \ln \left( u^2 \, Q_i + T \right)
  \; - \;
  \ln \left( T \, \frac{( P_{ijk} + R_{ij})}{P_{ijk}} \right)
 \right]
\notag\\
&\quad {} \quad {}\quad {}  + 
 \int_{{\Gamma}^{+}}  \,
 \frac{du}{u^2 \, P_{ijk} \, Q_i - R_{ij} \, T}  
   \; \eta \left( \frac{(R_{ij} + Q_i)}{Q_i}, \, (Q_{i} + T ) \right)
\notag\\
&\quad {} \quad {}\quad {}  + 
\left.
 \int_{{\Gamma}_{-}}  \,
 \frac{du}{u^2 \, P_{ijk} \, Q_i - R_{ij} \, T}  
 \;  \eta \left( T, \, \frac{(P_{ijk} + R_{ij})}{P_{ijk}} \right)
\right\}
\label{eqcas2bb}
\end{align}
Again eq. (\ref{eqcas2bb}) has a structure very similar to eqs. 
(\ref{eqcas3eclatesoustrait}), (\ref{eqcas4eclatesoustrait}) and 
(\ref{eqcas5eclatesoustrait}) and (\ref{eqcas2aa}). The cut of
$\ln ( u^2 \, P_{ijk} + (R_{ij} + Q_i + T))$ runs towards $\infty$ inside the
``south-east" quadrant, yet the branch point which this cut originates from 
may lie in the``north-east" quadrant. Accordingly the contour 
$\widehat{(0,1)}^{+}$ stretched
from 0 to 1 shall wrap the branch point and finite arc of cut 
from above inside this quadrant.
A mirror situation holds for the cut of
$\ln ( u^2 \, Q_i + T)$ which runs towards $\infty$ inside the
north-east quadrant yet with the branch point possibly
lying in the``south-east" quadrant. In the latter case the contour 
$\widehat{(0,1)}_{-}$ stretched from 0 to 1 shall be  wrap the branch point 
and finite arc of cut possibly located in the ``south-east" quadrant, 
from below inside that quadrant.

\vspace{1.0cm}

\noindent
{\bf 2.(c)} $\Im(\Delta_3) < 0$, $\Im(\Delta_2^{\{i\}}) > 0$, 
$\Im(\Delta_1^{\{i,j\}}) < 0$
\begin{align}
\hspace{2em}&\hspace{-2em}L_4^4(\Delta_3,\Delta_2^{\{i\}},\Delta_1^{\{i,j\}},\tD_{ijk}) \notag \\
&= - \, 
\left\{ \; 
i \, \int_0^{+\infty} \frac{du}{u^2 \, P_{ijk} \, Q_i + R_{ij} \, T} 
\right.
\notag\\
&\;\;\;\;\;\;\;\;\;\;\;\;\;\;\;\;\;\;\;\;\;\;\;\;\;\;
\left[ 
 \ln \left( \frac{- u^2 \, P_{ijk} + R_{ij}}{- u^2 \, P_{ijk}} \right) 
 - 
 \ln \left( \frac{-u^2 \, P_{ijk} +(R_{ij}+Q_i+T)}{-u^2 \, P_{ijk} + T} \right) 
\right]  
\notag \\
&\quad {} \;\;\;\;\;\;\; +
i \, \int_0^{+\infty} \frac{du}{u^2 \, P_{ijk} \, Q_i + R_{ij} \, T} \, 
\notag\\
&\;\;\;\;\;\;\;\;\;\;\;\;\;\;\;\;\;\;\;\;\;\;\;\;\;\;
\left[ 
 \ln 
 \left( 
  \frac{- u^2 \, (P_{ijk} + R_{ij} + Q_i) + T}{-u^2 \, P_{ijk} + T} 
 \right) 
 - 
 \ln \left( \frac{-u^2 \, Q_i + T}{T} \right) 
\right] 
\notag \\
&\quad {} \;\;\;\;\;\;\; +
 i \, \int_0^{+\infty} \frac{du}{u^2 \, P_{ijk} \, Q_i + R_{ij} \, T} \, 
\left[ 
 \ln \left( \frac{R_{ij}}{-u^2 \, P_{ijk} + R_{ij}} \right) 
 - 
 \ln \left( \frac{Q_i}{Q_i + T} \right) 
\right] 
\notag \\
&\quad {} \;\;\;\;\;\;\; - \;\;\,
 \int_0^{+\infty} \frac{du}{u^2 \, P_{ijk} \, Q_i - R_{ij} \, T} \, 
\left[ 
 \ln \left( \frac{-R_{ij}}{u^2 \, P_{ijk}} \right) 
 - 
 \ln \left( \frac{Q_i}{-T} \right) 
 \right] 
\notag \\
&\quad {} \;\;\;\;\;\;\; + \;\;\,
\int_1^{+\infty} \frac{du}{u^2 \, P_{ijk} \, Q_i - R_{ij} \, T}  
\notag\\
&\;\;\;\;\;\;\;\;\;\;\;\;\;\;\;\;\;\;\;\;\;\;\;\;\;
\left. 
\left[ 
 \ln 
 \left( 
  \frac{u^2 \, (P_{ijk} + R_{ij} + Q_i) + T}
  {u^2 \, P_{ijk} + (R_{ij} + Q_i + T)} 
 \right) 
 - 
 \ln \left( \frac{u^2 \, Q_i+T}{Q_i+T} \right) 
\right] 
\right\}
\label{eqcas6}
\end{align}
In this case, we have $\Im(P_{ijk}) < 0$, $\Im(R_{ij}) > 0$, $\Im(Q_i) < 0$, 
$\Im(Q_i+T) < 0$ and $\Im(T) > 0$. Furthermore, 
\begin{itemize}
\item
  $u^2 \, P_{ijk} - R_{ij} = u^2 \, \tD_{ijk}+ (1+u^2) \, \Delta_1^{\{i,j\}} - \Delta_2^{\{i\}}$ \\
  thus $\Im(u^2 \, P_{ijk} - R_{ij}) < 0$ when $u \in [0,+\infty[$, 
\item
$u^2 \, (P_{ijk}+R_{ij}+Q_i) - T = \tD_{ijk} \, u^2 + \Delta_3 \, (1+u^2)$ \\
thus $\Im(u^2 \, (P_{ijk}+R_{ij}+Q_i) - T) < 0$ when $u \in [0,+\infty[$, 
\item
$-u^2 \, P_{ijk} + T = -u^2 \, (\tD_{ijk}+\Delta_1^{\{i,j\}}) - \Delta_3$ \\
thus $\Im(-u^2 \,P_{ijk} + T) > 0$ when $u \in [0,+\infty[$, 
\item
$-u^2 \, P_{ijk} + R_{ij} + Q_i + T 
= - u^2 \, \tD_{ijk} - (1+u^2) \, \Delta_1^{\{i,j\}}$ \\ 
thus $\Im(-u^2 \, P_{ijk} + R_{ij} + Q_i + T) > 0$ when $u \in [0,+\infty[$,
\item
$u^2 \, Q_i - T =  (1+u^2) \, \Delta_3 - u^2 \, \Delta_2^{\{i\}}$ \\
thus $\Im(u^2 \, Q_i - T) < 0$ when $u \in [0,+\infty[$, 
\item
$u^2 \, (P_{ijk}+R_{ij}+Q_i) + T = u^2 \, \tD_{ijk} +  (u^2-1) \, \Delta_3$ \\
thus $\Im(u^2 \, (P_{ijk}+R_{ij}+Q_i) + T) < 0$ when $u \in [1,+\infty[$,
\item
$u^2 \, P_{ijk} + R_{ij} + Q_i + T 
= u^2 \, \tD_{ijk} + (u^2-1) \, \Delta_1^{\{i,j\}}$ \\
thus $\Im(u^2 \, P_{ijk} + R_{ij} + Q_i + T) < 0$ when $u \in [1,+\infty[$, 
\item
$u^2 \, Q_i + T = (u^2-1) \, \Delta_3  - u^2 \, \Delta_2^{\{i\}}$ \\
thus $\Im(u^2 \, Q_i + T) < 0$ when $u \in [1,+\infty[$. 
\end{itemize}
The implementation of the technics used in case {\bf 1.(b)} leads to:
\begin{align}
\hspace{2em}&\hspace{-2em}L_4^4(\Delta_3,\Delta_2^{\{i\}},\Delta_1^{\{i,j\}},\tD_{ijk}) \notag \\
&= - \, 
\left\{
 \int_{\widehat{(0,1)}^{+}_{1}} 
 \frac{du}{u^2 \, P_{ijk} \, Q_i - R_{ij} \, T}  
\right.
\notag\\
& \quad {} \quad {}\quad {} \quad {}\;
 \left[
  \ln \left( u^2 \, P_{ijk} + (R_{ij} + Q_i + T) \right)
  - 
  \ln \left( \frac{(Q_i + R_{ij})}{Q_{i}} (Q_i + T) \right)
 \right]
\notag\\
& \quad {} \quad {}\quad {}  +
 \int_{\widehat{(0,1)}^{+}_{2}} 
 \frac{du}{u^2 \, P_{ijk} \, Q_i - R_{ij} \, T} \, 
\notag\\
&\quad {} \quad{}\quad {} \quad{} \;
 \left[
  \vphantom{\frac{R_{ij} \, T}{Q_{i}}} 
  - 
  \ln \left( u^2 \, (P_{ijk} + R_{ij} + Q_i) + T \right) 
  + 
  \ln 
  \left( 
   T \, \frac{(P_{ijk} + R_{ij})}{P_{ijk}} \frac{(Q_i + R_{ij})}{Q_{i}} 
  \right)
 \right]
\notag\\
& \quad {} \quad {}\quad {}  +
 \int_{\widehat{(0,1)}^{+}_{3}} 
 \frac{du}{u^2 \, P_{ijk} \, Q_i - R_{ij} \, T}  
\notag\\
& \quad {} \quad {}\quad {} \quad {}\;
 \left[
  \ln \left( u^2 \, Q_i + T \right)
  \; - \;
  \ln \left( T \, \frac{(P_{ijk} + R_{ij})}{P_{ijk}} \right)
 \right]
\notag\\
& \quad {} \quad {}\quad {} 
  + \;
 \int_{0}^{1} 
 \frac{du}{u^2 \, P_{ijk} \, Q_i - R_{ij} \, T}  
\notag\\
& \quad {} \quad {}\quad {} \quad {}\;
 \left[ 
  \eta \left( \frac{(R_{ij}+Q_i)}{Q_i}, \, (Q_i+T) \right)
  \;\;\;\; - \;\;
  \eta 
  \left( 
   T \, \frac{(P_{ijk} + R_{ij})}{P_{ijk}}, \, \frac{(R_{ij} + Q_i)}{Q_{i}} 
  \right)
 \right]
\notag \\
&\quad {} \quad {}\quad {}  + 
 \int_{{\Gamma}^{+}}  \,
 \frac{du}{u^2 \, P_{ijk} \, Q_i - R_{ij} \, T}  
 \notag\\
& \quad {} \quad {}\quad {} \quad {}\;
 \left[ 
  \eta \left( \frac{(R_{ij}+Q_i)}{Q_i}, \; (Q_i+T) \right)
  \;\;\;\; - \;\;
  \eta 
  \left( 
   T \, \frac{(P_{ijk} + R_{ij})}{P_{ijk}}, \, \frac{(R_{ij} + Q_i)}{Q_{i}} 
  \right)
 \right] 
\notag\\
&\quad {} \quad {}\quad {}  + 
 \int_{{\Gamma}_{-}}  \,
 \frac{du}{u^2 \, P_{ijk} \, Q_i - R_{ij} \, T}  
\notag\\
& \quad {} \quad {}\quad {} \quad {}\;
\left. 
 \left[ 
  \eta \left( R_{ij} T, \, \frac{1}{P_{ijk}Q_i} \right)
  + \eta \left(  -R_{ij}, \, -T \right) 
  - \eta \left( P_{ijk}, \, Q_i \right)
\right]  
\right\}
\label{eqcas2cc}
\end{align}
Again eq. (\ref{eqcas2cc}) has a structure very similar to eqs. 
(\ref{eqcas3eclatesoustrait}), (\ref{eqcas4eclatesoustrait}) and 
(\ref{eqcas5eclatesoustrait}), (\ref{eqcas2aa}) and (\ref{eqcas2bb}). 
All three $u$-dependent logarithms have cuts running towards $\infty$ in the
south-east quadrant, yet the branch points which they respectively emerge from
may be located inside the ``north-east'' quadrant.
Accordingly the contours $\widehat{(0,1)}^{+}_{j}, \, j=1,2,3$ are stretched 
from 0 to 1 and may wrap the branch points and arcs of cuts from above 
inside the ``north-east" quadrant. These contours may be chosen distinct 
from each other so as to best fit the respective cuts.


\newpage
{\bf 2.(d)} $\Im(\Delta_3) < 0$, $\Im(\Delta_2^{\{i\}}) < 0$, 
$\Im(\Delta_1^{\{i,j\}}) < 0$

\begin{align}
\hspace{2em}&\hspace{-2em}L_4^4(\Delta_3,\Delta_2^{\{i\}},\Delta_1^{\{i,j\}},\tD_{ijk}) \notag \\
&=  
\left\{ 
\;\;\, i \int^{+\infty}_0 
\frac{du}{u^2 \, P_{ijk} \, Q_i + R_{ij} \, T} \, 
\right. 
\notag\\
&\;\;\;\;\;\;\;\;\;\;\;\;\;\;\;\;\;\;\;\;\;\;\;
\left[ 
 \ln \left( \frac{P_{ijk} + R_{ij}}{P_{ijk}} \right) 
 - 
 \ln 
 \left( 
  \frac{-u^2 \, (P_{ijk} + R_{ij} + Q_i) + T}{-u^2 \, P_{ijk} + T} 
 \right) 
\right] 
\notag \\
&\quad {} \;\;\;\;\;\;\; + 
i  \int^{+\infty}_0 \frac{du}{u^2 \, P_{ijk} \, Q_i + R_{ij} \, T} \, 
\notag\\
&\;\;\;\;\;\;\;\;\;\;\;\;\;\;\;\;\;\;\;\;\;\;\;\;\;
\left[ 
 \ln 
  \left( \frac{-u^2 \, P_{ijk} + (R_{ij} + Q_i + T)}{- u^2 \, P_{ijk} + T} \right)
  - 
  \ln \left( \frac{Q_i+T}{T} \right) 
\right] 
\notag \\
&\quad {} \;\;\;\;\;\;\; + 
\int^{+\infty}_1 \frac{du}{u^2 \, P_{ijk} \, Q_i - R_{ij} \, T} \, 
\left[ \ln \left( \frac{u^2 \, P_{ijk} + R_{ij}}{u^2 \, P_{ijk}} \right) 
 - 
 \ln \left( \frac{Q_i+T}{T} \right) 
\right] 
\notag \\
&\quad {} \;\;\;\;\;\;\; + 
\int^{+\infty}_1 \; 
\frac{du}{u^2 \, P_{ijk} \, Q_i - R_{ij} \, T} \, 
\notag\\
&\;\;\;\;\;\;\;\;\;\;\;\;\;\;\;\;\;\;\;\;\;\;\;\;\;
\left[ 
 \ln \left( \frac{u^2 \, (P_{ijk} + R_{ij})}{u^2 \, P_{ijk} + R_{ij}} \right) 
 - 
 \ln 
  \left( 
   \frac{u^2 \, (P_{ijk} + R_{ij} + Q_i) + T}
   {u^2 \, P_{ijk} + (R_{ij} + Q_i + T)} 
  \right) 
\right] 
\notag \\
&\quad {} \;\;\;\;\;\;\; + 
\left. 
\int^{1}_0 \;\;\;\;
\frac{du}{u^2 \, P_{ijk} \, Q_i - R_{ij} \, T} \, 
\left[ 
 \ln \left( \frac{P_{ijk} + R_{ij}}{P_{ijk}} \right) 
 - 
 \ln \left( \frac{u^2 \, Q_i + T}{T} \right) 
\right] 
\right\}
\label{eqcas8}
\end{align}
In this case, we have $\Im(P_{ijk}) < 0$, $\Im(P_{ijk} + R_{ij}) < 0$, 
$\Im(Q_i) > 0$, $\Im(Q_i+T) > 0$ and $\Im(T) > 0$. Furthermore, 
\begin{itemize}
\item
$u^2 \, (P_{ijk}+R_{ij}+Q_i) - T = u^2 \, \tD_{ijk} + (1+u^2) \, \Delta_3$ \\
thus $\Im(u^2 \, (P_{ijk} + R_{ij}+Q_i) - T) < 0$ when $u \in [0,+\infty[$, 
\item
$u^2 \, P_{ijk} - T = u^2 \, (\tD_{ijk}+\Delta_1^{\{i,j\}}) + \Delta_3$ \\
thus $\Im(u^2 \, P_{ijk} - T) < 0$ when $u \in [0,+\infty[$, 
\item
$-u^2 \, P_{ijk} + (R_{ij} + Q_i + T) 
= -u^2 \, \tD_{ijk} - (1+u^2) \, \Delta_1^{\{i,j\}}$ \\
thus $\Im(-u^2 \, P_{ijk} + R_{ij} + Q_i + T) > 0$ when $u \in [0,+\infty[$, 
\item
$u^2 \, P_{ijk} + R_{ij} 
= u^2 \, \tD_{ijk} + (u^2-1) \, \Delta_1^{\{i,j\}} +\Delta_{2}^{\{i\}})$ \\
thus $\Im(u^2 \, P_{ijk} + R_{ij}) < 0$ when $u \in [1,+\infty[$, 
\item
$u^2 \, (P_{ijk}+R_{ij}+Q_i) + T = u^2 \, \tD_{ijk} + (u^2-1) \, \Delta_3$ \\
thus $\Im(u^2 \, (P_{ijk}+R_{ij}+Q_i) + T) < 0$ when $u \in [1,+\infty[$,
\item
$u^2 \, P_{ijk} + R_{ij} + Q_i + T 
= u^2 \, \tD_{ijk} +  (u^2-1) \, \Delta_1^{\{i,j\}}$ \\
thus $\Im(u^2 \, P_{ijk} + (R_{ij} + Q_i +T)) < 0$ when $u \in [1,\infty[$, 
\item
$u^2 \, Q_i + T = - (1-u^2) \, \Delta_3 - u^2 \, \Delta_2^i$ \\
thus $\Im(u^2 \, Q_i + T) > 0$ when $u \in [0,1]$.
\end{itemize}
After the use of the tricks developped in case {\bf 1.(b)}, the following alternative expression is obtained:
\begin{align}
\hspace{2em}&\hspace{-2em}L_4^4(\Delta_3,\Delta_2^{\{i\}},\Delta_1^{\{i,j\}},\tD_{ijk}) \notag \\
&= - \, 
\left\{
 \int_{\widehat{(0,1)}^{+}_{1}} 
 \frac{du}{u^2 \, P_{ijk} \, Q_i - R_{ij} \, T}  
\right.
\notag\\
& \quad {} \quad {}\quad {} \quad {}\;
 \left[
  \;\;\; 
  \ln \left( u^2 \, P_{ijk} + (R_{ij} + Q_i + T) \right)
  \;\; - 
  \ln \left( \frac{(Q_i + R_{ij})}{Q_{i}} (Q_i + T) \right)
 \right]
\notag\\
& \quad {} \quad {}\quad {}  +
 \int_{\widehat{(0,1)}^{+}_{2}} 
 \frac{du}{u^2 \, P_{ijk} \, Q_i - R_{ij} \, T} \, 
\notag\\
&\quad {} \quad{}\quad {} \quad{} \;
 \left[
  \vphantom{\frac{R_{ij} \, T}{Q_{i}}} 
  - 
  \ln \left( u^2 \, (P_{ijk} + R_{ij} + Q_i) + T \right) 
  + 
  \ln 
  \left( 
   T \, \frac{(P_{ijk} + R_{ij})}{P_{ijk}} \frac{(Q_i + R_{ij})}{Q_{i}} 
  \right)
 \right]
\notag\\
& \quad {} \quad {}\quad {}  +
 \int_{0}^{1} 
 \frac{du}{u^2 \, P_{ijk} \, Q_i - R_{ij} \, T}  
\notag\\
& \quad {} \quad {}\quad {} \quad {}\;
 \left[
  \;\;\; \quad {} \quad {}
  \ln \left( u^2 \, Q_i + T \right)
  \quad {}\quad {} \quad {} - \quad {} \quad {}
  \ln \left(T \, \frac{\left( P_{ijk} + R_{ij} \right)}{P_{ijk}} \right)
 \right.
\notag\\
& \quad {} \quad {}\quad {} \quad {}
  + 
  \left.
  \eta \left( \frac{R_{ij}+Q_i}{Q_i}, \; (Q_i+T) \right)
  \;\;\;\; - \;\;
  \eta 
  \left( 
   T \, \frac{(P_{ijk} + R_{ij})}{P_{ijk}}, \, \frac{(R_{ij} + Q_i)}{Q_{i}} 
  \right)
 \right]
\notag \\
&\quad {} \quad {}\quad {}  + 
 \int_{{\Gamma}^{+}}  \,
 \frac{du}{u^2 \, P_{ijk} \, Q_i - R_{ij} \, T}  
 \notag\\
& \quad {} \quad {}\quad {} \quad {}\;
 \left[ 
  \;\;
  \eta \left( \frac{(R_{ij}+Q_i)}{Q_i}, \; (Q_i+T) \right)
   - 
  \eta 
  \left( 
   T \, \frac{(P_{ijk} + R_{ij})}{P_{ijk}}, \, \frac{(R_{ij} + Q_i)}{Q_{i}} 
  \right)
 \right.
 \notag\\
& \quad {} \quad {}\quad {} \quad {} \quad {} \quad {}\quad {} \quad {}
 \quad {} \quad {}\quad {} \quad {} \quad {} \quad {}\quad {} \quad {}
\left.
 \left. 
 - \quad {}\quad {}
  \eta \left( T, \; \frac{(P_{ijk} + R_{ij})}{P_{ijk}} \right)
 \quad {}\quad {} 
\right] 
\right\}
\label{eqcas2dd}
\end{align}
Eq. (\ref{eqcas2dd}) shares the structure common to eqs. 
(\ref{eqcas3eclatesoustrait}), (\ref{eqcas4eclatesoustrait}), 
(\ref{eqcas5eclatesoustrait}), (\ref{eqcas2aa}), (\ref{eqcas2bb}) and 
(\ref{eqcas2cc}) as well. The cuts of 
$\ln(u^2 \, P_{ijk} + (R_{ij} + Q_i) + T))$ and
$\ln(u^2 \, (P_{ijk} + R_{ij} + Q_i) + T)$ in the half plane $\{\Re(u) > 0\}$
both run towards $\infty$ in the ``south-east" quadrant", whereas the
contours $\widehat{(0,1)}^{+}_{1,2}$ stretched from 0 to 1 shall wrap the 
branch points and cuts of 
$\ln(u^2 \, P_{ijk} + (R_{ij} + Q_i + T))$ and
$\ln(u^2 \, (P_{ijk} + R_{ij} + Q_i) + T)$ respectively, from above in the
``north-east" quadrant in case the corresponding branch points lie in this 
quadrant; the two contours may be chosen distinct from each other so as to best
fit the respective finite arcs of cuts partly slashing the ``north-east" 
quadrant from the branch points.

\subsection{Synthesis}

As anticipated the number of integral contributions is profuse in a
case-dependent way from (\ref{eqcas1}) to (\ref{eqcas8}). A common structure 
can however be achieved by means of case-dependant contour deformations of the 
real interval $[0,1]$ supplemented by extra pole residue contributions weighted 
by case-dependant combinations of $\eta$ functions. 
Can this common structure be 
a starting point to recombine terms further and reduce the number of
contributions, as could be done for the three-point function in the general
complex mass case treated according to the ``indirect way"?

\vspace{0.3cm}

\noindent
In the case of the three-point function case, we 
could first cast the integrals weighting the sum over the $\bar{b}_{j}^{\{i\}}$ 
as one-dimensional contour integrals of a common type along some case-dependent contour 
defor\-mations of the interval $[0,1]$ which was used in the real mass case. 
Then, after appropriate changes of variables absorbing the corresponding factor 
$\bar{b}_{j}^{\{i\}}$ in each of these contour integrals, we were able to 
concatenate 
these rescaled contour integrals into a single contour integral. Lastly, the 
compound contour of the latter was deformed in its turn into exactly the 
interval $[0,1]$ involved in the real mass case. This resulted in a
simplification which proved to coincide with the one coming out via the 
``direct way". One may wonder whether the 
formal unification of the profuse diversity of expressions obtained for the 
four-point function with general complex masses could, at least
partially, be exploited in a similar way following a similar programme. 
This quest appears  much more complicated for the four-point function, all the
more so as we
already faced an issue in the reduction of the number of dilogarithms involved
in the expression of the four-point function for the real mass case using the 
present approach, compared with  't Hooft and Veltman's approach.
Nevertheless as already discussed in the end of sec.~\ref{sectthreepoint}, the 
dilogarithms obtained after performing the last integration are the same for all 
the $8$ cases and are similar to those of the real mass case. Here also, the 
discussion about the number of dilogarithms generated (cf.\ subsec.~\ref{P1-discfourpdilog} 
of \cite{paper1}) compared to ref.~\cite{tHooft:1978jhc} still holds and the 
solutions which will be found to counteract this proliferation of dilogarithms 
in the real mass case will be able to apply without modifications.

\section{Summary and outlook}

In this article we presented an extension of the novel approach developed in a companion article (cf.~\cite{paper1}) for the computation of one-loop 
three- and four-point functions in the general complex mass case. The method naturally proceeds in terms of 
algebraic kinematical invariants involved in reduction algorithms and applies 
to general kinematics beyond the one relevant for one-loop collider processes, 
it thereby offers a potential application to the calculation of processes 
at two-loop using one-loop (generalised) $N$-point functions as building 
blocks. 
This novel approach enables a smooth extension 
to the complex masse case for the generalised one-loop building blocks expressed in terms of dilogarithms. 
Nevertheless, in the case of a two-loop computations, the analyticity of the one-loop integrand with respect of the two extra Feynman parameters has to be carefully studied.
For sake of pedagogy, the method was exposed on ``ordinary" 
  three- and four-point functions in four dimensions in the real mass case in a companion article \cite{paper1}. The complex mass case has been studied hereby.
It can be extended in respect to the space-time dimension to tackle the infrared divergent case. Let us advertised it briefly. 

\vspace{0.3cm}

\noindent
In a third companion paper we extend the presented framework to the case
where some vanishing internal masses cause infrared soft and/or collinear
divergences. The method extends in a straightforward way, once
a few intermediate steps and tools are appropriately adapted.  

\vspace{0.3cm}

\noindent
The question of the proliferation of dilogarithms in the expression of the four-point function
computed in closed form with the present method comes up in the same terms as in the real mass case. It 
requires some extra work to be
better apprehended, in order to counteract it. This issue will be addressed in a
future article.

\vspace{0.3cm}

\noindent
The last goal is to provide the generalised one-loop building blocks 
entering as integrands in the computation two-loop three- and four-point 
functions by means of an extra numerical double integration.

\section*{In memoriam}

Various ideas and techniques used in this work were initiated by Prof. Shimizu
after a visit to LAPTh. He explained us his ideas about the numerical 
computation of scalar two-loop three- and four-point functions, he shared his 
notes partly in English, partly in Japanese with us and he encouraged us to 
push this project forward. J.Ph. G. would like to thank Shimizu-sensei for 
giving him a taste of the Japanese culture and for his kindness.

\section*{Acknowledgements}

We would like to thank P. Aurenche for his support along this project and for a careful reading of the manuscript.

\appendix

\section{General case for the second kind integral $J(\nu)$ \label{appendJ}} 

This appendix extends the results of appendix \ref{P1-appendJ} of ref. \cite{paper1}
concerning the second kind integral $J(\nu)$ given by:
\begin{equation}
  J(\nu) 
  = 
  \int^{+\infty}_0  
   \frac{d \xi}{\left(\xi^{\nu}+A \right) \, \sqrt{\xi^{\nu}+B}}
  \label{eqdefj1}
\end{equation}
because new cases appear which were not covered in this reference. 
In what follows $A$ and $B$ are assumed dimensionless and complex valued, the
signs of their real parts are unknown, and, contrary to the real mass case, the signs of their imaginary parts 
may or may not be the same. 
When no internal masses are vanishing it arises for 
$\nu = 2$ whereas infrared divergent cases regularised
by dimensional continuation beyond $n=4$ involve non integer $\nu$.
Anticipating our next paper on infrared divergent 
case, these various situations are treated all at once here,
specifying $\nu$ at will in the result. The integral need not be
computed in closed form and shall instead be 
recast in an alternative, more handy form cleared from any radical. 
So let us distinguish two cases according to the signs of the imaginary parts of $A$ and $B$.

\subsubsection*{1) $\Im(A)$ and $\Im(B)$ of the same sign}

Whenever $\Im({A})$ and $\Im({B})$ have the same
sign, the use of the celebrated Feynman ``trick" is justified and leads to:
\begin{equation}\label{feynmantrick}
\frac{1}{\left(\xi^{\nu}+A \right) \, \sqrt{\xi^{\nu}+B}}
=
\frac{1}{2} \, 
\int^1_0 
\frac{d x \, x^{-\frac{1}{2}}}{(\xi^{\nu}+(1-x) \, A + x \, B)^{3/2}}
\end{equation}
$J(\nu)$ is readily rewritten as:
\begin{equation}
J(\nu) 
= 
\frac{1}{2} \; 
\int^{+\infty}_0 d \xi \, 
\int^1_0 
\frac{d x \, x^{-\frac{1}{2}}}{(\xi^{\nu}+(1-x) \, A + x \, B)^{3/2}}
\label{eqdeffuncj1}
\end{equation}
Then the $\xi$ integration is performed first, using eq. (\ref{P1-eqFOND1}) of
appendix \ref{P1-ap2} of ref. \cite{paper1}. Performing the change of variable $z = \sqrt{x}$ in the result 
obtained yields:
\begin{equation}
J(\nu) 
= 
\frac{1}{\nu} \, 
B \left( \frac{3}{2}- \frac{1}{\nu},\frac{1}{\nu} \right) \, 
\int^1_0 dz \, \left( (1-z^2) \, A + z^2 \, B \right)^{-3/2 + 1/\nu}
\label{eqdeffuncj2}
\end{equation}
In particular for $\nu=2$:
\begin{equation}
J(2) = \int^1_0 \frac{dz}{(1-z^2) \, A + z^2 \, B}
\label{eqdeffuncjp2}
\end{equation}

\subsubsection*{2) $\Im(A)$ and $\Im(B)$ of opposite signs}

This more annoying case can be met when the internal masses are complex.
Naively reproducing the previous argument would again lead to eq. 
(\ref{eqdeffuncj2}). 
However the derivation of the Feynman ``trick" (\ref{feynmantrick}) 
{\em assumes} $\Im(A)$ and $\Im(B)$ to have the same sign (whenever the signs 
of their respective real values is undetermined, which is the case at hand): 
its use is illegitimate whenever $\Im(A)$ and $\Im(B)$ have opposite signs.
We shall first recast the r.h.s. of eq. (\ref{eqdefj1}) so that 
the imaginary parts of both factors in the denominator of the integrand have 
the same sign:
\begin{equation}
J(\nu) 
= 
- \int^{+\infty}_0 
\frac{d \xi}{\left( -\xi^{\nu}-A \right) \,\sqrt{\xi^{\nu}+B}}
\label{eqdeffuncj0p}
\end{equation}
{\em Then} we can apply the Feynman ``trick" to eq. (\ref{eqdeffuncj0p}):
\begin{align}
J(\nu) 
&= - \, \frac{1}{2} \; 
 \int^{+\infty}_0 d \xi \, 
 \int^1_0  
 \frac{dx \, x^{-\frac{1}{2}} }
 {((2 \, x - 1) \, \xi^{\nu}-(1-x) \, A + x \, B)^{3/2}} 
\label{eqdeffuncj30}
\end{align}
We again intend to perform the $\xi$ integration first, yet the task is a 
little more tricky than for (\ref{eqdeffuncj1}). In order to use eq. 
(\ref{P1-eqFOND1}) of ref. \cite{paper1} 
we shall factor out a fractional power of $(2 \, x -1)$ which is not always 
positive when $x$ spans $[0,1]$, so that some care is required.
Introducing $S_B = \sign(\Im(B))$, $S_x^{\prime} = \sign(2 \, x-1)$ 
and an infinitesimal parameter $0 < \lambda \ll 1$ we 
have\footnote{This comes from the splitting of $\ln (ab)$ with $a$ real 
not necessarily $>0$ and $b$ is complex non real, for which \cite{tHooft:1978jhc}
\[
\ln(ab) 
= 
\ln(a - i \, \lambda \, S_b) + \ln(b), \;\; 
S_b = \sign \left( \Im(b) \right)
\]
}:
\[
((2 \, x - 1) \, \xi^{\nu}-(1-x) \, A + x \, B)^{3/2} 
= 
(2 \, x - 1 - i \, S_B \, S_x^{\prime} \, \lambda)^{3/2} \,
\left[ 
 \xi^{\nu} + \frac{x \, B - (1-x) \, A}{2 \, x - 1} 
\right]^{3/2}
\] 
so that:
\begin{align}
J(\nu)
&=  
- \, \frac{1}{2} \; 
\int^1_0 \frac{dx}{\sqrt{x}} \, 
\frac{1}{(2 \, x - 1 - i \, S_B \, S_x^{\prime} \, \lambda)^{3/2}} \, 
\int^{\infty}_0 
\frac{d \xi}
{\left(\xi^{\nu} + \frac{x \, B - (1-x) \, A}{2 \, x - 1} \right)^{3/2}}
\label{eqdeffuncj3}
\end{align}
The $\xi$ integration performed using eq. (\ref{P1-eqFOND1}) of ref. \cite{paper1} yields:
\begin{align}
J(\nu) 
& = 
- \, \frac{1}{2 \, \nu} \, 
B \left( \frac{3}{2}-\frac{1}{\nu},\frac{1}{\nu} \right) 
\nonumber\\
&\quad {} 
\times 
\int^1_0 \frac{dx}{\sqrt{x}} \, 
\frac{1}{(2 \, x - 1 - i \, S_B \, S_x^{\, \prime} \, \lambda)^{3/2}} \, 
\left( \frac{2 \, x - 1}{(x-1) \, A + x \, B} \right)^{3/2-1/\nu}
\label{eqdeffuncj4}
\end{align}
Some care is required again to split the fraction raised to the non 
integer power $3/2-1/\nu$ into a fraction of powers:
\begin{equation}
\left( \frac{2 \, x - 1}{(x-1) \, A + x \, B} \right)^{3/2-1/\nu} 
= 
\frac{(2 \, x - 1 + i \, \lambda \, S_B)^{3/2-1/\nu}}
{((x-1) \, A + x \, B)^{3/2-1/\nu}}
\label{eqtempfj1}
\end{equation}
Eq. (\ref{eqdeffuncj4}) can be written as:
\begin{align}
J(\nu) 
&= - \, \frac{1}{2 \, \nu} \, 
B \left( \frac{3}{2}-\frac{1}{\nu},\frac{1}{\nu} \right) 
\notag \\
&\int^1_0 \frac{dx}{\sqrt{x}} \, 
\frac{(2 \, x - 1 + i \, S_B \, \lambda)^{3/2-1/\nu}}
{(2 \, x - 1 - i \, S_B \, S_x^{\prime} \, \lambda)^{3/2}} \, 
\left( (x-1) \, A + x \, B \right)^{-3/2+1/\nu}
\label{eqdeffuncj41}
\end{align}
We now split the range of integration in $x$ in two parts : 
$0 \le x \le 1/2$ and $1/2 \le x \le 1$, so that in each sub-range, 
$2 \, x - 1$ has a definite sign. $J(\nu)$ can be written as :
\begin{align}
J(\nu) 
&= - \, \frac{1}{2 \, \nu} \, 
B \left( \frac{3}{2}-\frac{1}{\nu},\frac{1}{\nu} \right) \notag \\ 
&\quad {} \times 
\left[ 
 e^{- i \, S_B \, \pi/\nu} \, 
 \int^{1/2}_0 dx \, x^{-1/2} \, (1-2 \, x)^{-1/\nu} \, 
 (B \, x + (x-1) \, A)^{-3/2+1/\nu} 
\right. 
\notag \\
&
\;\;\;\;\;\;\;\;\;\;\;
+ 
\left. 
 \int^1_{1/2} dx \, x^{-1/2} \, (2 \, x - 1)^{-1/\nu} \, 
 (B \, x + (x-1) \, A)^{-3/2+1/\nu} 
\right]
\label{eqdeffuncj5}
\end{align}
With the help of the Euler changes of variables $\sqrt{x - 2 \, x^2} = x \, t$ 
in the first integral and $\sqrt{2 \, x^2- x} = x \, t$ in the second integral 
of eq. (\ref{eqdeffuncj5}), we recast $J(\nu)$ into:
\begin{align}
J(\nu) 
&= - \, \frac{1}{\nu} \, 
B \left( \frac{3}{2}-\frac{1}{\nu},\frac{1}{\nu} \right) \notag \\ 
&\quad {} \times  
\left[ 
 e^{- i \, S_B \, \pi/\nu} \, 
 \int^{+\infty}_0  dt \; t^{1-2/\nu} \, 
 \left( B  - (t^2+1) \, A \right)^{-3/2+1/\nu} 
\right. 
\notag \\ 
&
\;\;\;\;\;\;\;\;\;\;\;
+ 
\left. 
 \int^{1}_{0}  dt \; t^{1-2/\nu} \, 
\left( B + (t^2-1) \, A \right)^{-3/2+1/\nu} 
\right]
\label{eqdeffuncj6}
\end{align}
Finally, we trade $t$ for $z=1/t$ so that $J(\nu)$ becomes :
\begin{align}
J(\nu) 
&= - \, \frac{1}{\nu} \, 
B \left( \frac{3}{2}-\frac{1}{\nu},\frac{1}{\nu} \right) \notag \\ 
&\quad {} \times  
\left[ 
 e^{- i \, S_B \, \pi/\nu} \, 
 \int^{+\infty}_0  dz \, 
 \left( B \, z^2  - (1+z^2) \, A \right)^{-3/2+1/\nu} 
\right. 
\notag \\ 
&
\;\;\;\;\;\;\;\;\;\;\;
+ 
\left. 
 \int^{+\infty}_{1}   dz \, 
 \left( B \, z^2 + (1-z^2) \, A \right)^{-3/2+1/\nu} 
\right]
\label{eqdeffuncj7}
\end{align}
In particular for $\nu =2$, $J(2)$ becomes:
\begin{align}
J(2) 
&=  
\left[ 
 i \, S_B \, \int^{+\infty}_0  \frac{dz}{B \, z^2  - (1+z^2) \, A} 
 - \int^{+\infty}_{1}   \frac{dz}{B \, z^2 + (1-z^2) \, A} 
\right]
\label{eqdeffuncj8}
\end{align}
Note that the two integrals of the right hand size of eq.
(\ref{eqdeffuncj7}) are well defined because 
$\Im(B \, z^2  - (1+z^2) \, A)$ and $\Im(B \, z^2 + (1-z^2) \, A)$
never vanish in the respective $z$ ranges of integration 
thus the branch cuts (poles for $3/2-1/\nu$ integer) of the
integrands lie away from the integration ranges. We will elaborate a
little more about their location in the complex $z$ plane below.

\vspace{0.3cm}

\noindent
The two cases 1) vs. 2) disentangled above can be reunified by seeing 
eq. (\ref{eqdeffuncj7}) as an analytic continuation in $A$ of eq. 
(\ref{eqdeffuncj2}) which possibly requires a deformation of the contour 
$[0,1]$ originally drawn along the real axis in eq. (\ref{eqdeffuncj2}). 
The normalisation factor in $J(\nu)$ is irrelevant in 
the following discussion, we drop it (apart the overall minus sign) to simplify the expressions.
\begin{align}
J(\nu) 
&=
{} - e^{- i \, S_B \, \pi/\nu} \, 
 \int^{+\infty}_0 dz \, 
 \left( B \, z^2  - (1+z^2) \, A \right)^{-3/2+1/\nu} 
\notag \\ 
&\;\;\;\;\;\;\;\;\;\;\;\;\;\;\;\;\;\;\;\;
-  
 \int^{+\infty}_{1} dz \,
 \left( B \, z^2 + (1-z^2) \, A \right)^{-3/2+1/\nu} 
\label{eqdeffuncj7renorm}
\end{align}
$J(\nu)$ can be alternatively written:
\begin{align}
J(\nu) 
&= \left\{ \int_0^{- \, i \, S_{B} \, \infty}  + \int_{+\infty}^{1} \right\} 
\, dz \,
\left( B \, z^2 + (1-z^2) \, A \right)^{-3/2+1/\nu} 
\label{eqdeffuncj7renormbis}
\end{align}
The function $(B \, z^2 + (1-z^2) \, A)^{-3/2+1/\nu}$ of the complex
variable $z$ has two discontinuity cuts supported respectively by either of 
the two branches of the hyperbola $\{\Im(B \, z^2 + (1-z^2) \, A) = 0\}$.
Let us label ${\cal C}_{+}$ the cut relevant\footnote{The other cut ${\cal C}_{-}$ is the symmetric of 
${\cal C}_{+}$ under parity: located in the left half plane 
$\{\Re(z) < 0\}$ it is irrelevant here.}
for our concern. ${\cal C}_{+}$ lies in the right half $z$-plane 
$\{\Re(z) > 0\}$. It originates at the point $z_{+} = (A/(A-B))^{1/2}$
and slashes the right half plane through the quadrant 
$\{ \Re(z) > 0, \, S_{B} \, \Im(z) > 0 \}$ away to $\infty$. 
In case $z_{+}$ belongs to the quadrant 
$\{ \Re(z) > 0, \, S_{B} \, \Im(z) < 0 \}$ this cut crosses the 
real interval $[0,1]$ at  $z_{c} = [\Im(A)/\Im(A-B)]^{1/2}$ (cf. appendix \ref{cut} for more details).

\vspace{0.3cm}

\noindent
The integration contour in the r.h.s. of eq. (\ref{eqdeffuncj7renormbis}) 
can be closed by drawing an arc $\widehat{(0,1)}$ between 0 and 1 so as to
reformulate $J(\nu)$ as a contour integral along $\widehat{(0,1)}$ according 
to the Cauchy theorem, the extra arc at $\infty$ also involved by the 
Cauchy theorem to close the contour yields a vanishing contribution 
$\sim {\cal O}(1/R)$ where $R \to + \infty$ is ``$|z|$ on the contour at 
$\infty$".
\begin{description}
  \item[(i)] If ${\cal C}_{+}$ entirely belongs to the quadrant 
$\{ \Re(z) > 0, \, S_{B} \, \Im(z) > 0 \}$ this extra
arc $\widehat{(0,1)}$ can be taken along the real segment $[0,1]$. 
\item[(ii)] However if $z_{+}$ belongs to the quadrant 
$\{ \Re(z) > 0, \, S_{B} \, \Im(z) < 0 \}$ the extra arc 
$\widehat{(0,1)}$ shall wrap the bit of ${\cal C}_{+}$ from slightly before 
$z_{c}$ around $z_{+}$ and back to slightly after $z_{c}$
inside $\{ \Re(z) > 0, \, S_{B} \, \Im(z) < 0 \}$ as if 
${\cal C}_{+}$ were locally sinking the contour away from the 
real segment $[0,1]$ inside this quadrant as pictured on figure \ref{contour}.
\end{description}
In either case:
\begin{equation}
J(\nu) 
= \int_{\widehat{(0,1)}} dz \, 
\left( B \, z^2 + (1-z^2) \, A \right)^{-3/2+1/\nu} 
\label{eqdeffuncj7renormter} 
\end{equation}
is the argued analytic continuation in $A$ of eq. (\ref{eqdeffuncj2}).

\section{Basic integrals in terms of dilogarithms and
logarithms: $K$-type integrals}\label{appF}

This appendix comes in addition to appendix \ref{P1-appF} of ref. \cite{paper1}.
The computations of the various $N$-point functions in closed form can be
reduced to the calculation of integrals of simple types. The $K$-type is 
of the form
\[
K 
=  
\int^a_b du \, 
\frac{\ln ( A \, u^2 + B) - \mbox{``subtracted term''}}{u^2 - u_0^2}
\]
where $u_0^2 \neq - B/A$. 
In the case of complex masses, $A$ is complex and 
the complex quantity $B$ has a non vanishing 
imaginary part yet with $\Im(A \, u^2 +B)$
keeping a cons\-tant sign while $u$ spans the real interval $[a,b]$. 
The general complex mass case involves the contour $[0,1]$ as 
well as the two other contours $[0,+\infty[$ and $[1,+\infty[$.

\vspace{0.3cm}

\noindent
In the complex mass case, the situation is less diverse than in the real mass case. 
The parameters $u_0^2$ in the $K$-type integrals 
are generically complex with a non infinitesimal imaginary part, in which case 
the poles in the integrands are well off the contour of integration 
thus the calculation can be formulated using either a vanishing or non vanishing
``subtracted term'' it does not matter. One may choose to use $K$-type integrals
with a ``subtracted term'' equal to 0 so as to have the simplest possible
expressions, or instead e.g.\ equal to $\ln(A \, u_0^2 + B)$ so as to involve
similar building blocks as for the real mass case, cf.\ below: this thereby 
minimises the number of encoded functions in practical numerical 
implementations.
\vspace{0.3cm}

\noindent
This appendix often makes use of the identity
\begin{align}
\ln(z) &= \ln(-z) + i \, \pi \, S(z) \;\; , \;\; S(z) \equiv \sign(\Im(z))
\label{eqdeflnzlnmz}
\end{align}

\subsection{$a=0$, $b=1$}

\noindent
The calculations of $N$-point functions are formulated 
so as to be expressed in terms of quantities of the form: 
\begin{equation}
K^C_{0,1}(A,B,u_0^2) = \int^1_0 du \, 
\frac{\ln(A \, u^2 + B) - \ln(A \, u^{2}_0 + B)}{u^2 - u_0^2}
  \label{eqdefkcab1}
\end{equation}
where $A$ and $B$ are now plain complex numbers yet with $\Im(A \, u^2 +B)$
keeping a cons\-tant sign while $u$ spans the real interval $[0,1]$. 
The logarithms may now be conveniently split as
\begin{align}
\ln (A \, u^2+B) 
&= \ln(A) + \ln(u \; - \baru) + \ln(u \; + \baru) + \eta(A,-\baru^2)
\label{eqarrange3}\\
\ln(A \, u^{2}_0 + B) 
&= \ln(A) + \ln(u_0 - \baru) + \ln(u_0+\baru) 
\notag \\
&\quad {} \quad {}\quad {}\quad {}\quad {}\quad {}
+ \eta(u_0-\baru,u_0+\baru) + \eta(A,u^2_0-\baru^2)
\label{eqarrange4}
\end{align}
where $\baru = \sqrt{-B/A}$ and 
the function $\eta(z_1,z_2)$ is given by eq.~(\ref{P1-eqdefeta01}) in appendix \ref{P1-appF} of \cite{paper1}.
Substituting identities (\ref{eqarrange3}), (\ref{eqarrange4}) into eq. 
(\ref{eqdefkcab1}), and proceeding along the same line as for the real 
mass case, we get:
\begin{align}
K^C_{0,1}(A,B,u_0^2) 
&= \frac{1}{2 \, u_0} 
\left\{ 
 \calf(u_0,\baru) + 
 \left[
  \eta \left( A,-\baru^2  \right) - \eta \left( A,u^2_0-\baru^2  \right)
  \vphantom{\ln \left( \frac{\Lambda - \baru}{1 - \baru} \right)}
 \right.
\right.
\notag\\
& \quad{} \quad{} \quad{}\quad{} \quad{} \quad{}
\quad{} \quad{} \quad{} \quad{} \quad{}\quad{} \quad{} \quad{}
\left.
 \left.
  - \eta(u_0-\baru,u_0+\baru) 
 \vphantom{\ln \left( \frac{\Lambda - \baru}{1 - \baru} \right)}
 \right]
 \, \ln \left( \frac{u_0-1}{u_0+1} \right) 
\right\} 
\label{eqdefkrab4}
\end{align}
where $\calf(y,z)$ is given by eq. (\ref{P1-eqdefcalf2a}) of ref. \cite{paper1}. 

\subsection{$a=1, b=+\infty$}\label{g2}

The computation of $N$-point functions in the complex mass case 
also involves integrals of the following kind:
\begin{align}
K^C_{1,\infty}(A,B,u_0^2) 
&= \int^{\infty}_1 du \, 
\frac{\ln(A \, u^2 + B) - \ln(A \, u^{2}_0+ B)}{u^2 - u_0^2}
\label{eqdefkrab5}
\end{align}
where $A$, $B$ and $u_0^2$ are complex numbers such that $\Im(A \, u^2 + B)$ 
keeps a constant sign while $u$ spans the range $[1,\infty[$ along the real 
axis. 
Logarithms can be split as in identities (\ref{eqarrange3}), (\ref{eqarrange4}) 
above, and the partial fraction decomposition of $1/(u^2-u_0^2)$ proceeds as 
in the real mass case.
We wish to conveniently handle the various terms resulting from the 
partial fraction decomposition separately. Yet the latter individually diverge
logarithmically at large $u$, although the integral in eq. (\ref{eqdefkrab5}) 
converges. We therefore introduce a regularisation procedure
by means of ``large $u$" cut-off $\Lambda$. We then recombine individually
divergent terms $\propto \ln^2(\Lambda)$ and $\propto \ln(\Lambda)$, so as to 
make them respectively cancel among each other explicitly. We then take the 
limit $\Lambda \to + \infty$. With the above definitions of $u_0$ and $\baru$, 
the regularised splitting of $K^C_{1,\infty}(A,B,u_0^2)$ reads:
\begin{align}
K^C_{1,\infty}(A,B,u_0^2) 
&= \frac{1}{2 \, u_0} \, 
\left\{
 \lim_{\Lambda \rightarrow + \infty} E(\Lambda) + 
 \left[ 
  \vphantom{\ln \left( \frac{\Lambda - \baru}{1 - \baru} \right)} 
  \eta \left( A,1-\baru^2  \right) -  \eta \left( A,u^2_0-\baru^2  \right) 
 \right. 
\right. 
\notag \\
&\quad {}\quad {}\quad {}\quad {}\quad {}\quad {}\quad {}\quad {}
 - 
\left.
 \left.
  \vphantom{\ln \left( \frac{\Lambda - \baru}{1 - \baru} \right)} 
  \eta \left( u_0-\baru,u_0+\baru \right) 
 \right] \, 
 \left[ \ln \left( 1 + u_0 \right) - 
 \ln \left( 1 - u_0 \right) \right] 
\frac{}{}
\right\}
\label{eqdefkrab6}
\end{align}
with
\begin{align}
E(\Lambda) 
&= \int^{\Lambda}_{1} du \, 
\left[ \frac{1}{u-u_0} - \frac{1}{u+u_0} \right] \, 
\notag \\
&\quad {} \quad {} \quad {} \;\; \times
\left[ 
 \ln (u-\baru) - \ln(u_0 - \baru) + \ln(u+\baru) - \ln(u_0+\baru) 
\right]
\label{eqdefelambda1}
\end{align}
Introducing the quantity
\begin{equation}
R^{\Lambda}(y,z)  = \int^{\Lambda}_1 du \frac{\ln (u-y) - \ln(z-y)}{u-z}
\label{eqdefrlamda1}
\end{equation}
$E(\Lambda)$ reads in terms of $R^{\Lambda}$:
\begin{align}
E(\Lambda) 
&= 
R^{\Lambda}(u_0,\baru) + R^{\Lambda}(u_0,-\baru) - 
R^{\Lambda}(-u_0,\baru) - R^{\Lambda}(-u_0,-\baru) 
\notag \\
&\quad {} + 
i \, \pi \, 
\left[ S(u_0+\baru) + S(u_0-\baru) \right] \, 
\ln \left( \frac{\Lambda+u_0}{1+u_0} \right)
\label{eqdefelambda2}
\end{align}
The computation of $R^{\Lambda}(y,z)$ 
proceeds along the same line as $R^{\prime}(y,z)$ in the appendix \ref{P1-appF} of ref.\ \cite{paper1} and we get:
\begin{align}
R^{\Lambda}(y,z)  
&= 
\dilog \left( \frac{1-z}{y-z} \right) - 
\dilog \left( \frac{\Lambda-z}{y-z} \right) 
\notag \\
&\quad {} + 
\eta \left( 1-y, \frac{1}{z-y} \right) \, 
\ln \left( \frac{1-z}{y-z} \right) - 
\eta \left( \Lambda-y, \frac{1}{z-y} \right) \, 
\ln \left( \frac{\Lambda-z}{y-z} \right) 
\label{eqcalcrlambda1}
\end{align}
We use the identity relating $\dilog(z)$ and $\dilog(1/z)$ and we note that
\begin{equation}
\eta \left( \Lambda-y, \frac{1}{z-y} \right) 
= 
\eta \left( 1- \frac{y}{\Lambda}, \frac{1}{z-y} \right)
\label{eqdiffetafunc1}
\end{equation}
We rewrite eq. (\ref{eqcalcrlambda1}) as:
\begin{align}
R^{\Lambda}(y,z)  
&= \dilog \left( \frac{1-z}{y-z} \right) + 
\eta \left( 1-y, \frac{1}{z-y} \right) \, \ln \left( \frac{1-z}{y-z} \right) + 
\frac{\pi^2}{6} 
\notag \\
&\quad {}
- \dilog \left( \frac{y-z}{\Lambda-z} \right)  \,+\,
\frac{1}{2} 
\left[ \ln(\Lambda) + \ln \left( \frac{1-z/\Lambda}{z-y} \right) \right]^2 
\notag \\
&\quad {} 
- \eta \left( 1-\frac{y}{\Lambda}, \frac{1}{z-y} \right) \, 
\left[ \ln(\Lambda) + \ln \left( \frac{1-z/\Lambda}{y-z} \right) \right]
\label{eqcalcrlambda2}
\end{align}
For fixed $y$ and $z$, when $\Lambda$ is large enough 
$ \eta( 1-y/\Lambda, 1/(z-y)) = 0$. Dropping all terms which vanish when 
$\Lambda \to \infty$, eq. (\ref{eqcalcrlambda2}) can be rewritten as:
\begin{align}
R^{\Lambda}(y,z)  
&= \dilog \left( \frac{1-z}{y-z} \right) + 
\eta \left( 1-y,\frac{1}{z-y} \right) \, \ln \left( \frac{1-z}{y-z} \right) + 
\frac{\pi^2}{6}
\notag \\ 
&\quad {} + 
\frac{1}{2} \, \ln^2(z-y) - \ln(\Lambda) \, \ln(z-y) +
\frac{1}{2} \, \ln^2(\Lambda)
  \label{eqcalcrlambda3}
\end{align}
Substituting eq. (\ref{eqcalcrlambda3}) into eq. (\ref{eqdefelambda2}), we get:
\begin{align}
E(\Lambda) 
&= \quad{}
\dilog \left( \frac{u_0-1}{u_0-\baru} \right) + 
\eta \left( 1-\baru, \frac{1}{u_0-\baru} \right) \, 
\ln \left( \frac{u_0-1}{u_0-\baru} \right) 
\notag \\
&\quad {} + 
\dilog \left( \frac{u_0-1}{u_0+\baru} \right) + 
\eta \left( 1+\baru, \frac{1}{u_0+\baru} \right) \, 
\ln \left( \frac{u_0-1}{u_0+\baru} \right) 
\notag \\
&\quad {} - 
\dilog \left( \frac{u_0+1}{u_0+\baru} \right) - 
\eta \left( 1-\baru, \frac{-1}{u_0+\baru} \right) \, 
\ln \left( \frac{u_0+1}{u_0+\baru} \right) 
\notag \\
&\quad {} - 
\dilog \left( \frac{u_0+1}{u_0-\baru} \right) - 
\eta \left( 1+\baru, \frac{-1}{u_0-\baru} \right) \, 
\ln \left( \frac{u_0+1}{u_0-\baru} \right) 
\notag \\
&\quad {} + 
\frac{1}{2} \, 
\left( 
 \ln^2(u_0-\baru) + \ln^2(u_0+\baru) - \ln^2(-u_0-\baru) - \ln^2(-u_0+\baru) 
\right) 
\notag \\
&\quad {} - 
\ln(\Lambda) \, 
\left( 
 \ln(u_0-\baru) + \ln(u_0+\baru) - \ln(-u_0-\baru) - \ln(-u_0+\baru) 
\right) 
\notag \\
&\quad {} + 
i \, \pi \, 
\left( S(u_0+\baru) + S(u_0-\baru) \right) \, 
\left[ 
 \ln(\Lambda) - \ln \left( 1+u_0 \right)  
\right]
\label{eqeqdefelambda3}
\end{align}
Using eq. (\ref{eqdeflnzlnmz}), the sums of logarithmic terms in eq.
(\ref{eqeqdefelambda3}) can be expressed in terms of the sign function $S$:
\begin{align}
&\ln(u_0-\baru)+\ln(u_0+\baru)-\ln(-u_0-\baru)-\ln(-u_0+\baru) 
\notag \\
& \quad {} \quad {}\quad {}\quad {}\quad {} \quad {}\quad {}
= i \, \pi \, \left( S(u_0-\baru)+S(u_0+\baru) \right) 
  \label{eqrelsumlog1}\\
&\ln^2(u_0-\baru)+\ln^2(u_0+\baru)-\ln^2(-u_0-\baru)-\ln^2(-u_0+\baru) 
\notag \\
&\quad {}\quad {}\quad {}\quad {}\quad {}\quad {}\quad {}
= 2 \, \pi^2 + 2 \, i \, \pi \, 
\left[ S(u_0-\baru) \, \ln(u_0-\baru) + S(u_0+\baru) \, \ln(u_0+\baru) \right] 
\label{eqrelsumlog2}
\end{align}
Substituting eqs. (\ref{eqrelsumlog1}), (\ref{eqrelsumlog2}) and
(\ref{eqeqdefelambda3}) in (\ref{eqdefkrab6}), we get:
\begin{align}
K^C_{1,\infty}(A,B,u_0^2) &= \frac{1}{2 \, u_0} \, 
\left\{ 
 \quad {}
 \dilog \left( \frac{u_0-1}{u_0-\baru} \right) + 
 \eta \left( 1-\baru, \frac{1}{u_0-\baru} \right) \, 
 \ln \left( \frac{u_0-1}{u_0-\baru} \right) 
\right.
\notag \\
&\quad {} \quad {} \quad {} \quad {}+ 
 \dilog \left( \frac{u_0-1}{u_0+\baru} \right) + 
 \eta \left( 1+\baru, \frac{1}{u_0+\baru} \right) \, 
 \ln \left( \frac{u_0-1}{u_0+\baru} \right) 
\notag \\
&\quad {} \quad {}\quad {} \quad {} - 
 \dilog \left( \frac{u_0+1}{u_0+\baru} \right) - 
 \eta \left( 1-\baru, \frac{-1}{u_0+\baru} \right) \, 
 \ln \left( \frac{u_0+1}{u_0+\baru} \right) 
\notag \\
&\quad {} \quad {} \quad {} \quad {}- 
 \dilog \left( \frac{u_0+1}{u_0-\baru} \right) - 
 \eta \left( 1+\baru, \frac{-1}{u_0-\baru} \right) \, 
 \ln \left( \frac{u_0+1}{u_0-\baru} \right) 
\notag \\
&\quad {} \quad {} \quad {} \quad {} + 
 i \, \pi \, S(u_0-\baru) \, 
 \left[ \ln(u_0-\baru)-\ln(u_0+1) \right] 
\notag \\
&\quad {} \quad {} \quad {} \quad {}+ 
 i \, \pi \, S(u_0+\baru) \, 
 \left[ \ln(u_0+\baru)-\ln(u_0+1) \right] + 
 \pi^2 
\notag \\
&\quad {} \quad {}\quad {} \quad {} +  
\left[ 
 \eta \left( A,1-\baru^2  \right) -  \eta \left( A,u^2_0-\baru^2  \right) - 
 \eta \left( u_0-\baru,u_0+\baru \right) 
\right] 
\notag \\
&\quad {} \quad {} \quad {} \quad {}  \quad {} \times 
\left.
 \left[ \ln \left( 1 + u_0 \right) - \ln \left( 1 - u_0 \right) \right]
 \vphantom{\ln \left( \frac{\Lambda - \baru}{1 - \baru} \right)}
\right\}
\label{eqdefkrab7}
\end{align}
The dilogarithms of eq. (\ref{eqdefkrab7}) 
are - up to an overall sign minus - the same than those appearing in the 
$\calf$ function (c.f. eq. (\ref{P1-eqdefcalf2a}) of ref. \cite{paper1}). We thus force the 
appearance of $\calf$ by introducing the necessary extra $\eta$ functions 
and, noting that $\eta (1+u_0,1/(1-u_0)) = 0$ we rewrite eq.(\ref{eqdefkrab7}) 
as:
\begin{align}
K^C_{1,\infty}(A,B,u_0^2) 
&= \frac{1}{2 \, u_0} 
\left\{ 
 - \calf(u_0,\baru) + i \pi \, S(\baru)  
 \left[ 
  \ln \left( \frac{u_0+1}{u_0-\baru} \right) 
  - 
  \ln \left( \frac{u_0+1}{u_0+\baru} \right) 
 \right] + \pi^2
\right.
\notag \\
&\qquad \quad \quad {} + 
i \pi \, S(u_0-\baru) \, \eta \left( u_0+1,\frac{1}{u_0-\baru} \right)
\notag \\
&\qquad \quad \quad {} + 
i \pi \, S(u_0+\baru) \, \eta \left( u_0+1,\frac{1}{u_0+\baru} \right) 
\notag \\
&\qquad \quad \quad {} + 
 \left[ 
  \eta \left( A,1-\baru^2  \right) -  
  \eta \left( A,u^2_0-\baru^2  \right) - 
  \eta \left( u_0-\baru,u_0+\baru \right) 
 \right] 
\notag \\
&\qquad \qquad \quad {} \times
\left.
 \ln \left( \frac{1+u_0}{1-u_0} \right)
\right\}
\label{eqdefkrab8}
\end{align}
Eq.~(\ref{eqdefkrab8}) is obtained by noting that, for any two complex
numbers $a$ and $b$:
\begin{equation}
  \eta(a,b)-\eta(-a,-b) = - i \, \pi \, \left[ S(a) + S(b) \right]
  \label{eqdefdiffeta}
\end{equation}
Rearranging the logarithmic terms  and noting that $(S(u_0+\baru)-S(\baru)) \, \eta(u_0+1,1/(u_0+\baru))$ 
as well as $(S(u_0-\baru)+S(\baru)) \, \eta(u_0+1,1/(u_0-\baru))$ always vanish, we end up with:
\begin{align}
K^C_{1,\infty}(A,B,u_0^2) 
&= \frac{1}{2 \, u_0} \, 
\left\{ 
 \vphantom{\ln \left( \frac{\Lambda - \baru}{1 - \baru} \right)}
 - \calf(u_0,\baru) 
+ i \, \pi 
\left[ 
  S(\baru) \, \left[ \ln \left( u_0+\baru \right) - \ln \left( u_0-\baru \right) \right] - i \pi
\right]
\right.
\notag \\
&\qquad \quad \quad {}  
+ 
 \left[ 
  \eta \left( A,1-\baru^2  \right) -   
  \eta \left( A,u^2_0-\baru^2  \right) - 
  \eta \left( u_0-\baru,u_0+\baru \right) 
 \right]
 \notag \\
&\qquad \qquad \quad {} 
\left.
\times  \ln \left( \frac{1+u_0}{1-u_0} \right) 
\right\}
\label{eqdefkrab9}
\end{align}

\subsection{$a=0,b=+\infty$}

The computation of $N$-point functions in the complex mass case
also involves integrals of the following third kind:
\begin{align}
K^C_{0,\infty}(A,B,u_0^2)
&= \int^{\infty}_0 du \, 
\frac{\ln(A \, u^2 + B) - \ln(A \, u^{2}_0 + B)}{u^2 - u_0^2}
\label{eqdefkrab3}
\end{align}
where $A$, $B$ and $u_0^2$ are complex numbers such that $\Im(A \, u^2 + B)$ 
keeps a constant sign while $u$ spans the range $[0,\infty[$ along the real 
axis. Under the assumption made, $K^C_{0,\infty}$ can be split as:
\begin{align}
K^C_{0,\infty}(A,B,u_0^2)&= K^C_{0,1}(A,B,u_0^2) + K^C_{1,\infty}(A,B,u_0^2)
\label{eqdefkrab31}
\end{align}
with the expressions of $K^C_{0,1}(A,B,u_0^2)$ and $K^C_{1,\infty}(A,B,u_0^2)$ computed 
above in eqs. (\ref{eqdefkrab4}) and (\ref{eqdefkrab9}) respectively. In the 
sum, the $\calf$ contribution drops out so that 
$K^C_{0,\infty}(A,B,u_0^2)$ contains only logarithmic terms:
\begin{align}
K^C_{0,\infty}(A,B,u_0^2) 
&= \frac{1}{2 \, u_0} \, 
\biggl\{  
 i \, \pi \, \left[ S(\baru) \, 
 \left( \ln \left( u_0+\baru \right) - \ln \left( u_0-\baru \right) \right)  
 - i \pi \right. 
\notag \\
&\qquad \quad \quad {} + \left.
  \left(  \eta(A,1-\baru^2)
   - \eta \left( A,u^2_0-\baru^2  \right) - 
 \eta \left( u_0-\baru,u_0+\baru \right) \right) \, S(u_0) \right] 
\notag \\
&\qquad \quad \quad {} + 
\left[ \eta(A,-\baru^2) - \eta(A,1-\baru^2) \right] \, \ln \left( \frac{u_0-1}{u_0+1} \right)
\biggr\}
\label{eqdefkrab32}
\end{align}

\noindent
By assumption the sign of $\Im(A \, u^2 + B))$ is constant 
when $u$ spans the range $[0,+\infty[$, which means that $S(A)=S(B)$. As $S(B/A) = S(1 + B/A)$
then $\eta(A,B/A) = \eta(A,1+B/A)$ or in terms of $\baru$ $\eta(A,-\baru^2) = \eta(A,1-\baru^2)$, 
so the eq.~(\ref{eqdefkrab32}) simplifies:
\begin{align}
K^C_{0,\infty}(A,B,u_0^2) 
&= \frac{i \, \pi}{2 \, u_0} \, 
\biggl[  
 S(\baru) \, 
 \left[ \ln \left( u_0+\baru \right) - \ln \left( u_0-\baru \right) \right] 
 - i \pi 
\notag \\
&\qquad \quad \quad {} + 
 \left[ 
  \eta \left( A,-\baru^2  \right) - \eta \left( A,u^2_0-\baru^2  \right) - 
  \eta \left( u_0-\baru,u_0+\baru \right) \right] \, S(u_0) 
\biggr]
\label{eqdefkrab33}
\end{align}
A comment is in order here. 
We used the ``trick" (\ref{eqdefkrab31}) to obtain eq. 
(\ref{eqdefkrab32}) in an economical way. One shall be cautious that 
practical calculations, especially
of four-point functions with general complex masses, involve 
$K^C_{0,\infty}(A,B,u_0^2)$ and $K^C_{1,\infty}(A^{\prime},B^{\prime},u^{\prime \, 2}_0)$ where the
arguments $(A,B,u_0^2)$ differ from $(A^{\prime},B^{\prime},u^{\prime \, 2}_0)$ so that no cheap
simplification can be made. A closer look reveals though that some pairs
of $K^C_{0,\infty}(A,B,u_0^2)$ and $K^C_{1,\infty}(A^{\prime},B^{\prime},u^{\prime \, 2}_0)$ may be
combined using Cauchy's theorem into analytic continuations of some 
$K^C_{0,1}(A^{\prime},B^{\prime},u^{\prime \, 2}_0)$ defined by contour integrals along 
some deformations $\widehat{(0,1)}$ of the segment $[0,1]$ designed to wrap 
the cuts of the logarithms $\ln(A^{\prime} \, u^2 + B^{\prime})$. In this
respect see also the discussion at the end of appendix \ref{appendJ}.

\section{Prescription for the imaginary part of $\dets$: general complex mass case}\label{ImofdetS}

\noindent
This appendix extends the appendix \ref{P1-ImofdetS} of ref. \cite{paper1} for the complex mass case.
Let us recap the result found in \cite{paper1} (cf. eq. \ref{P1-e7-0}) 
\begin{align}
\det \left[ {\cal S} + i \lambda E \right]
&= 
\dets + i \lambda  \, B \, \dets
\nonumber\\
&=
\dets + i \lambda \, (-1)^{N-1} \detg 
\label{e7-0}
\end{align}
with
$E_{ij} = 1 \;\; \mbox{for all} \; i,j = 1, \cdots, N$.
It holds whether $\lambda$ is 
infinitesimal or finite: it thus tells the sign of $\Im(\dets)$ also
for the particular complex mass case where the imaginary parts of all internal 
masses squared would be equal; however it is not enough to extract the sign of 
the imaginary part of $\dets$ in the general complex mass case. This general
case is addressed below, and contains the one in ref. \cite{paper1} as a particular subcase.
Let us note, in this appendix, $Q_R = \Re(Q)$ and $Q_I = \Im(Q)$ for any complex number $Q$.

\vspace{0.3cm}

\noindent
As shown in the appendix~\ref{P1-detsdetg} (cf.\ eq.~(\ref{P1-e10-0}) of ref.\ \cite{paper1}), the determinant 
of the $\cals$ matrix can be expressed in term of the determinant of the Gram 
matrix $G^{(N)}$ obtained by singling out the line and column $N$ of the $\cals$ 
matrix, as well as the matrix of cofactors of $G^{(N)}$ $\mbox{Cof} \left[ G^{(N)} \right]$: 
\begin{eqnarray}
\dets 
& = &
(-1)^{N-1} 
\left\{ 
 {\cal S}_{NN} \, \detg 
 +
 V^{(N) \, T} \cdot \mbox{Cof} \left[ G^{(N)} \right] \cdot V^{(N)}
\right\}
\label{e10-0}
\end{eqnarray}
The matrix $G^{(N)}$ is {\em real}, since all internal masses appearing in 
$\cals$ cancel among one another in the expression of $G^{(N)}$, 
all the imaginary parts are thus located in ${\cal S}_{NN}$ and the $V^{(N)}$ and more
precisely, the imaginary part of $\dets$ is {\em linear} in the 
latter and given by:
\begin{align}
\Im \left[ \dets \right]
&=
(-1)^{N-1} \, 
\left\{ 
\cals_{NN \, I} \, \detg 
\right.
\nonumber\\
&\qquad \qquad \qquad 
 \left.
 {} + \; 
 2 \, V^{(N) \, T}_R
 \cdot \mbox{Cof} \left[ G^{(N)} \right] \cdot 
 V^{(N)}_I
\right\}
\label{e10}
\end{align}
Notice that because of the definition of the vector $V^{(N)}$ (cf.\ eq.~(\ref{eqVJA3})), 
the components of its imaginary part are just the difference of the imaginary 
parts of two masses squared: 
$V_{I \, j}^{(N)} = - \, (m^{2}_{I \, j} \, - \, m^{2}_{I \, N})$ for  
$j=1, \cdots, N-1$. Furthermore, we have that ${\cal S}_{NN \, I} = - \, 2 \, m^{2}_{I \, N}$.
In the particular case where all masses squared
have the same\footnote{In particular when all masses squared are real, the 
Feynman contour prescription effectively provides a common infinitesimal 
$\Im$ part $- \, \lambda$.} imaginary part $m_{I}^{2}$ which is negative, we have:
$
\cals_{NN} = - 2 \, m_{I}^{2}, \;\; 
V^{(N)}_I = 0
$
and we recover the result (\ref{P1-e7-0}) of \cite{paper1}.
However in the general complex mass case, the sign of 
$\Im \left[ \dets \right]$ is a more complicated function 
of the imaginary parts of the masses squared and of the Gram matrix whose sign 
depends on the kinematics. 
 
\vspace{0.3cm}

\noindent
According to appendix~\ref{P1-detsdetg} of \cite{paper1} the subtraction of the line 
and the column $N$ leads from the $\cals$ matrix to the block matrix $\widehat{\cals}^{(N)}$ written as:
  \begin{align}
    \widehat{\cals}^{(N)}
      &=
      \left[
       \begin{array}{ccc}
         {} - G^{(N)} & | & V^{(N)} \\
        --     & + & - \\
        V^{(N) \, T}  & | & \cals_{NN} \\
      \end{array}
      \right]
      \label{e7}
  \end{align}
This matrix can be decomposed into its real and imaginary parts $\widehat{\cals}^{(N)}_R$ 
and $\widehat{\cals}^{(N)}_I$. Due to the fact that the Gram matrix $G^{(N)}$ is real, the blocks constituting the last two matrices are:
\begin{align}
  \widehat{\cals}^{(N)}_R
    &=
    \left[
     \begin{array}{ccc}
       {} - G^{(N)} & | & V^{(N)}_R \\
      --     & + & - \\
      V^{(N) \, T}_R  & | & \cals_{NN \, R} \\
    \end{array}
    \right]
    \label{e70} \\
  \widehat{\cals}^{(N)}_I
    &=
    \left[
     \begin{array}{ccc}
       {} 0 & | & V^{(N)}_I \\
      --     & + & - \\
      V^{(N) \, T}_I  & | & \cals_{NN \, I} \\
    \end{array}
    \right]
    \label{e71}
\end{align}
Let us note $b_{[R] \,j}, j=1, \cdots, N$ the reduction 
coefficients 
solving the equation:
\begin{equation}
  {\cal S}_{R} \cdot b_{[R]}  = \left[
    \begin{array}{c}
      1 \\
      \vdots \\
      1
    \end{array}
  \right]
  \label{eqsolsrbr}
\end{equation}
and 
$B_{[R]} = \sum_{j=1}^{N} b_{[R] \, j}$. 
With the help of eqs.~(\ref{P1-inv-inv1}), (\ref{P1-inv-inv2}) and (\ref{P1-eqsolbbar}) in appendix~\ref{P1-detsdetg} of \cite{paper1}, the solution of eq.~(\ref{eqsolsrbr}) 
reads\footnote{The solution is given for non vanishing $\detg$, for special cases see ref.~\cite{paper1}. Let us remind that $\bbar_{[R]} = b_{[R]} \, \det \left( \cals_{R} \right)$.}:
\begin{align}
  B_{[R]} &= (-1)^{N-1} \, \frac{\detg}{\det (\cals_R)} \label{eqdefbr0} \\
  b_{[R] \, j} &= (-1)^{N-1} \, \frac{1}{\det (\cals_R)} \, \left( \text{Cof} \left[ G^{(N)} \right] \cdot V^{(N)}_R
 \right)_j
  \label{eqdefbrj0}
\end{align}
we wrote this solution in a way such that it is well behaved in the case where $\detg = 0$, 
cf.\ appendix~\ref{P1-detsdetg} of \cite{paper1}.
Putting the expressions for $V_{I}^{(N)}$ and ${\cal S}_{NN \, I}$ into eq.~(\ref{e10}) 
and using the eqs.~(\ref{eqdefbr0}) and (\ref{eqdefbrj0}), we get an appealing form 
for the imaginary part of $\dets$:
\begin{align}
  \Im \left[ \dets \right] &= - 2 \, \det \left( \cals_R \right) \, \sum_{j=1}^{N} m^{2}_{I \, j} \, b_{[R] \,j} \notag \\
  &= - 2 \, \sum_{j=1}^{N} m^{2}_{I \, j} \, \bbar_{[R] \,j}
  \label{eqimdets1}
\end{align}
When the $m_{I \, j}^{2}$ are all equal eq. (\ref{eqimdets1}) reduces to 
$ (-1)^{N-1} \, (- 2 \, m^{2}_{I}) \, \detg$ whose sign is readily that 
obtained from $\detg$.
In our convention $m^2_I$ is negative and so the sign of the imaginary part of $\dets$ in this case
  is the same as the one appearing in the real mass case as it should be (cf. appendix \ref{P1-ImofdetS} of \cite{paper1}).
When the 
$m_{I \, j}^{2}$ are unequal, the sign of eq. (\ref{eqimdets1})
is not explicit and may differ from the sign of $(-1)^{N-1} \, \detg$ depending on the 
kinematics, if the various $b_{[R] \, j}$ happen to have different signs. 

\vspace{0.3cm}

\noindent
Let us stress that
the $b_{[R]}$ are {\em not} the real parts of 
the $b$, {\em nor} is $\det ({\cal S}_{R})$ the real part of 
$\det( {\cal S})$ in general, namely 
\[
\Re(\dets) - \det ( {\cal S}_{R}) 
\propto 
\left( V_{I}^{(N) \, T} \cdot \mbox{Cof}[G^{(N)}] \cdot V_{I}^{(N)} \right)
\]
Yet, when $V_{I}^{(N)} = 0$ regardless of ${\cal S}_{NN \, I}$,
\[
  b = \frac{\det ( {\cal S}_{R})}{\dets} \, b_{[R]}
\]

\section{Location of cuts in the complex mass case}\label{cut}

Let us consider $\ln(A \, z^2 + B)$ for two complex numbers $A$ and
$B$ whose imaginary parts are non vanishing.
We borrow from appendix \ref{appF} the function $S(z) \equiv \sign(\Im(z))$.
We want to determine what are the conditions on $A$ and $B$ 
in order that one of the cuts of the logarithm crosses the real axis between $0$ and $1$.
A necessary condition for that is given by the fact that the imaginary part of $A \, z^2 + B$ changes its sign when $z$ spans the real segment $[0,1]$, 
which translates into $S(A+B) \ne S(B)$. This last inequality implies that $S(A) \ne S(B)$. 
For any complex $Q$ let us note 
$Q_R \equiv \Re(Q)$ and $Q_I \equiv \Im(Q)$.
With this new notation, the two conditions on the signs of the imaginary parts of $A$, $B$ and $A+B$ reads:
  \begin{equation}
    0 < - \, \frac{B_I}{A_I} < 1
    \label{eqnesscond1}
  \end{equation}
The logarithm considered has two cuts 
located where the two following conditions are simultaneously fulfilled:
\begin{eqnarray}
\Im(A \, z^2 + B) & = & 0
\label{c1}\\
\Re(A \, z^2 + B) &\leq& 0
\label{c2}
\end{eqnarray}
The branch points are located where the last inequality saturates i.e. at
$z_{\pm} = \pm \, (-B/A)^{1/2}$. Eq. (\ref{c1}) reads:
\[
A_I \, (z^2)_R + A_R \, (z^2)_I + B_I = 0
\] 
which is solved in $(z^2)_R$ parametrically in $(z^2)_I$ by:
\begin{equation}\label{e3}
(z^2)_R 
= 
- \, \frac{1}{A_I} \left[ A_R \, (z^2)_I + B_I \right]
\end{equation}
Substituting eq. (\ref{e3}) into the l.h.s. of ineq. (\ref{c2}) we get
\footnote{We could have chosen just as well $(z_{-}^2)_I$ instead of $(z_{+}^2)_I$ 
in eq.~(\ref{e4}) but, in the rest, we will consider only the cut in the half plane $z_R > 0$.}:
\begin{eqnarray}
\mbox{l.h.s. (\ref{c2})}
& = &
- \, S(A) \, \frac{|A|^2}{|A_I|} \left[ (z^2)_I - (z_{+}^2)_I \right]
\label{e4}
\end{eqnarray}
Since $(z^2)_I = 2 \, z_R \, z_I$, the cut in the half complex plane 
$\{z_R \geq 0\}$ slashes through $\infty$ in the quadrant 
$\{z_R > 0, S(A) \, z_I > 0\}$ whatever the sign of $z_{+ \, I}$.
This cut crosses the real segment $[0,1]$ at
the point $z_c = (- B_I/A_I)^{1/2}$
if and only if the branch point $z_{+}$ is {\em not} in the quadrant 
slashed through to $\infty$ by the cut i.e. if and only if $S(A) \, z_{+ \, I} < 0$, i.e.
if and only if $S(A) \, (z_{+}^2)_{I} < 0$. The latter condition reads explicitly:
\begin{equation}
S(A) \left[ B_I \, A_R - B_R \, A_I \right] < 0
  \label{eqcondnecsuf}
\end{equation}
So, the conditions necessary and sufficient for the cut in the half complex plane $\{z_R \geq 0\}$
to cross the real segment $[0,1]$ are given by eqs. (\ref{eqnesscond1}) and (\ref{eqcondnecsuf}).

\vspace{0.3cm}

\noindent
Although the goal of this appendix has been reached, more details about
the nature of the support of the cut and its parametrisation in term of $z_I$ are given in the rest of this appendix.

\vspace{0.3cm}

\noindent
Eq. (\ref{c1}) alternatively reads:
\begin{equation}\label{e5}
A_I \, (z_R)^2 + 2 \, A_R \, z_I \, z_R + 
\left[ B_I - A_I \, (z_I)^2 \right] = 0
\end{equation} 
It is the Cartesian equation of a hyperbola since the coefficients
of $(z_R)^2$ and $(z_I)^2$ are opposite. The two branches of hyperbola are
symmetric to each other w.r.t. the origin as (\ref{e5}) is invariant under the
parity transformation $z_R \to - \, z_R, z_I \to - \, z_I$.
Let us solve eq. (\ref{e5}) in  $z_R$ parametrically in $z_I$. 
The discriminant $\Delta^{\prime}$ given by
\[
\Delta^{\prime} = |A|^2 \, (z_I)^2 - B_I \, A_I > 0
\]
is manifestly $>0$ for $z_I$ spanning all $\mathds{R}$, so that the two roots are 
real for all $z_I$ real. The product of these roots
given by
\[
\Pi = - \, \left[ (z_I)^2 + z_c^2 \right] < 0
\] 
indicates that the two roots have opposite signs. Let us label them
$\overline{z}_{R \, \xi}$, $\xi = \pm$ as:
\begin{eqnarray}
\overline{z}_{R \, \xi} 
= \frac{- S(A) \, A_R \, z_I + \xi \sqrt{\Delta^{\prime}}}{|A_I|}
\label{e6}
\end{eqnarray}
and for all $z_I$ real we have:
\[
\overline{z}_{R \, -} < 0 < \overline{z}_{R \, +}
\]
We now focus on $\overline{z}_{R \, +}$. 
The variation of $\overline{z}_{R \, +}$ with $z_I$ is captured by computing
\begin{align}
\frac{\partial}{\partial z_I} \, \overline{z}_{R \, +}
&= 
\frac{1}{|A_I|} 
\left[
 - S(A) \, A_R + \frac{|A|^2 \, z_I}{\sqrt{\Delta^{\prime}}}
\right]
  \label{eqderivzrplus}
\end{align}
whose sign is given by
\begin{equation}
- S(A) \, A_R \, \sqrt{\Delta^{\prime}} + |A|^2 \, z_I
\label{e7a}
\end{equation}
Let us compute
\begin{eqnarray}
{\cal D} 
& = &
- A_R^2 \, \Delta^{\prime} + |A|^4 \, (z_I)^2
\nonumber\\
& = & 
 A_I^2 \, |A|^2 \,
\left[
 (z_I)^2 - z_{I \, *}^2
\right]
\label{e8a}\\
\mbox{with} \;\;
z_{I \, *}
& = &
\frac{|A_R|}{|A|} \, z_c
\label{e8b}
\end{eqnarray}
so that 
\[
{\cal D} \; \mbox{is} \; 
\left\{
 \begin{array}{clcl}
   > 0 & \mbox{when} \; (z_I)^2 > z_{I \, *}^2 & : &
  \mbox{$z_I$ imposes its sign to} \; 
  (\partial \overline{z}_{R \, +}/\partial z_I) \\
   < 0&  \mbox{when} \; (z_I)^2 < z_{I \, *}^2 & : &
  \mbox{$- \, S(A) \, A_R$ imposes its sign to} \; 
  (\partial \overline{z}_{R \, +}/\partial z_I) 
 \end{array}
\right.
\]
Thus,\begin{itemize}
\item
if $- \, S(A) \, A_R < 0$, 
$\overline{z}_{R \, +} \searrow$ for $z_I < z_{I \, *}$, 
$\overline{z}_{R \, +}$ reaches its minimum 
$(\overline{z}_{R \, +})_{min}$ at $z_I = z_{I \, *}$, 
$\overline{z}_{R \, +} \nearrow$ for $z_I > z_{I \, *}$;
\item
if $- \, S(A) \, A_R > 0$, 
$\overline{z}_{R \, +} \searrow$ for $z_I < - \, z_{I \, *}$, 
$\overline{z}_{R \, +}$ reaches its minimum 
$(\overline{z}_{R \, +})_{min}$ at $z_I = - \, z_{I \, *}$, 
$\overline{z}_{R \, +} \nearrow$ for $z_I > - \, z_{I \, *}$
\end{itemize}
As can be checked explicitly, the minimum $(\overline{z}_{R \, +)_{min}}$
takes the same analytic expression in both cases, and the latter is given by:
\begin{equation}\label{e9}
(\overline{z}_{R \, +})_{min}
=
\frac{|A_I|}{|A|} \, z_c 
\end{equation}
which is manifestly in $]0, z_c[$ ($z_c < 1$ being also the value of 
$\overline{z}_{R \, +}$ when $z_I=0$).

\vspace{0.3cm}

\noindent
When $z_I$ is $< 0$ and large, 
\[
\overline{z}_{R \, +} 
\sim 
- \, \frac{S(A) \, A_R + |A|}{|A_I|} \, z_I 
\;\; \to + \infty
\]
When $z_I$ is $> 0$ and large, 
\[
\overline{z}_{R \, +} 
\sim 
+ \, \frac{- \, S(A) \, A_R + |A|}{|A_I|} \, z_I 
\;\; \to + \infty
\]
Let us show that $z_I \, \overline{z}_{R \, +}$ is a monotonously growing 
function of $z_I$. The derivative w.r.t. $z_I$ is given by
\begin{eqnarray}
\lefteqn{\frac{\partial}{\partial z_I} (z_I \, \overline{z}_{R \, +} )}
\nonumber\\
& = &
\frac{1}{|A_I|} 
\left\{
 - \, S(A) \, A_R \, z_I + \sqrt{\Delta^{\prime}} +
 z_I 
 \left[ 
  - \, S(A) \, A_R  + \frac{|A|^2 \, z_I}{\sqrt{\Delta^{\prime}}}
 \right]
\right\}
\nonumber\end{eqnarray}
whose sign the same as 
$(\Delta^{\prime} + |A|^2 \, z_I^2) - 
2 \, S(A) \, A_R \, z_I \, \sqrt{\Delta^{\prime}}$, which is $>0$ since
\begin{eqnarray}
\lefteqn{\left( \Delta^{\prime} + |A|^2 \, z_I^2 \right)^2 -
4 \, A_R^2 \, z_I^2\, \Delta^{\prime}}
\nonumber\\
&=& 
A_I^2 
\left[ 4 \, |A|^2 z_I^4 - 4 B_I \, A_I \, z_I^2 + B_I^2 \right]
\nonumber\\
& >& 0    \quad {}\quad {}\quad {}\quad {}\quad {}\quad {} q.e.d.
\nonumber
\end{eqnarray}
Therefore, the r.h.s. of eq. (\ref{e4}) is $- \, S(A) \, \times$ a monotonously
growing function of $z_I$ while $z_I$ spans all $\mathds{R}$. It vanishes once
i.e. at the branch point $z_{+}$. Its sign is $S(A)$ when $z_I < z_{+ \, I}$ 
and $- \, S(A)$ when $z_I > z_{+ \, I}$. The cut corresponds to the arc such
that this sign is ``$-$".

\bibliographystyle{unsrt}
\bibliography{../biblio,../publi}

\end{fmffile}

\end{document}